\documentclass[traditabstract,longauth]{aa}
\usepackage{natbib}
\usepackage{wasysym}
\usepackage{txfonts}
\usepackage{rotating}
\usepackage{pdflscape}
\newcommand{\corot}{{CoRoT}}
\newcommand{\kepler}{\emph{Kepler}}
\newcommand{\Ctone}{{CoRoT-1b}}

\newcommand{\Ctthree}{{CoRoT-3b}}
\newcommand{\Ctseven}{{CoRoT-7b}}

\newcommand{\Ctten}{{CoRoT-10b}}
\newcommand{\Ctsv}{{CoRoT-17b}}
\newcommand{\Cttwt}{{CoRoT-22b}}
\newcommand{\Cttwf}{{CoRoT-24b}}
\newcommand{\Cttwfv}{{CoRoT-25b}}
\newcommand{\Cttws}{{CoRoT-26b}}
\newcommand{\Cttwh}{{CoRoT-28b}}

\newcommand{\Rs}{\ensuremath{R_{\star}}}

\newcommand{\Rp}{\ensuremath{R_{\rm p}}}
\newcommand{\Msun}{\ensuremath{M_{\odot}}}             

\renewcommand{\degr}{\ensuremath{^\circ}}

\newcommand{\MJ}{\ensuremath{M_{\rm Jup}}}

\newcommand{\RE}{\ensuremath{R_{\oplus}}}

\newcommand{\totalLC}{177\,454}
\newcommand{\totaltargets}{163\,665}
\newcommand{\totaldw}{101\,083}
\newcommand{\totalV}{61\,174}
\newcommand{\totalEBfilt}{1653}
\newcommand{\totalEBfiltper}{1561}
\newcommand{\totalCB}{616}
\newcommand{\totalmono}{92}
\newcommand{\totaleccB}{243}   

\newcommand{\totalCandnoP}{557}
\newcommand{\totalCandfilt}{594} 
\newcommand{\totalCand}{626} 
\newcommand{\totalfeat}{4123}
\newcommand{\totalunres}{193}
\newcommand{\totalunresfl}{119}

\newcommand{\totalfu}{406}
\newcommand{\totalbin}{2269}
\newcommand{\totalGh}{211}
\newcommand{\totalPls}{115}
\newcommand{\totalFd}{499}
\newcommand{\totalFA}{824} 

\begin{document}

\title{Planets, candidates, and binaries from the CoRoT/Exoplanet programme\thanks{The CoRoT space mission, launched on December 27th 2006, has been developed and is operated by CNES, with the contribution of Austria, Belgium, Brazil , ESA (RSSD and Science Programme), Germany and Spain.}}

   \subtitle{The CoRoT transit catalogue}

\author{
M. Deleuil\inst{\ref{lam}}
\and S. Aigrain \inst{\ref{oxford}}
\and C. Moutou\inst{\ref{lam}} 
\and J. Cabrera\inst{\ref{dlr}}
\and F. Bouchy\inst{\ref{lam},\ref{Geneve}}
\and H.J., Deeg\inst{\ref{iac},\ref{UnivCan}}
\and J.-M. Almenara\inst{\ref{lam},\ref{Geneve}} 
\and G. H\'ebrard\inst{\ref{ohp},\ref{iap}}
\and A. Santerne\inst{\ref{lam}}
\and R. Alonso\inst{\ref{UnivCan}} 
\and A.S. Bonomo\inst{\ref{inaf}}
\and P. Bord\'e\inst{\ref{bord}} 
\and Sz. Csizmadia\inst{\ref{dlr}}
\and A. Erikson\inst{\ref{dlr}}
\and M. Fridlund\inst{\ref{leiden},\ref{Onsala}}
\and D. Gandolfi\inst{\ref{torino},\ref{heild}}
\and E. Guenther\inst{\ref{tls}}
\and T. Guillot\inst{\ref{oca}}
\and P. Guterman\inst{\ref{lam}} 
\and S. Grziwa\inst{\ref{inst18}}
\and A. Hatzes\inst{\ref{tls}}
\and A. L\'eger\inst{\ref{ias}}
\and T. Mazeh\inst{\ref{telaviv}}
\and A. Ofir\inst{\ref{goet},\ref{rehovot}}
\and M. Ollivier\inst{\ref{ias}}
\and M. P\"atzold\inst{\ref{inst18}}
\and H. Parviainen\inst{\ref{iac},\ref{UnivCan}}
\and H. Rauer\inst{\ref{dlr},\ref{zaa}}
\and D. Rouan\inst{\ref{lesia}}
\and J. Schneider\inst{\ref{luth}}
\and R. Titz-Weider\inst{\ref{dlr}}
\and B. Tingley\inst{\ref{aarhus}}
\and J. Weingrill\inst{\ref{sri},\label{aip}}
             }

\institute{
Aix Marseille Universit\'e, CNRS, CNES, LAM (Laboratoire d'Astrophysique de Marseille) UMR 7326, 13388, Marseille, France, \email{magali.deleuil@lam.fr} \label{lam}
\and Department of Physics, Denys Wilkinson Building Keble Road, Oxford, OX1 3RH\label{oxford}
\and Institute of Planetary Research, German Aerospace Center, Rutherfordstrasse 2, 12489 Berlin, Germany\label{dlr}
\and Observatoire de Gen\`ve, Universit\'e de Gen\`eve, 51 Ch. des Maillettes, CH-1290 Sauverny, Switzerland\label{Geneve}
\and Instituto de Astrofisica de Canarias, E-38205 La Laguna, Tenerife, Spain\label{iac}
\and Universidad de La Laguna, Dept. de Astrof\'\i sica, E-38200 La Laguna, Tenerife, Spain\label{UnivCan}
\and Observatoire de Haute Provence, 04670 Saint Michel l'Observatoire, France\label{ohp}
\and Institut d'Astrophysique de Paris, 98bis boulevard Arago, 75014 Paris, France\label{iap}
\and Osservatorio Astrofisico di Torino, Strada Osservatorio, 20, 10025 Pino Torinese, Italy\label{inaf}
\and Laboratoire d'astrophysique de Bordeaux, Univ. Bordeaux, CNRS, B18N, all\'ee Geoffroy Saint-Hilaire, 33615 Pessac, France\label{bord}
\and Leiden Observatory, University of Leiden, PO Box 9513, 2300 RA, Leiden, The Netherlands\label{leiden}
\and Department of Earth and Space Sciences, Chalmers University of Technology, Onsala Space Observatory, 439 92, Onsala, Sweden\label{Onsala} 
\and Dipartimento di Fisica, Universit\'a di Torino, via Pietro Giuria 1, I-10125, Torino, Italy\label{torino}
\and Landessternwarte K\"onigstuhl, Zentrum f\"ur Astronomie der Universit\"at Heidelberg, K\"onigstuhl 12, 69117, Heidelberg, Germany\label{heild}
\and Th\"uringer Landessternwarte, Sternwarte 5, Tautenburg 5, D-07778 Tautenburg, Germany\label{tls}
\and Observatoire de la C\^ote d'Azur, Laboratoire Cassiop\'ee, BP 4229, 06304 Nice Cedex 4, France\label{oca}
\and Rheinisches Institut f\"ur Umweltforschung an der Universit\"at zu K\"oln, Aachener Strasse 209, 50931, Germany\label{inst18} 
\and Institut d'Astrophysique Spatiale, Universit\'e Paris-Sud 11 \& CNRS (UMR 8617), B\^at. 121, 91405 Orsay, France\label{ias}
\and School of Physics and Astronomy, Tel Aviv University, Tel Aviv, Israel\label{telaviv} 
\and Department of Earth and Planetary Sciences, Weizmann Institute of Science, Rehovot, 76100, Israel\label{rehovot}
\and Institut f\"ur Astrophysik, Georg-August-Universit\"at, Friedrich-Hund-Platz 1, 37077 G\"ottingen, Germany\label{goet}
\and Zentrum f{\"u}r Astronomie und Astrophysik, TU Berlin, Hardenbergstra{\ss}e 36, D-10623 Berlin\label{zaa}
\and LESIA, Obs de Paris, Place J. Janssen, 92195 Meudon cedex, France\label{lesia}
\and LUTH, Observatoire de Paris, CNRS, Universit\'e Paris Diderot; 5 place Jules Janssen, 92195 Meudon, France \label{luth}
\and Department of Physics and Astronomy, Aarhus University, 8000 Aarhus C, Denmark\label{aarhus}
\and Space Research Institute, Austrian Academy of Science, Schmiedlstr. 6, A-8042 Graz, Austria\label{sri} 
\and Leibniz-Institut f\"ur Astrophysik Potsdam (AIP), An der Sternwarte 16, 14482 Potsdam, Germany\label{aip}
}
\date{Received 28 April 2017 / accepted 05 February 2018}

  \abstract
{
  The \corot\ space mission observed \totaltargets\ stars over 26 stellar fields in the faint star channel. The exoplanet teams detected a total of \totalfeat\ transit-like features in the \totalLC\ light curves. We present the complete re-analysis of all these detections carried out with the same softwares so that to ensure their homogeneous analysis. Although the vetting process involves some human evaluation, it also involves a simple binary flag system over basic tests: detection significance, presence of a secondary, difference between odd and even depths, colour dependence, V-shape transit, and duration of the transit. We also gathered the information from the large accompanying ground-based programme carried out on the planet candidates and checked how useful the flag system could have been at the vetting stage of the candidates.\\
From the initial list of transit-like features, we identified and separated \totalFA\ false alarms of various kind, \totalbin\ eclipsing binaries among which \totalCB\ are contact binaries and \totalEBfilt\ are detached ones, 37 planets and brown dwarfs, and \totalCandnoP\ planet candidates. We provide the catalogue of all these transit-like features, including false alarms.  For the planet candidates, the catalogue gives not only their transit parameters but also the products of their light curve modelling: reduced radius, reduced semi-major axis, and impact parameter, together with a summary of the outcome of follow-up observations when carried out and their current status. For the detached eclipsing binaries, the catalogue provides, in addition to their transit parameters, a simple visual classification.\\
Among the planet candidates whose nature remains unresolved, we estimate that eight (within an error of three) planets are still to be identified. After correcting for geometric and sensitivity biases, we derived planet and brown dwarf occurrences and confirm disagreements with \kepler\ estimates, as previously reported by other authors from the analysis of the first runs: small-size planets with orbital period less than ten days are underabundant by a factor of three in the \corot\ fields whereas giant planets are overabundant by a factor of two. These preliminary results would however deserve further investigations using the recently released \corot\ light curves that are corrected of the various instrumental effects and a homogeneous analysis of the stellar populations observed by the two missions.}


   \maketitle
%

\section{Introduction}

The \corot\ space mission \citep{Baglin2006} operated from January 2007 to October 2012, with the two core science goals of discovering transiting exoplanets and probing the structure of stars through asteroseismology. During this period, the instrument photometrically monitored \totaltargets\ targets distributed over 26 stellar fields in two opposite regions in the galactic plane. It collected \totalLC\ light curves lasting between 21 and 152 days with some targets covered in two or more 
separate light curves. Their analysis provided a few thousand transit events that went through a complete screening process to identify astrophysical false positives (such as eclipsing binaries mimicking a planetary transit). The transit candidates are first subjected to a detailed light curve analysis, exploiting the excellent photometric precision and long, almost uninterrupted coverage characteristic of a space-based transit survey such as \corot. Surviving candidates are then included in an extensive programme of ground-based follow-up observations, aiming to weed out as many of the remaining false positives as possible, and to establish the planetary nature and measure masses for genuine planets. The complete screening process for each candidate can last more than a year, but it provides a useful insight into the nature and relative frequency of the false positive scenarios. 

Initially, individual papers were used to publish lists of transit candidates and eclipsing binaries, as well as with results from the follow-up observations from the fields IRa01 \citep{Carpano2009,Moutou2009}, LRc01 \cite{Cabrera2009}, LRa01 \citep{Carone2012} SRc01 \citep{Erikson2012}, and LRa03 and SRa03 \citep{Cavarroc2012}, with a comparison between predicted and observed rates of false positives from the first three long runs (IRa01 to LRa01) given by \cite{Almenara2009}. However, this procedure was discontinued in favour of collating all the candidates from all the stellar fields, and their status at the end of the follow-up programme, in a single location. This is the purpose of the present paper, which also provides a general summary of the results and a basic analysis of the statistical properties of the candidates and false positives. To ensure consistency, and because both the light curve analysis and the ground-based follow-up have improved over time, the fields  published in the aforementioned papers are also included in this study. 
\begin{figure*}[ht]
\label{fig:fields}
\begin{center} 
\includegraphics[width=9cm]{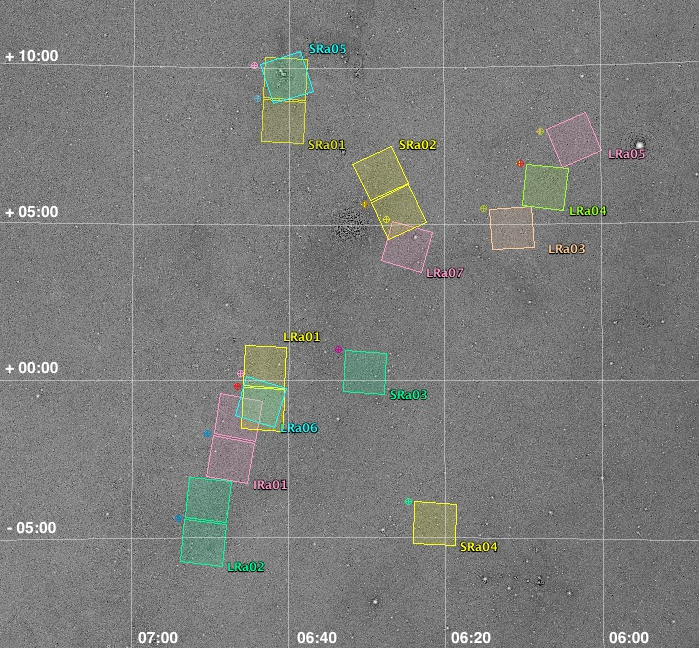}\includegraphics[width=9cm]{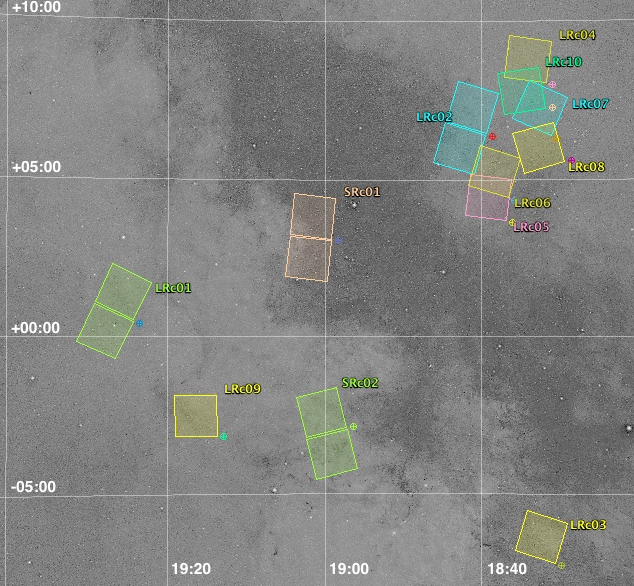}
\caption{Position of all the exoplanet fields observed by \corot\ in the Galactic anti-centre (left) and centre (right) directions.}
\end{center}
\end{figure*}

The remainder of this paper is structured as follows. The strategy of observation and the stellar fields that have been observed are summarised in Section~\ref{sec:fields}.  Section~\ref{sec:detvet} gives details of the methods used to detect transit events and vet them on the basis of their light curves only. The properties of the surviving candidates and the ground-based follow-up observations are reported in Section~\ref{sec:fup}. In Section~\ref{sec:hindsight}, we use the results of the follow-up programme to assess the effectiveness of the candidate vetting based on the light curves only. Candidates whose nature remains unsolved are discussed in Section~\ref{sec:unsolved}, while Section~\ref{sec:occurrence} presents some very simplified estimates of the occurrence of different kinds of planets based on the \corot\ results, and a comparison to published planet occurrence results. Our summary and conclusions are given in Section~\ref{sec:concl}.


\section{The \corot\ exoplanet mission profile}
A complete description of the mission profile and observations could be found in \cite{Baglin2016} but to make reading easier, we provide a quick description in the following subsections.

\label{sec:fields}
\subsection{Photometry in the exoplanet chanel}
\label{ssec:gen}

\corot's focal plane was equipped with four CCDs, each covering $1.3 \times 1.3$\degr. The exoplanet and seismology observations, which targeted stars of very different brightnesses, took place side by side, with two CCDs dedicated to each of the scientific objectives. The breakdown of the first data processing unit (DPU1), which occurred in March 2009, caused the loss of one CCD in each of the exoplanet and seismology channels, reducing the field of view by half. In the exoplanet channel, the satellite's on board processing and telemetry capacity enabled the observation of up to 6000 stars per CCD. Each was assigned a pixellised photometric aperture at the start of each run, which was selected automatically from a library of 254 pre-defined masks, so as to optimise the signal to noise ratio of the integrated flux \citep{Llebaria2006}. For these stars, photometry was carried out on board and only light curves were downloaded to Earth. In addition, twenty $10 \times 15$ pixel windows were used on each CCD to provide sky reference images and monitor the background level. A further 80 such windows, known as imagettes, were assigned to selected targets of interest, and were downloaded as pixel-level data to enable a finer photometric analysis on the ground. Nominally, the targets in the exoplanet channel have magnitude $11 \leq r \leq 16$, but a number of brighter stars were also observed, despite being saturated. Most of these were assigned an imagette to enable their photometry to be optimised on the ground. 

A prism was located in the optical path of \corot's exoplanet channel, so that each star produced a `mini-spectrum' on the focal plane. For stars with magnitude $r \leq 15$, the photometric aperture was divided along detector column boundaries into three regions corresponding approximately to the red, green and blue parts of the visible spectrum, and three-colour light curves were extracted and transmitted to Earth. These could then be summed together on the ground to give a `white' light curve. For stars with $r > 15$, only white light curves were extracted and no colour information is available. 

\subsection{Observation programme}
\label{ssec:obs}
\begin{figure}[h]
\begin{center} 
\includegraphics[height=9.cm,width=9cm]{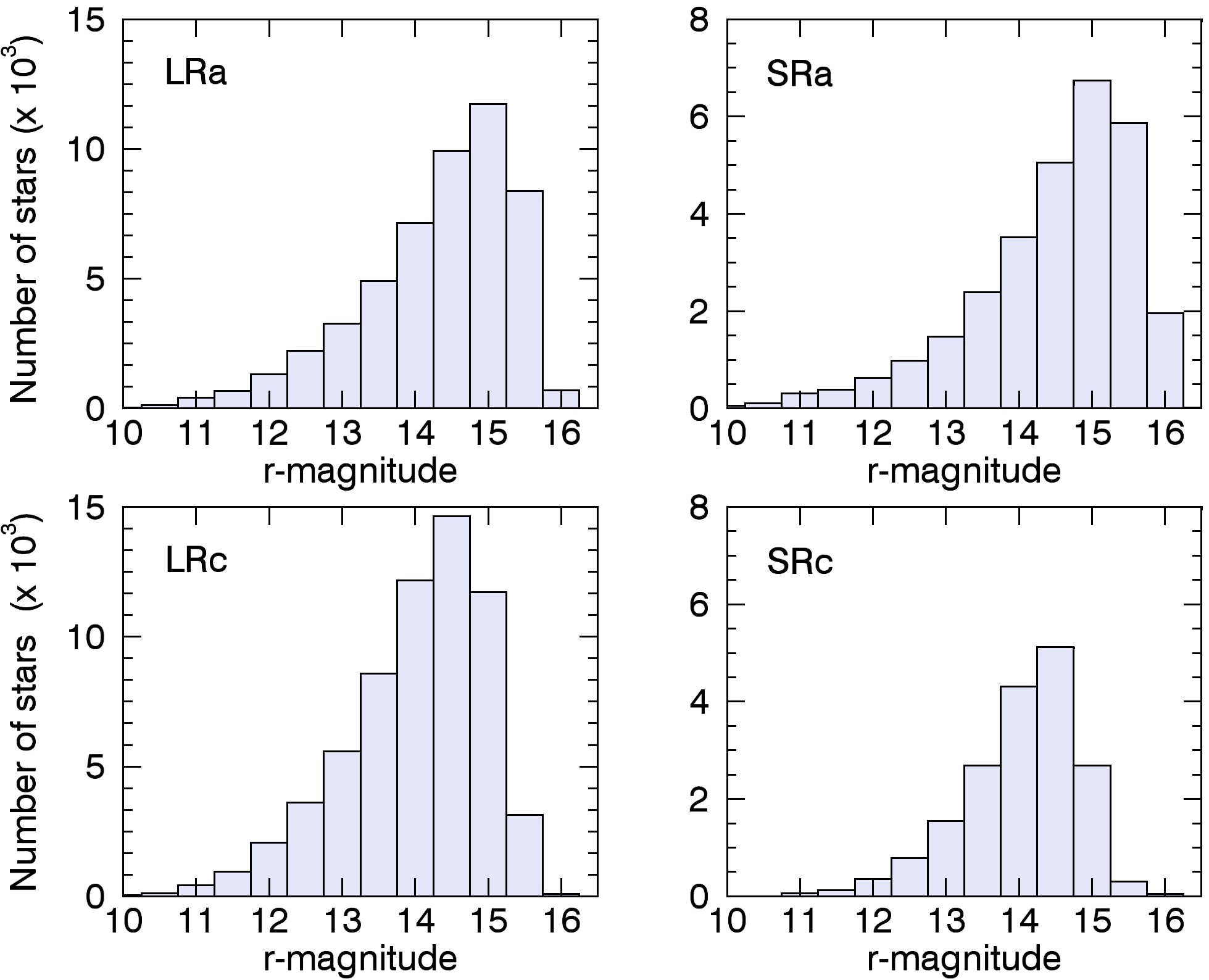}
\end{center}
\caption{{\bf Distribution of r-magnitude for the stars observed in the long runs and short runs in the anticentre (top) and centre (bottom).}
}
\label{fig:rmagDistrib}
\end{figure}

The target fields accessible to \corot\ were restricted to two circles of radius $\sim 10 \degr$ located in the Ecliptic plane and separated by 180\degr, known as the \corot\ continuous viewing zones (CVZs) or CoRoT Eyes. Within the CVZs, continuous observations for up to nearly six months were possible while maintaining the amount of light scattered by the Earth hitting the detector at or below an acceptable level. The two CVZs were centred on zero declination and right ascension $6^{\rm h}~50^{\rm m}$ and $18^{\rm h}~50^{\rm m}$, corresponding approximately to the Galactic anti-centre and centre directions, respectively. The telescope switched between those two directions twice a year (in April and October). 

At the beginning of the mission, the observations mainly consisted of one long run (LR), lasting about 140 days, and one shorter run (SR), lasting between 20 and 30 days, per half-year. The exact duration and the number of pointings per year were flexible, however, and were adapted later in the mission to account for the evolution of the scientific requirements of both the exoplanet and the stellar physics programmes. After the break-down of DPU1, the observation strategy was changed, resulting in two runs of intermediate duration per six-month season, to compensate for the lower star counts per pointing. This flexibility also made it possible to re-observe the same field after few years. This was done, for example, in January 2012, returning to the field in which the transiting super-Earth \Ctseven\ was initially discovered in 2008 \citep{Leger2009}. This enabled a more precise determination of that planet's radius by scheduling simultaneous observations with \corot\ and HARPS \citep{Barros2014,Haywood2014}. Similarly, the SRc03 pointing was designed to re-observe a single transit of CoRoT-9b \citep{Deeg2010}. Consequently, it lasted only five days, and only 652 targets were effectively photometrically measured, making it unsuitable for transit searches but sufficient to secure simultaneous Spitzer observations and further update the CoRoT-9b physical parameters \citep{Bonomo2017, Lecavelier2017}. Although it is listed in Table~\ref{tab:fields}, the SRc03 field was then excluded from this study.  

The location of all 26 exoplanet fields observed during the mission, from January 2007 to October 2012, is shown in Fig.~\ref{fig:fields}, and their details are listed in Table~\ref{tab:fields}. In a number of cases, there is some overlap between successive fields, albeit with a different orientation. Some targets were thus observed twice or even three times a few months or years apart, with a slightly different instrumental configuration. As a consequence, the photometric mask used to perform the on-board photometric measurements typically varied from one observation to the next, changing the contamination of the aperture (the fraction of measured flux coming from other stars in the vicinity of each target). Table~\ref{tab:fields} also provides an estimate of the  median photometric precision at $R = 14.0$. For these estimates, we applied a one-hour duration non-linear filter to all light curves in a given run, ignoring obvious outliers flagged by the \corot\ data reduction pipeline such as SAA crossing. The scatter on each light curve was then estimated on these detrended light curves as the median from the median of a running window of three hours duration  \citep{Hoaglin1983}. Finally the noise at $R = 14.0$ was calculated as the median of the scatter of all stars in the range 13.9 to 14.1 in R-mag.

\begin{table*}
\begin{center}
\caption{Summary of the \corot\ runs. Column 2 gives the number of CCD, column 6 the number of stars monitored during the pointing, column 7 the number of those targets that were observed in this field only, column 8 the number of those targets observed in this field and another, column 9 the number of those targets  observed in this field and two others, column 10 the number of targets classified as dwarfs, column 10 the pipeline version used for the candidates and EB analysis, column 11 the CDPP measured at mag-R=14 on a 2 hours timescale (see text). The last line gives total counts without duplication for the whole mission. }
\medskip
\scalebox{0.8}{
\begin{tabular}{lccrcrrrrrrrrcc}
\hline
Field  &   CCD  & Start date & Duration &  Overlap &  Targets & Targ. $\sharp$ 1 & Targ. $\sharp$ 2 & Targ. $\sharp$ 3 &  Dwarfs & Dwarfs & FGKM & FGKM  & Pipeline & noise @R=14 \\
          &                 & (dd/mm/aa)   &  (days)   &                 &             &                             &                            &                           &  (IV/V) &  (V) & (IV/V) &  (V)  &   Version   & (10$^{-4}$)  \\
\hline
IRa01 & 2 & 06/02/2007 & 54.3    & LRa01/LRa06 & 9921   & 8216  & 821 & 884 & 6550 & 4507 & 4017 & 2683 & 2.1 & 2.73  \\
LRa01 & 2 & 23/10/2007 &  131.5 & IRa01/LRa06 & 11448 & 11448 & 0    & 0     & 8961 & 5593 & 4907 & 3150 & 2.1b & 2.87 \\
SRa01 & 2 & 21/03/2008 & 23.4 & SRa05 & 8190 & 5822 & 2368 & 0 & 4218 & 2252 & 2173 &  989 & 2.1 & 3.27  \\
SRa02 & 2 & 11/10/2008 & 31.8 & LRa07 & 10305 & 10305 & 0 & 0 & 7990 & 4247 & 4770 & 2372 & 2.1b & 4.47 \\
LRa02 & 2 & 16/11/2008 & 114.7 &               & 11448 & 11448 & 0 & 0 & 9410  & 5940 & 6292 & 4048 & 2.1b & 4.06  \\
LRa03 & 1 & 03/10/2009  & 148.3 &               & 5329 & 5329 & 0 & 0 & 3862  & 2537 & 2793 & 1811 & 2.2 & 3.65   \\
SRa03 & 1 & 05/03/2010 & 24.3 &               & 4169 & 4169 & 0 & 0 & 3038  & 1670 & 1856 &  950 & 2.2 &  3.19 \\
LRa04 & 1 & 29/09/2010 & 77.6 &               & 4262 & 4262 & 0 & 0 & 2967 & 1910 & 2128 & 1354 & 2.2 & 8.10  \\
LRa05 & 1 & 21/12/2010 & 90.5 &               & 4648 & 4648 & 0 & 0 & 3332 & 1918 & 2624 & 1466 & 2.2 & 4.39  \\
SRa04 & 1 & 07/10/2011  & 52.3 &               & 5588 & 5588 & 0 & 0 & 3840 & 2103 & 3500 & 1886 & 3.0 &  4.00  \\
SRa05 & 1 & 01/12/2011  & 38.7 & SRa01 & 4213 & 4213 & 0 & 0 & 2452 & 1271 & 1106 &  514 & 3.0 & 4.27  \\
LRa06 & 1 & 12/01/2012     & 76.7 & LRa01/IRa01 & 5724 & 1356 & 3484 & 884 & 947 &    601 & 701 &  449 & 3.2 &  3.66   \\
LRa07 & 1 & 04/10/2012 & 29.3 & SRa02 & 4844 & 4390 & 454 & 0 & 3173 & 1540 & 1936 &  926 & 3.3 &   4.62 \\
\hline
SRc01 & 2 & 13/04/2007 & 25.6 &       -        & 7015 & 7015 & 0 & 0  & 4484 & 2560 & 3039 & 1790 & 2.1 & 3.51  \\
LRc01 & 2 & 16/05/2007 & 142.1 &               & 11448 & 11448 & 0 & 0  & 4922 & 2995 &  4632 & 2805 & 2.1 & 3.80 \\
LRc02 & 2 & 15/04/2008 & 145 & LRc06/LRc05 & 11448 & 11448 & 0 & 0 & 6239 &  4283 & 5324 & 3732 & 2.1 & 3.44 \\
SRc02 & 2 & 15/09/2008 & 20.9 &               & 11448 & 11448 & 0 & 0 & 3477 & 1782 & 1765 &  651& 2.1 & 3.70\\
LRc03 & 1 & 03/04/2009 & 89.2 &               & 5724 & 5724 & 0 & 0 & 3639 & 1839 & 2753 & 1168 & 2.1 & 4.60 \\
LRc04 & 1 & 07/07/2009 & 84.2 & LRc10 & 5724 & 5724 & 0 & 0 & 4200 & 2695 & 3987 & 2635 & 2.2 & 3.34 \\
LRc05 & 1 & 08/04/2010 & 87.3 & LRc06 & 5724 & 5724 & 0 & 0 & 2456 & 1673 & 1951 & 1332 & 2.2 & 3.38 \\
LRc06 & 1 & 08/07/2010 & 77.4 & LRc02/LRc05 & 5724 & 3836 & 1880 & 8 & 2029 & 1311 & 1709 & 1149 & 2.2 & 3.55 \\
LRc07 & 1 & 08/04/2011 & 81.3 & LRc08/LRc10 & 5724 & 3953 & 1771 & 0 & 1784 & 1182 & 1631 & 1107 & 3.0 & 4.20  \\
SRc03 & 1 & 01/07/2011 & 20.9 & LRc02/LRc06 & 652 & 85 & 559 & 8 & 0 & & & & -  & - \\
LRc08 & 1 & 08/07/2011 & 83.6 & LRc07/LRc10 & 5724 & 5724 & 0 & 0 & 2658 & 1793 & 2488 & 1670 & 3.0 & 4.21 \\
LRc09 & 1 & 12/04/2012 & 83.6 &               & 5724 & 5724 & 0 & 0 & 2630 & 1780 & 2444 & 1649 & 3.0 & 3.56 \\
LRc10 & 1 & 09/07/2012   & 83.5 & LRc04/LRc07 & 5286 & 4618 & 668 & 0 & 1825 & 1192 & 1628 & 1130 & 3.2 & 4.18\\
\hline
Total &  &  &  & & 163665 & 150768 & 12005 & 892 & 101083 & 61174 &  72154 & 43416  \\
\hline
\end{tabular}
}
\label{tab:fields}      
\end{center}
\tablefoot{
The SRc03 field was excluded from this study, because of its very short duration and the limited number of stars observed.  
LR stands for long run, SR for short run and IR for initial run. The next letter, `a' or `c', means anticentre or centre direction respectively.
}
\end{table*}
\begin{figure*}[h!]
\begin{center} 
\includegraphics[height=7.0cm]{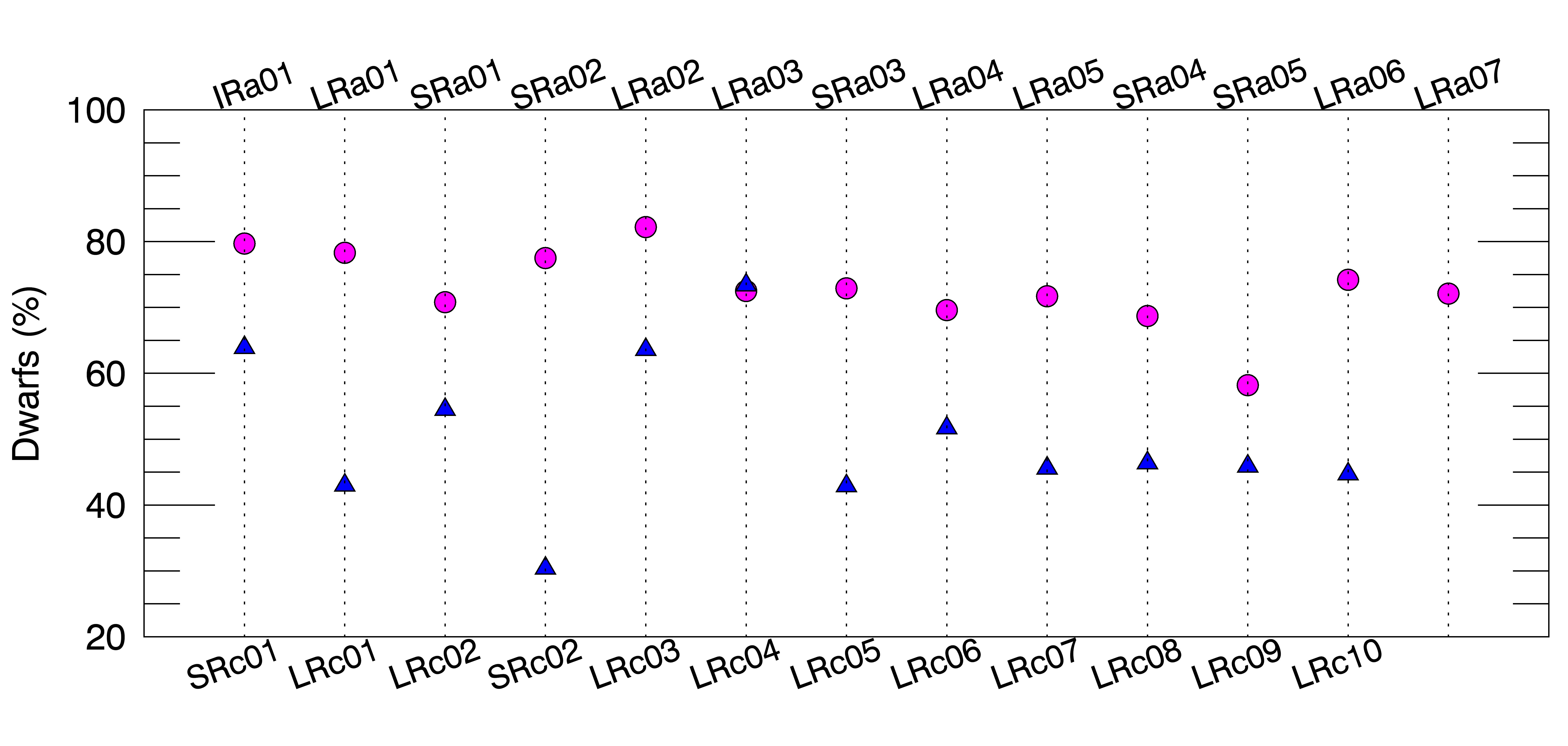}
\end{center}
\caption{Percentage of stars classified as dwarfs (class V and IV) in the various fields (blue triangle: centre; pink circle: anti-centre).}
\label{fig:dwarf_frac}
\end{figure*}

The targets observed in each exoplanet field were selected to maximise the number of main-sequence stars with spectral type F or later, and with relatively uncontaminated apertures. This involved estimating the spectral type and luminosity class of all potential \corot\ targets in the magnitude range $10.5 \leq r \leq 16$, as well as the contamination within the \corot\ aperture. The latter was calculated prior to launch, taking into account fainter background contaminants and using an ideal generic photometric aperture \citep{Llebaria2006}. The resulting estimate of the fraction of the measured flux coming from neighbouring stars was re-evaluated later for the targets actually observed, taking into account the actual photometric mask used for the observation and the in-flight measured point-spread functions (PSFs).

Where possible, the spectral classification was initially made using dedicated, ground-based multi-colour photometric observations carried out prior to the instrument launch \citep{Deleuil09}. However, due to the large area of the continuous viewing zones, this was possible only for some pre-selected regions  corresponding to the approximate fields of the first few long runs (those planned during the mission's initial nominal lifetime of 3.5 years, which were known before the launch). 

The location of the short runs was primarily driven by the stellar physics programme, and for those we relied on existing photometric catalogues for the exoplanet target classification and selection. In some cases, such as SRa01 and SRa02, the target selection was performed without any stellar classification information. Furthermore, the mission lifetime was extended by three additional years in 2009, and again in 2012 (although this was ultimately cut short by the failure of the second data processing unit, DPU2). This led to the selection of new long and short run pointings, for which no dedicated ground-based observations were available, and once again we had to rely on published catalogue information. 

The final release of the \corot\ light curves \citep{Chaintreuil2016b} \footnote{\begin{footnotesize} http://cdsbib.u-strasbg.fr/cgi-bin/cdsbib?2014yCat....102028C 
\end{footnotesize}} was accompanied by a complete update of the database of spectral classifications for all potential exoplanet targets in the continuous viewing zones (ExoDat), which is described in \cite{Damiani2016}. To overcome the incomplete coverage of the dedicated ground-based observing programme, the final version of ExoDat is based on the PPMXL catalogue, which combines USNO-B1.0 and the 2MASS catalogues \citep{Roeser2010}. This results in a homogeneous, magnitude limited coverage of the entire continuous viewing zones. The spectral classification was performed using the same methodology as presented in \cite{Deleuil09}, but the dwarf-giant separation in the $(J,J - K_s)$ colour-magnitude diagrams was adjusted for each field, and made use of reddening maps from the {\it Planck} mission. When tested against synthetic Galactic populations generated using the {\it Galaxia} code \citep{Robin2003}, the classification appears accurate to about half a spectral class for late type stars. Even though the uncertainties can be large for individual stars \citep[see e.g.][]{Gazzano2010}, the classification is statistically reliable, and gives a good overview of the stellar properties in the various fields, as well as an estimate of the total number of main-sequence stars usable for transit searches in each case. 

The target selection made use of the version of the spectral classification that was available in the input catalogue at the time the observations were prepared. The priorities in the selection process have evolved slightly from one run to another, but for the exoplanet programme the main criteria have been, in decreasing order of priority: 1) F, G, K, or M-type dwarfs with $r \leq 16$, and with a contamination rate less than 10\%; 2) F, G, K or M-type dwarfs with $r \leq 14$, regardless of contamination rate; 3) K-type giants with $r \leq 14$, and with a  contamination rate less than 10\%; 4) A-type dwarfs with $r \leq 14$, and with a contamination rate less than 10\%. For the target selection process, stars with luminosity class V or IV were treated as dwarfs because the boundary between the two classes is not very precise: our priority was to avoid missing potential good targets. While some fields are crowded, none had sufficiently high dwarf counts to use up all the available apertures, even when including luminosity class IV stars. Consequently, the remaining available photometric apertures were allocated to stars with a lower priority flag, typically stars with a much higher contamination rate up to 30\% or even more, or to stars specifically selected by stellar physics programmes.This ad-hoc and evolving selection process resulted in a non-homogeneous distribution of target magnitudes from one field to another  (Fig.~\ref{fig:rmagDistrib}).

\begin{figure}[h]
\begin{center} 
\includegraphics[height=9.0cm,width=7.5cm]{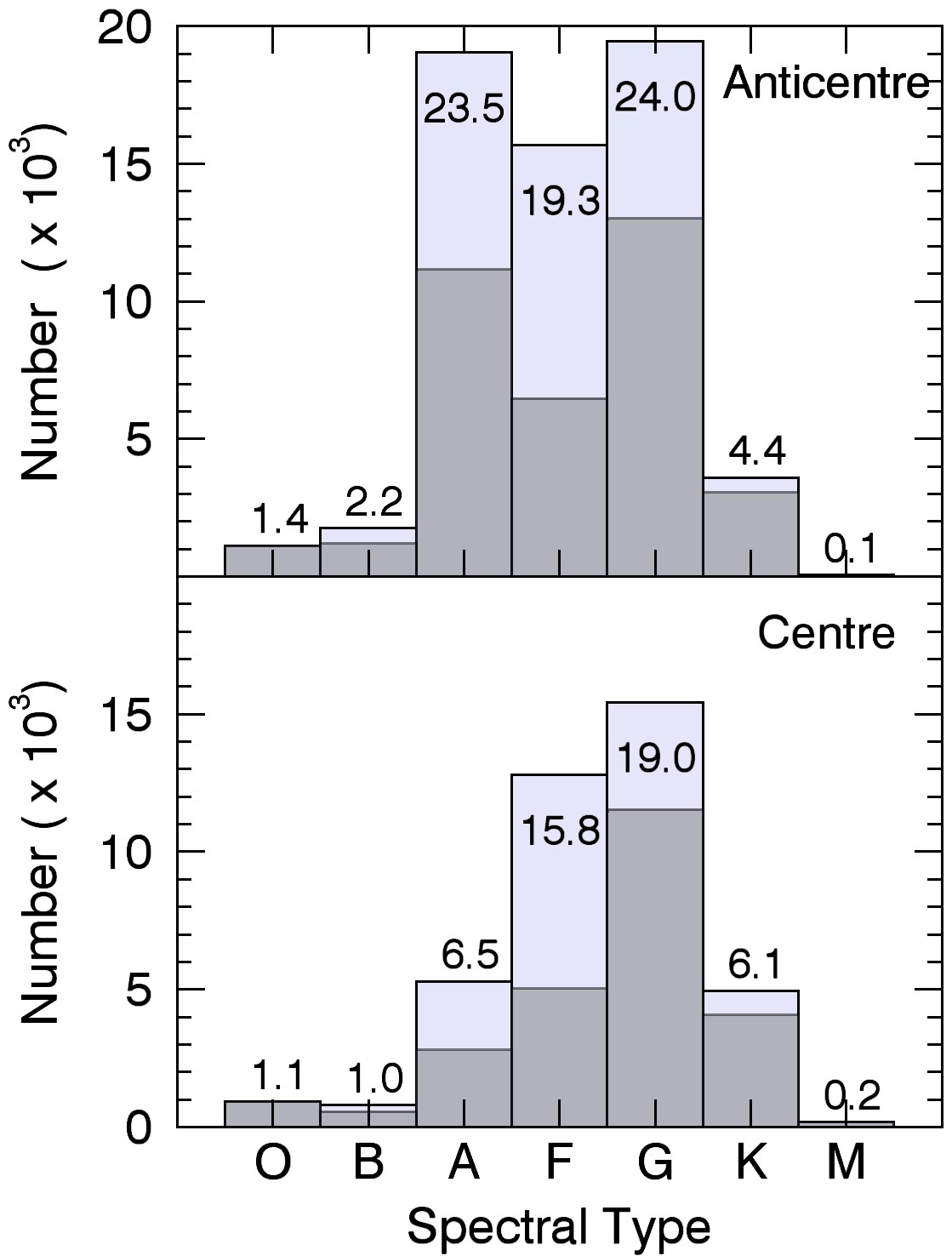}
\end{center}
\caption{Distribution of targets classified as dwarfs (class IV and V) over A, F, G, K, and M spectral types, observed in the exoplanet anticentre (top) and centre (bottom) fields without duplication. The numbers at the top of the bars give the percentage of dwarfs in each spectral type and for each direction.}
\label{fig:dwarf_fields}
\end{figure}

Table~\ref{tab:fields} provides a summary of the CoRoT runs in terms of targets.  A target that has been observed more than one time is counted in Column 7 in the run with the longest duration and not in the shortest run(s). For the latest, re-observed targets appear in Columns 8 or 9 depending whether they have been re-observed one or two times. Among the \totaltargets\ targets, 12\, 005 were observed twice and 1\,784 three times, providing a total of \totalLC\  light curves obtained through on-board photometry or from imagettes, that is the complete photometric window time-series downloaded and processed on the ground \citep[see][]{Barros2014}.

According to the updated spectral classification \citep{Damiani2016}, \totalV\ of these targets are identified as luminosity class V stars. This number increases to \totaldw\ when also including luminosity class IV, showing that dwarf and subgiant stars represent the majority of the targets observed by \corot. There is however a significant difference in dwarfs counts depending on the pointing direction: they  account for 48.9\% of the observed targets in the Galactic centre fields, and 74.8\% in the anti-centre ones.  These numbers drop to 30.4\% and 44.4\% respectively, when we consider only stars of luminosity class V (see Table~\ref{tab:fields}).  As expected the dwarf counts are more homogeneous in the anti-centre fields than in the centre ones (Fig.~\ref{fig:dwarf_frac}). In the latter, there are noticeable differences between fields to another, with dwarf counts as low as 30\% in SRc02, for example, but as high as 73\% in LRc04.

Figure~\ref{fig:dwarf_fields} shows for dwarfs only how these targets distribute over the main spectral types in the two directions. Among them, 72\,154 have spectral type F, G, K, or M, and are thus best suited for transit detection due to their small stellar radii.  They represent 44\% of the total number of targets observed by \corot. The largest spectral type group among the remaining dwarfs are A-stars, which account for 14.9\% of the total number of stars observed.

\section{Transit candidate detection and vetting}
\label{sec:detvet}
\begin{figure*}[h]
\begin{center} 
\includegraphics[height=10cm,width=13.0cm]{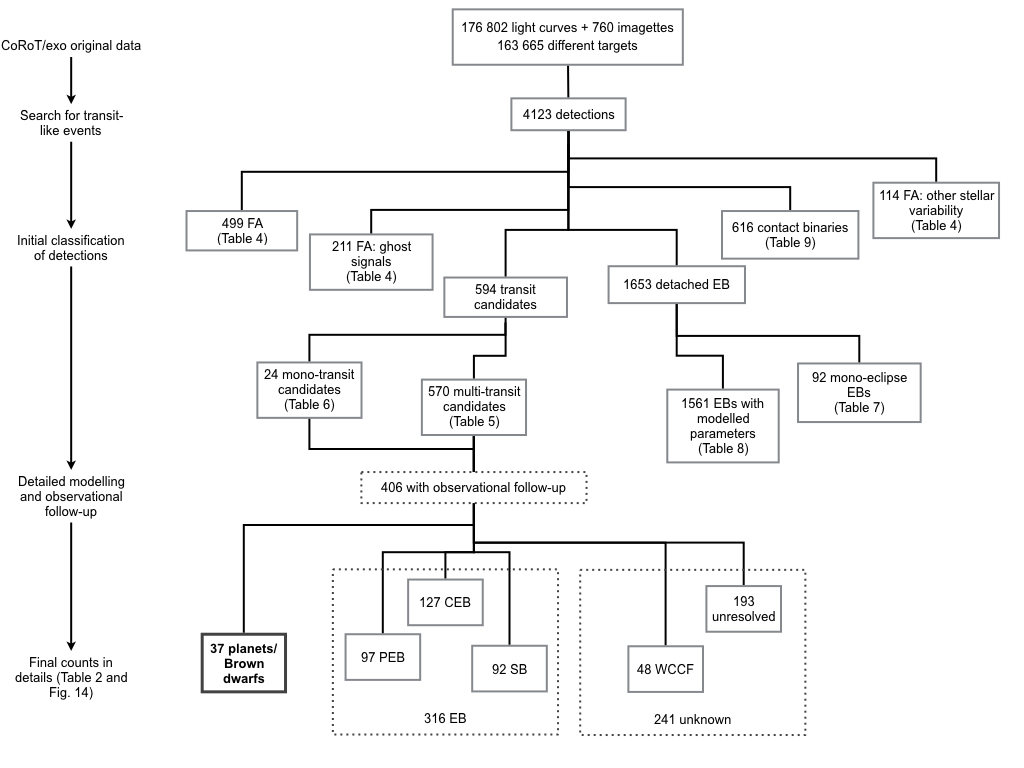}
\end{center}
\caption{Flow-chart of the processing steps and their resulting counts over the various categories described in Sect.~\ref{sec:detvet} and Sect.~\ref{sec:fup}. We note that the first count gives the total number of transit-like features detected in the \corot\ light curves. The others numbers are corrected from duplications. FA stands for false alarms.}
\label{fig:flowcounts}
\end{figure*}

Whatever the spectral classification of each target or its photometric behaviour, 
all the available light curves (including those extracted from imagettes), were searched for transit-like signals. The
light curves used in the present paper to produce the final version of
the transit candidate and eclipsing binary catalogues were produced using
different versions of the \corot\ pipeline, as listed in
Table~\ref{tab:fields}. The basic processing steps in the \corot\
pipeline are described in \citet{Auvergne2009} and the differences
between successive versions are listed in the documentation of the
\corot\ archive \citep{Chaintreuil2016b} \footnote{\begin{footnotesize}http://idoc-corotn2-public.ias.u-psud.fr/jsp/doc/CoRoT$\_$N2$\_$versions$\_$30sept2014.pdf
\end{footnotesize}}.
The changes between the successive versions are mostly minor, the
  most noteworthy being a significant improvement in the jitter
  correction from version 3.0 onwards. We note that the photometric noise  \citep{Asensio-Torres2016} as 
well as the spatial variation
  of the background has increased over \corot's lifetime due to the ageing of the CCDs, which led to a gradual increase of the dark current and decrease of the charge transfer efficiency \citep{Ollivier2016}. This has been corrected in Version N2-4.4 of the pipeline, but data processed using this pipeline version only became available shortly before the submission of the present paper, too late to be incorporated in the analysis. Thus we caution that the light curves used here have sub-optimal background correction, which might affect the reported transit depths, especially for the later runs. As a guide along the various processing steps that are described in the following sections and the object counts that resulted, we refer the reader to the flow-chart in Fig.~\ref{fig:flowcounts}.

\subsection{Transit detection}
\label{ssec:det}

There was no official \corot\ pipeline at mission level for transit
detection and light curve analysis. As described in previous run
report papers \citep[see e.g.][]{Carpano2009,Cabrera2009}, once the
science-grade (N2) light curves of a given run were released to the
co-investigators and associated scientists, the transit search was carried
out in parallel by different teams, who use different methods to
filter the light curves and detect the transits. The methods
implemented by the different teams to identify transit signals were
presented in \cite{Erikson2012}. We do not describe them again
here as there has been no major change in the detection algorithms since
then. On the other hand, efforts have been made to improve the
pre-filtering and detrending of the light curves
\citep{Ofir2010,Grziwa2012,Bonomo2012}. As the transit search for each
run was carried out as soon as it was released, the methods used for
transit detection and the initial vetting of the candidates by the
individual team have evolved significantly from the first to the last
run. We also note that not all teams analysed every run.

Each team produced a list of transit candidates for every run they processed, and most also reported the obvious eclipsing
binaries they had identified in the process. The candidate lists produced by
the different teams were then combined, and each candidate discussed
individually, in order to come to a consensus on the likely nature of
the signal: bona-fide transit candidate, astrophysical false alarm, or
instrumental false alarm. The plots and assessments used to inform this discussion evolved over the lifetime of the
mission, eventually settling into the transit vetting, modelling and candidate
flagging procedure described in the rest of this section, but manual
inspection of the light curve and discussion of each candidate during
regular teleconferences remained an integral part of the candidate
vetting process throughout the mission.

In this manner, the combined detection teams identified a total of 
\totalCand\ significant transit-like events, of which just over 600 were deemed potentially 
worthy of follow-up observations after discarding obvious instrumental and astrophysical false alarms. 
Before being provided to the follow-up team, these
candidates were given a priority ranking ranging from 1 to 4, on the
basis of the likelihood that the transit event was indeed of planetary
origin. This priority ranking was conferred by the detection teams. The
follow-up team then modified the priorities to take into account the
magnitude of the star, as radial velocity precision is mostly limited
by photon noise. These intermediate priority rankings were intended solely to help
organise the follow-up process, the results of which have been
incorporated into the candidate catalogues, so they are not reported here.

For the present paper, we went back to the full list of transit-like events reported by one or more of the detection teams, including cases ultimately deemed by them to be false alarms, and systematically re-analysed them, in order to produce a homogeneous transit candidate catalogue. Each of the detection teams provided initial estimates for the transit parameters, namely the period, depth, duration, and epoch of the transits. When a given detection was reported by more than one team, these estimates sometimes differed somewhat from each other, as they depend on the pre-processing of the light curve and the specifics of the transit detection algorithm used. Additionally, there was considerable variation in the number and nature of the checks which were performed by the different teams to identify false alarms, such as grazing and diluted eclipsing binaries. To overcome this limitation and produce a coherent catalogue, we systematically vetted all the candidates by performing a uniform set of semi-automated checks, modelling all the transits in a consistent manner, and producing a number of plots for each candidate used as assessment tools. This was done using a purpose-written software package developed in Oxford and written in {\sc Python}. This package was initially developed to help prioritise transit candidates and optimise the follow-up. It has been used in this manner since 2010, although it has evolved somewhat since it was first used. For the present paper, we re-ran the latest version of the code on all the transit-like events identified since the start of the mission, resulting in a homogeneous set of transit parameter estimates. The main lines of this analysis are described in the following sections. All results, either plots or tables, are provided online, 
as html pages per field and per candidate.

\begin{figure}[h]
\begin{center} 
\includegraphics[height=5.5cm,width=9cm]{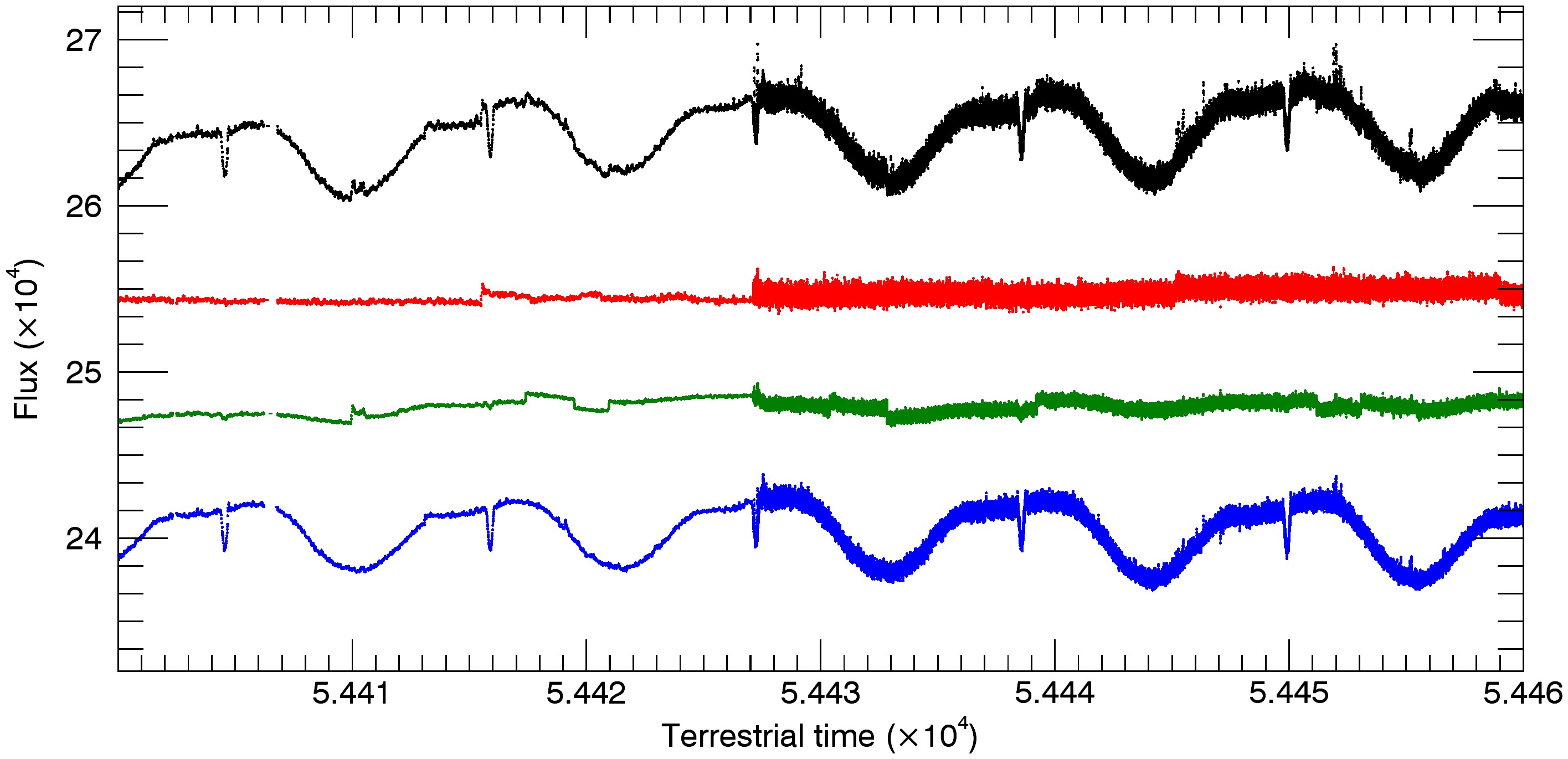}
\includegraphics[height=4cm,width=4cm]{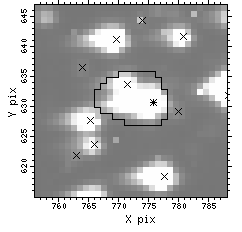}
\end{center}
\caption{Top: Light curves in white, red, green, and blue (with the according colour code) of a false positive. Bottom: The CoRoT image of the field around the target indicated by a star symbol (CID: 102779171, r-mag=13.84) with the shape of the photometric mask overploted and the location of nearby stars indicated with crosses. }
\label{fig:BEB}
\end{figure}

\begin{figure}[h]
\begin{center} 
\includegraphics[height=6.0cm,width=8.7cm]{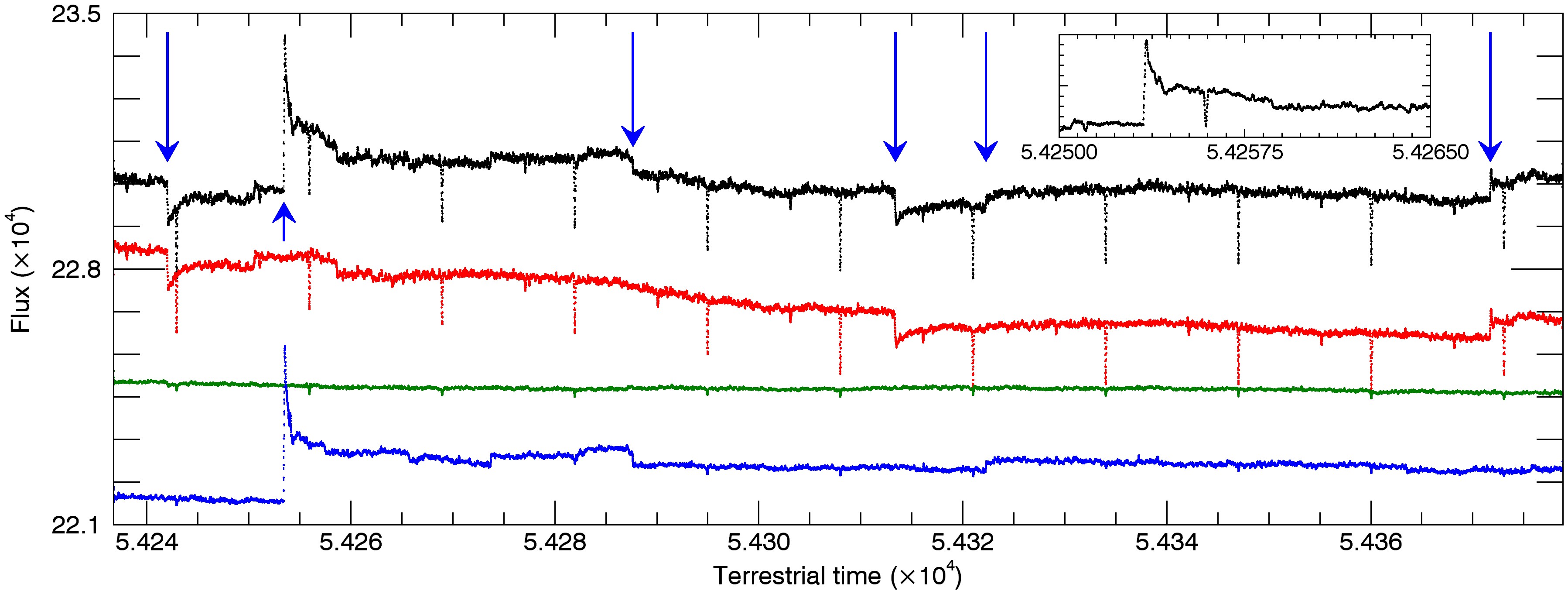}
\end{center}
\caption{Light curves of an eclipsing binary that displays a series of discontinuities indicated by blue arrows. The colour code corresponds to the three colour light curves, with the resulting white light curve plotted in black. We note that the colour light curves have been shifted by a constant in order to avoid too large a scale on the y axis. The top inset is a magnification of a portion of the light curve where the discontinuity in flux is not a simple step but a sudden increase of the flux, followed by an exponential decrease as could be generated by the impact of a proton on the CCD.}
\label{fig:hotpixels}
\end{figure}

\begin{figure}[h]
\begin{center} 
\includegraphics[height=6cm,width=6cm]{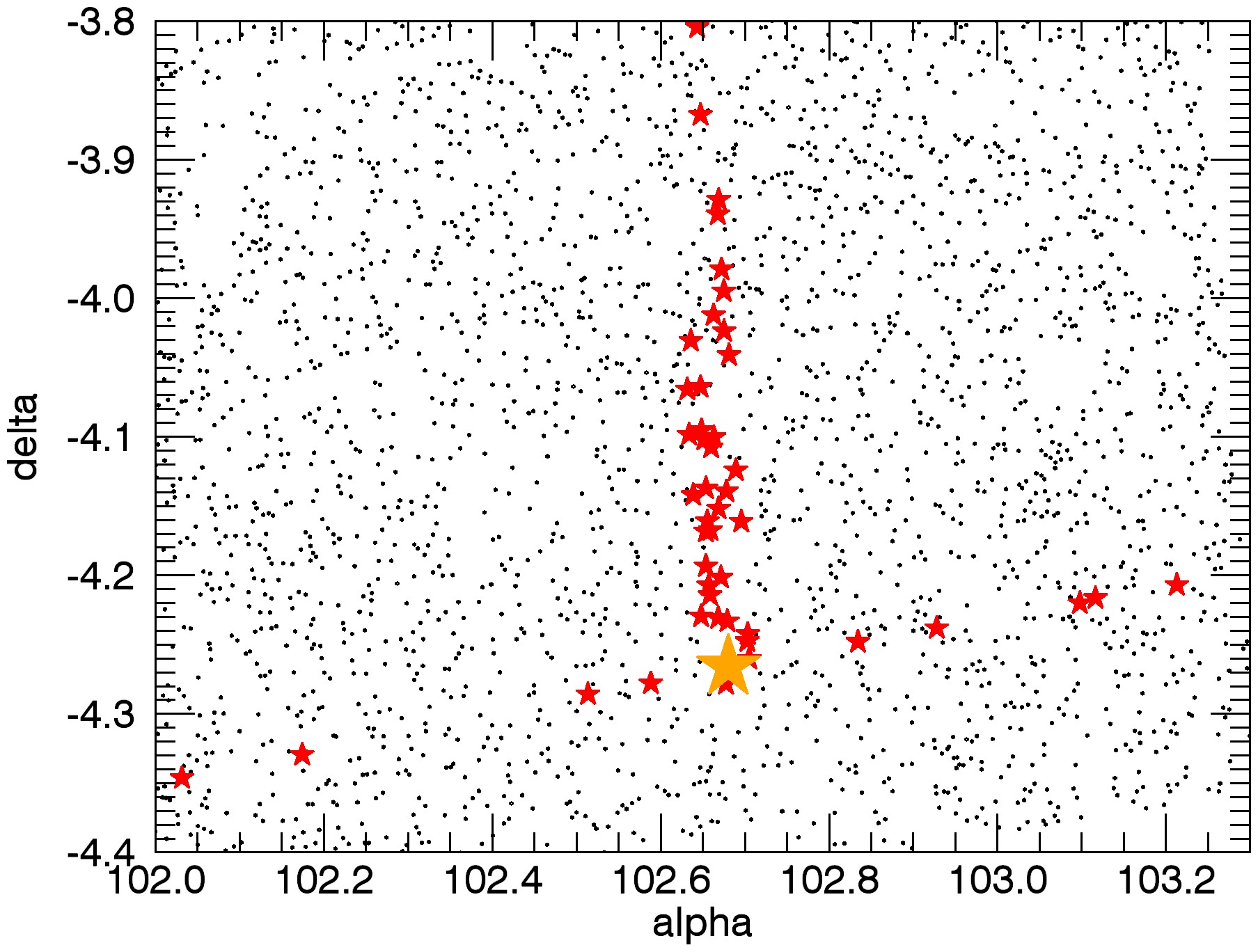}
\end{center}
\caption{Enlargement of one CoRoT CCD in the LRa02. Among the targets (black dots), the light curve of some of them (red stars) contains the imprints of a bright periodic variable star,  V 741 Mon, whose position is indicated by the orange star symbol.  }
\label{fig:VMon}
\end{figure}

\begin{figure}[h]
\begin{center} 
\includegraphics[height=6cm,width=6cm]{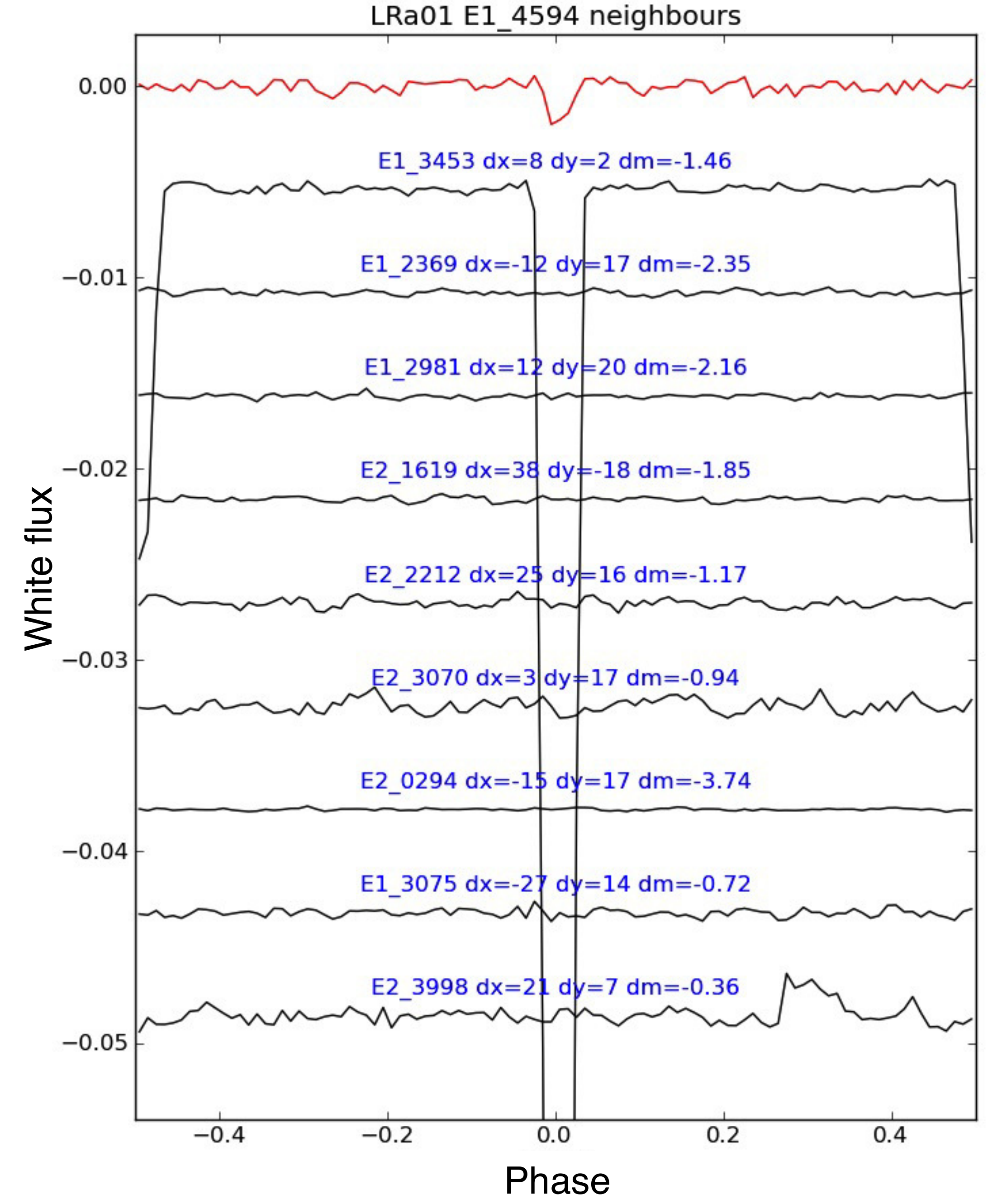}
\end{center}
\caption{Example of a clear `ghost' false positive.  
The light curves of the candidate, LRa01\_E1\_4594 (red line at the top),  and its closest neighbours are all phase-folded at the period of the detected transit signal. In this case, the transits detected in LRa01\_E1\_4594 are those that occur on the nearby eclipsing binary, LRa01\_E1\_3453, whose primary and secondary eclipses are clearly visible in its light curve.}
\label{fig:ghost_example}
\end{figure}

\subsection{Initial transit candidate vetting}
\label{ssec:vet}

Various instances of eclipsing stellar systems which can mimic a planetary transit are the  astrophysical false positives to be chased for. The on-board  photometry did not provide photo-centre curves that could have been used to assess the transit source location within the large photometric mask. Instead, for the brightest targets, CoRoT provided  light curves in the three coloured bandpasses.  They were used to identified obvious cases where a background eclipsing binary is causing the transit signals. Figure~\ref{fig:BEB} shows such an exemple: for this bright target (r-mag = 13.84) neither the transits observed in the white light curves or the stellar activity signal appear in the red and green light curves but both are clearly visible in the blue. The source of the two is likely the nearby faint contaminant (r-mag = 15.31) whose flux is enclosed in the photometric mask.

In addition to these well-known stellar configurations, the other main sources of false detections, are the following two phenomena, which require careful
treatment. The first phenomenon is the `hot-pixels', which produce
sudden discontinuities in the light curve and can cause spurious but
significant detections. These can be identified by inspecting individual
transit events and, where available, the three-colour photometry (as
hot pixels typically affect one of the three channels only) (see Fig.~\ref{fig:hotpixels}). The
second occurs when light from a bright eclipsing binary leaks over one
or more pixel columns either due to blooming effect that can occur for a very bright and saturated star, or due to smearing 
generated during the charges transfert. Depending on its brightness and its position on the CCD, a
bright EB can leave its photometric imprint in the light curve(s) of
other nearby target(s). In that case, the ghost transit signal
exhibits the same period, epoch, and duration as those of the
contaminating eclipsing binary, but the depth is shallower. An impressive example of such a
phenomenon occurred in LRa02: V 741 Mon, a well-known variable star of CVn type \citep{Renson2001}, far too bright to be selected for observation in the faint channel, left its imprints through both blooming and smearing in the light curves of 46 targets as transit-like signals with the same period of 1.143 days and epoch. As shown in Fig.~\ref{fig:VMon}, stars distant by more than 47 arcmin from V 741 Mon were contaminated.

To identify these `ghost' signals, we systematically inspected the light
curves of stars in the vicinity of each transit candidate (see
Fig.~\ref{fig:ghost_example}). We also produced plots of the N2 light curve of a candidate, both unfolded and folded at the period of the transits, including the coloured light curves when available, and magnifications of the individual transit events to help identify false alarms due to hot pixels and obvious EBs. The same plots were also produced after removing all variability on timescales longer than a day using the iterative non-linear filter of \cite{Aigrain2004}. We visually inspected these plots for each object, and obvious false alarms are excluded from the rest of the analysis described in Sections~\ref{ssec:model} and \ref{ssec:flags}. 
\begin{table*}
\begin{center}
\caption{Summary of the transit events per field. We note that these numbers are filtered out from duplications. The types of counts that are given in the columns 3-8  correspond to the candidates categories discussed in Sect. 4. Column 9 gives the number of candidates with follow-up observations. Column 10 gives the number of transit events identified as detached EB, and column 11 those identified as contact binary.}
\label{tab:nb_cands}
\begin{tabular}{lcccccrrrrr}
\hline
\hline
Field & \multicolumn{8}{c}{Candidates} &  EB &  CB \\
       &  Total  &  Planet  &  unres.  &  PEB  &  CEB  &  SB  &  WCCF  &  FU  &   &   \\
\hline
IRa01 & 39 & 2 & 13 & 1 & 9 & 9 & 5 & 27 & 98 & 15\\
LRa01 & 52 & 4 & 14 & 6 & 15 & 7 & 6 & 42 & 162 & 18\\
SRa01 & 8 & 0 & 4 & 1 & 1 & 0 & 2 & 4 & 77 & 12\\
SRa02 & 18 & 1 & 8 & 3 & 2 & 4 & 0 & 10 & 92 & 59\\
LRa02 & 40 & 3 & 11 & 1 & 7 & 10 & 8 & 37 & 130 & 39\\
LRa03 & 16 & 0 & 5 & 1 & 4 & 3 & 3 & 15 & 40 & 10\\
SRa03 & 11 & 3 & 5 & 0 & 0 & 2 & 1 & 7 & 41 & 3\\
LRa04 & 7 & 0 & 0 & 1 & 1 & 5 & 0 & 7 & 38 & 10\\
LRa05 & 19 & 0 & 5 & 7 & 1 & 5 & 1 & 11 & 43 & 9\\
SRa04 & 11 & 2 & 1 & 1 & 5 & 2 & 0 & 11 & 50 & 10\\
SRa05 & 8 & 1 & 0 & 2 & 3 & 1 & 1 & 7 & 35 & 57\\
LRa06 & 10 & 0 & 2 & 5 & 3 & 0 & 0 & 5 & 15 & 5\\
LRa07 & 5 & 0 & 3 & 0 & 0 & 1 & 1 & 3 & 42 & 25\\
Total Anticentre & 244 & 16 & 71 & 29 & 51 & 49 & 28 & 186 & 863 & 272  \\
\hline
SRc01 & 47 & 0 & 32 & 0 & 3 & 4 & 8 & 24 & 114 & 26\\
LRc01 & 42 & 4 & 7 & 8 & 14 & 9 & 0 & 36 & 109 & 37\\
LRc02 & 50 & 6 & 20 & 11 & 6 & 7 & 0 & 28 & 94 & 31\\
SRc02 & 16 & 0 & 3 & 0 & 5 & 2 & 6 & 13 & 117 & 67\\
LRc03 & 45 & 2 & 14 & 10 & 9 & 8 & 2 & 25 & 72 & 21\\
LRc04 & 29 & 0 & 11 & 7 & 8 & 3 & 0 & 17 & 50 & 17\\
LRc05 & 30 & 2 & 8 & 8 & 10 & 2 & 0 & 13 & 46 & 42\\
LRc06 & 18 & 1 & 6 & 6 & 4 & 0 & 1 & 7 & 35 & 23\\
LRc07 & 10 & 2 & 4 & 0 & 2 & 2 & 0 & 10 & 30 & 5\\
LRc08 & 14 & 3 & 6 & 3 & 0 & 1 & 1 & 10 & 39 & 27\\
LRc09 & 28 & 1 & 5 & 10 & 7 & 3 & 2 & 17 & 40 & 31\\
LRc10 & 21 & 0 & 6 & 5 & 8 & 2 & 0 & 20 & 44 & 17\\
Total Centre & 350 & 21 & 122 & 68 & 76 & 43 & 20 & 220 & 790 & 344  \\
\hline
 Grand Total & 594 & 37 & 193 & 97 & 127 & 92 & 48 & 406 & 1653 & 616 \\
\hline
\end{tabular}
\end{center}
\end{table*}

This visual inspection step may seem primitive, but it is relatively quick and very effective, reducing the number of targets under consideration from \totalfeat\ light curves with transit-like events to \totalCandfilt\ surviving planetary candidates. Of the discarded transit-like events, \totalFd\ were found to be false detections, some due to hot pixels, others to unconfirmed detection or spurious contamination, \totalGh\ clear ghosts, and at least \totalPls\ false alarms due to other forms of stellar variability, such as pulsations or rotational modulation of star spots. 
Among these \totalfeat\ transit-like events, we identified \totalEBfilt\  detached eclipsing binaries (EB) and \totalCB\ contact binaries. 
 Table~\ref{tab:nb_cands} gives an overview of the number of candidates and binaries detected in each pointing. It also  provides the number of candidates identified as different types of astrophysical false positives on the basis of a deeper analysis of their light curve or ground-based follow-up observations (which are discussed in more detail in Sections~\ref{ssec:fup_obs} \& \ref{ssec:fup_res}).  
Table~\ref{tab:discarded} reports the list of the transit features that we discarded as false alarms. We stress that this table is not exhaustive, since no attempt was made to identify all the ghost signals caused by each bright EB systematically for example. Only those that happened to be considered initially as transit planetary or eclipsing binary candidates are listed here, but we deemed it useful to record them nonetheless. 

For the remainder of this paper, the term candidates is used only to refer to the objects which passed the preliminary vetting steps, while those that were rejected at this step are referred to as false alarms or EBs. Of course, some of the candidates are in fact EBs identified in the second step of the analysis (see Sect. \ref{ssec:fup_obs}). These cases which were left as candidates, were originally included in the follow-up programme, before the light curves vetting tests were fully set up.  We thus keep the term EBs in the following sections, to designate `obvious' EBs identified at the preliminary vetting stage. This visual inspection step also helps identify cases where some error has crept into the transit properties reported by the detection teams, which need to be corrected manually before the transits can be modelled in detail.

\subsection{Transit candidate modelling}
\label{ssec:model}

For the surviving transit candidates, we fitted the light curve with a simple transit model assuming a zero eccentricity. 
We first fitted a linear trend to the region around each individual transit, to remove stellar variability, phase-fold the resulting sections of light curve, and perform a global transit fit using the formalism of \cite{Mandel2002}. The fitted parameters are the period $P$, the time of transit centre $T_0$, the planet-to-star radius ratio $\Rp/\Rs$, the system scale $a/\Rs$, and the impact parameter $b$. We used a quadratic limb-darkening law, but fix the coefficients to $u_a=0.44$ and $u_b=0.23$, the values tabulated by \cite{Sing2010} for a $0.9\,\Msun$ star in the \corot\ bandpass. The signal-to-noise ratio of the transits is not sufficient, in most cases, to fit for the limb-darkening coefficients, and we opted to use a single set of values because of the large uncertainty in the stellar parameters. It is important to bear in mind, however, that the limb-darkening coefficients used may not be appropriate for some objects. The fit was performed using an implementation of the Levenberg-Marquart algorithm for non-linear least-squares regression adapted for {\sc Python} from the the {\sc Idl} program {\sc Mpfit}. 

Once these physical parameters have been obtained, the individual transit events are fitted, allowing only the time of transit centre to vary, and the ephemeris is refined using a linear fit to the times of transit centre. We repeated the process of variability removal, global and individual fits, until all the parameters have converged, meaning that their values change by less than their formal uncertainties (which are derived from the covariance matrix of the fit). At that point, we also computed the orbital inclination, stellar density and stellar radius, using the equations of \cite{Seager2003} and assuming a power-law stellar mass-radius relation with an index of 0.8. These were used to check if the transits are too long for their period, indicating a large (early type, or evolved) primary star, or a blended system. We note however that these calculations ignore limb-darkening, and are very approximate, particularly for grazing eclipses, where there is an almost complete degeneracy between the planet-to-star radius ratio, system scale and impact parameter. Therefore, the reported parameters for grazing events have very large uncertainties, and and we do not report stellar density estimates for such events, as they are essentially meaningless. Additional fits are made using simple trapezoidal or triangular models to obtain direct estimates of the transit depth, total transit duration and duration of totality (or `outer' and `inner' durations respectively). Finally, we also performed separate transit fits whose results were used to perform some basic tests intended to help identify the candidates most likely to be planetary (see Sect.~\ref{ssec:flags}) :
\begin{itemize}
    \item on the odd- and even-numbered transits separately: significant differences between the two  indicate that the transits-like events are caused by a near-equal mass eclipsing binary whose light is diluted by that of a third star (blended eclipsing binary), rather than by a planet.
    \item on the light curves in the three coloured bandpasses (where available): significant differences can indicate that the transit-like events are not grey -- and hence have a stellar origin. This is to be used with caution, though, as there is a degeneracy between actual colour, and spatial location along the dispersion direction of the \corot\ prism: different depths in different colour channels are actually more likely to be due to a faint star contaminating the blue or red end of the photometric aperture, than to a real colour difference. In such cases, the transit might be on the main target -- and hence might still be a genuine planetary event -- or might be on the contaminating star (blended eclipsing binary).
    \item on the light curve around phase 0.5, to check for a secondary eclipse, which would indicate a stellar origin for the transits.
\end{itemize} 
These additional transit fits were carried out with the period, epoch, impact parameter and system scale fixed to the values determined from the main transit fit, allowing only the radius ratio (i.e. the depth) to vary. For the secondary eclipse check, the limb-darkening parameters are set to zero and the initial estimate of the depth is set to a tenth of the transit depth. We note that we only checked for secondaries around phase 0.5 (for practicality reasons), so weak secondaries from eccentric binaries are missed.
 
 The resulting parameters of the \totalCandfilt\ planet candidates  are reported in Table~\ref{tab:candidates}, for all those which were not identified as obvious false positives at the preliminary vetting stage, and which displayed at least two transits in any given run. For the fitted parameters, we report the formal uncertainties (from the diagonal elements of the covariance matrix), standard error propagation is used to compute uncertainties for the derived parameters. We note that these errors do not account for correlations between the parameters, or for correlated noise in the data, so they should be taken as indicative only. 

Some of them were observed twice; for these we report the parameters derived from the light curve obtained in the pointing with the longest duration, but we also list the other runs in which they were observed. This number includes those (37) that have been confirmed as planets or brown dwarfs on the basis of subsequent ground-based observations.

A further 24 candidates, which displayed only one transit in any given run, are listed separately in Table~\ref{tab:mono}. Two single transit events (\corot\ IDs 102723949 and 102765275), initially discovered in IRa01 \citep{Carpano2009,Moutou2009}, were re-observed in the LRa01. This allowed us to determine their orbital period and they are thus now listed in Table~\ref{tab:candidates}. 

\subsection{Eclipsing binaries}
\label{ssec:EB}

A total of  \totalbin\ clear eclipsing binaries were identified at the preliminary vetting stage. 
Of those,  \totalEBfilt\ were sufficiently well-detached for the transit modelling described in Section~\ref{ssec:model} to converge, so their light curves were also modelled in the same way. Since the transit model used assumes a non-luminous companion, some of the fitted parameters such as $\Rp/\Rs$, $b$ and $a/\Rs$ are meaningless for EBs, but the modelling process does enable us to derive improved estimates of the ephemeris and primary eclipse depth and duration. Of this sample, 137 were observed twice or even three times (mostly in fields IRa01, LRa01, and LRa06, which have the strongest overlap, see Fig.~\ref{fig:fields}), leaving a total of \totalEBfilt\ unique detached EBs identified and characterised as a by-product of the exoplanet search. The multi-transit ones (\totalEBfiltper) are reported in Table~\ref{tab:EBs}, along with a rough (by eye) classification based on their phase-folded light curve. Indeed, in the first step of the vetting process, binaries were visually classified in four sub-classes: 
\begin{enumerate}
\item eclipsing binaries with distinct eclipses and a detected secondary eclipse at phase 0.5;
\item eccentric eclipsing binaries with distinct eclipses and a detected secondary eclipse not at phase 0.5;
\item eclipsing binaries with distinct eclipses but without detected secondary at phase 0.5;
\item contact binaries that present no clear eclipse but a near-sinusoidal modulation of their light curve. 
\end{enumerate}

In \totaleccB\ of these detached EB we identified a secondary, but not at phase 0.5, indicating an eccentric system. Three of those were observed twice, which improved the constraints on their orbital periods. \corot\ ID\,105499823 was observed in both LRc05 and LRc06, and its orbital period was determined thanks to the observation of two secondary eclipses during LRc05. For CID\,102768841, an orbital period of 54.138 days was derived from the 131.5 days of LRa01 observations, but only one eclipse was observed during the much shorter LRa06 (76.6 days). Among these detached EBs, \totalmono\ showed only one primary eclipse in a given run (Table~\ref{tab:monoEB}). A single eclipse of CID\,102586624 was observed in each of LRa01 and LRa06, which implies that its orbital period is greater than 132 days. 

Finally, Table~\ref{tab:CBs} provides the list of transit features we classified as contact binary. For these kind of binaries, we carried out no modelling but provide the ephemeris and the period. Sorting all these events as a function of their period allow us to identify a second round of ghosts. An independent compilation of eclipsing binaries in CoRoT data has been published by \cite{Klagyivik2016}. Their table contains 2290 likely eclipsing binaries of all types (contact or detached), which they used as input sample for a search for circumbinary planets.

\subsection{Flag system}
\label{ssec:flags}

Once the transit fits were complete, a number of tests were performed to check if the transit parameters are compatible with a planetary origin. The outcome of these tests were recorded in the form of six binary flags, which are also included in Table~\ref{tab:candidates} and \ref{tab:EBs}. The flags are:
\begin{itemize}
\item $F_{\rm det}$: low detection significance, set if the transit depth in the white light curve is less than five times the corresponding uncertainty;
\item $F_{\rm sec}$: secondary eclipse detected, set if the secondary eclipse depth (at phase 0.5) is more than three times the corresponding uncertainty;
\item $F_{\rm odd/even}$: odd/even depth differences, set if if the odd-to-even depth ratio is more than 1.1 at $3\sigma$ confidence level;
\item $F_{\rm col}$: strong colour dependence, set if the ratio of the deepest to the shallowest of the transits in the three colour channels is more than 1.5 at $3\sigma$ confidence level;
\item $F_{\rm long}$: transit too long, set if the best fit stellar radius is $>2\,R_\odot$ at $3\sigma$ confidence level;
\item $F_{\rm V}$: V-shaped transit, set if the best-fit transit model is grazing that is with a null inner duration estimate which means a bottom flat section is lacking.  
\end{itemize}
While the four first flags are directly related to the light curve analysis, the two last are associated to the physical parameters that have been derived assuming the star is a solar twin. We note that these flags are intended for a first, quick-look sorting of the candidates: they are by no means unequivocal, in the sense that a real planet could have one or more flag set, and many candidates, which were later found to be astrophysical false alarms, had none.

All the candidates (Table~\ref{tab:candidates}), including the planets, and the EBs (Table~\ref{tab:EBs}) have been yet blindly re-analysed with this new tool, even if their nature had already been elucidated using ground-based follow-up observations. This gives us an opportunity to learn whether we could have made better use of the transit modelling and flag system to prioritise the follow-up resources, and may be helpful in informing the candidate prioritisation strategies for future missions such as TESS and PLATO.

\begin{figure}[ht]
\begin{center} 
\includegraphics[height=8.5cm,width=6.0cm,angle=-90]{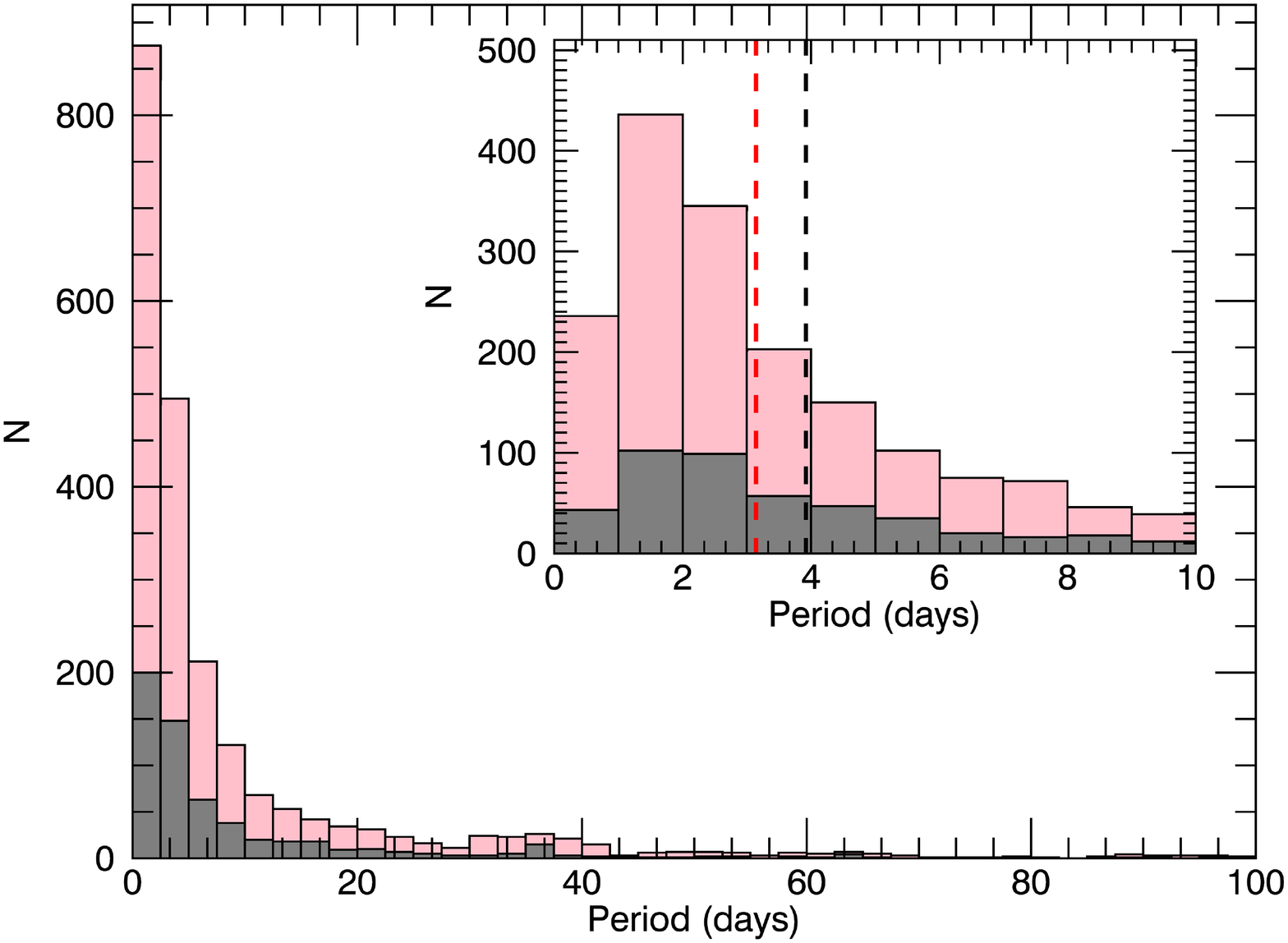}
\end{center}
\caption{Stacked histograms of the period distribution (in days) of EBs (pink) and candidates (grey). The dash lines give the median of each distribution. The inset shows an enlargement of the short end of the period range.
}
\label{fig:period_dist_all}
\end{figure}
%

\begin{figure}[ht]
\begin{center} 
\includegraphics[height=8.5cm,width=6.5cm,angle=-90]{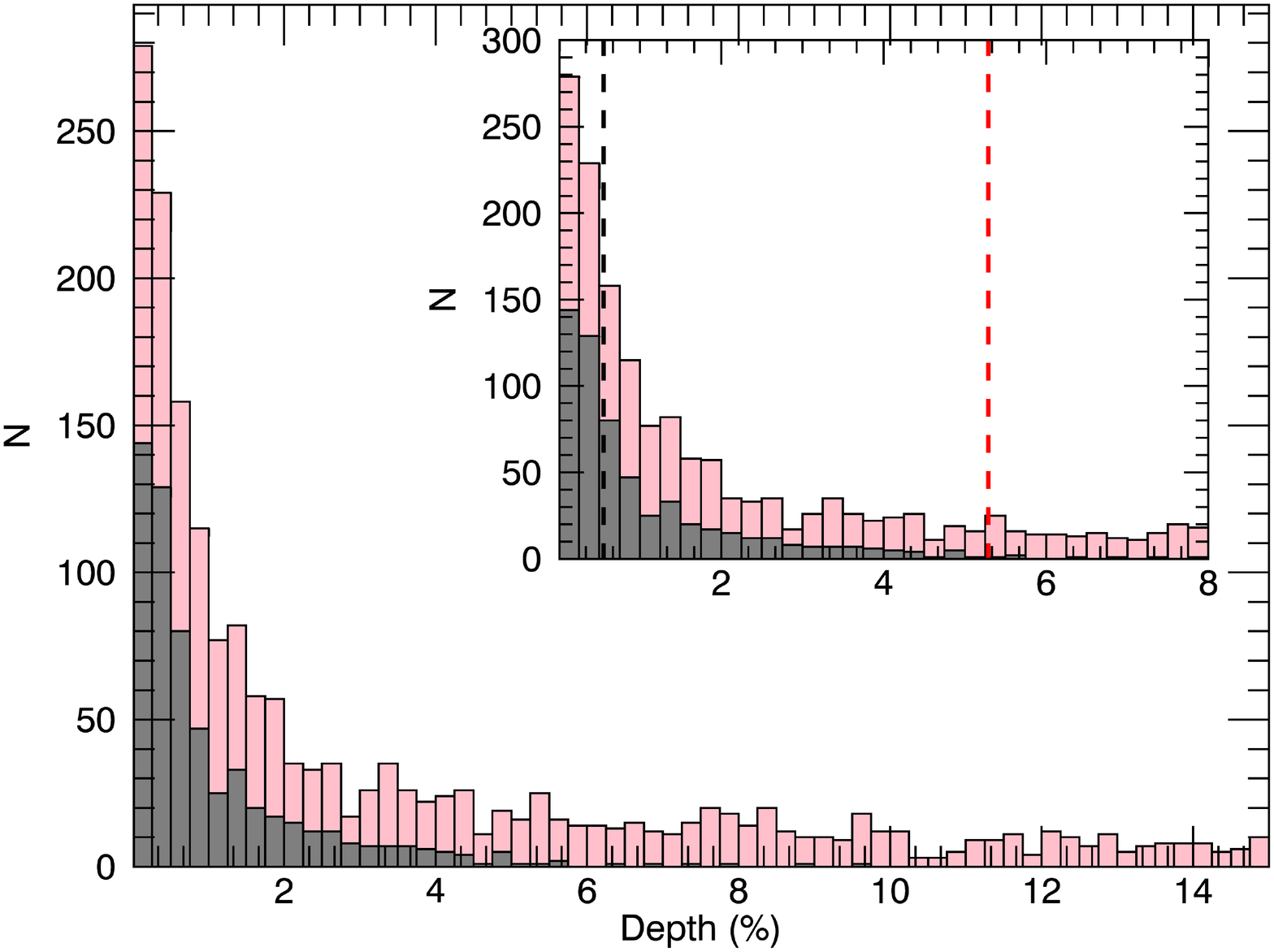}
\end{center}
\caption{Stacked histograms of the depth distribution (in \%) of EBs (pink) and candidates (grey). The inset shows an enlargement of the short end of the depth range and the dash lines give the median of each distribution. 
}
\label{fig:depth_dist_all}
\end{figure}

\section{Overview and follow-up observations of the candidates}
\label{sec:fup}

\subsection{Statistics of the candidates}
\label{ssec:cand_prop}

Figure~\ref{fig:period_dist_all} shows the period distribution of the candidates compared to that of the EBs. The two distributions are similar: both peak at 1.5 days, and their medians are 3.9 days and 3.1 days, respectively. Two thirds of both the candidates and the EBs have orbital periods shorter than ten days, and 90\% shorter than 25 days. In both cases, a handful do show orbital periods in excess of 100 days. Some of those are single transit events, but some were detected as periodic events during long runs, as was the case for CoRoT-9b \citep[e.g.]{Deeg2010}. 

The distributions of the transit depths, shown in Fig.~\ref{fig:depth_dist_all}, are more different. While both peak at 0.15\%, the medians depths are 0.54\% and 5.26\% for the candidates and EBs respectively. This is of course as expected, since the depth of the transits was one of the factors used in distinguishing EBs from possible planets. As a consequence, the distribution is truncated at large depths for the candidates, but not for the EBs. 

There is a noticeable spread in the number of candidates detected during each pointing, ranging from 8 in SRa01 to 50 in LRc02 (Table~\ref{tab:nb_cands}).
While the centre fields account for less dwarfs than the anticentre ones (33\,3351 against 38\,803), we find 6.3 $\pm$ 0.4 candidates per 1000 F, G, K, and M dwarfs surveyed in the anticentre fields and 10.5 $\pm$ 0.6 in the centre. This difference is also found in the number of planets with 0.41$\pm$0.1 $\permil$ and 0.63$\pm$0.1 $\permil$ respectively. By contrast, for the same targets sample, there is no such difference in the number of EBs detected in each pointing direction:  22.09$\pm$0.7 and 23.57$\pm$0.8 per thousand of the same population of targets, in the anticentre and centre directions respectively. The higher rate at which we detect planets in the centre fields might reflect some dependency on the properties of stellar populations located in opposite galactic directions. Assessing such a dependency would however require precise and complete parameters of the underlying stellar populations we are still lacking. 

This lack of a precise characterisation of the stellar population not only prevents a detailed assessment of the detection performance. Other limitations include the different duration of the \corot\ runs, their different noise properties due for example to different background levels or aging of the instrument, and the fact that the detections do not come from a single software, but from various ones that have been updated and optimised throughout the life of the mission. As a first attempt we checked any dependency with the run duration. Indeed, the median of runs duration is 54.3 days in the anticentre and 83.6 days in the centre.
We  calculated the Spearman's rank-order correlation between the number of candidates and the run duration. We found a correlation coefficient of 0.5626  with the evidence against a null hypothesis (p-value) 0f 0.34\%, indicating a weak but significant correlation between the number of candidates and the duration of the run. 

\begin{figure}[h]
\begin{center} 
\includegraphics[height=6.5cm,width=7.5cm]{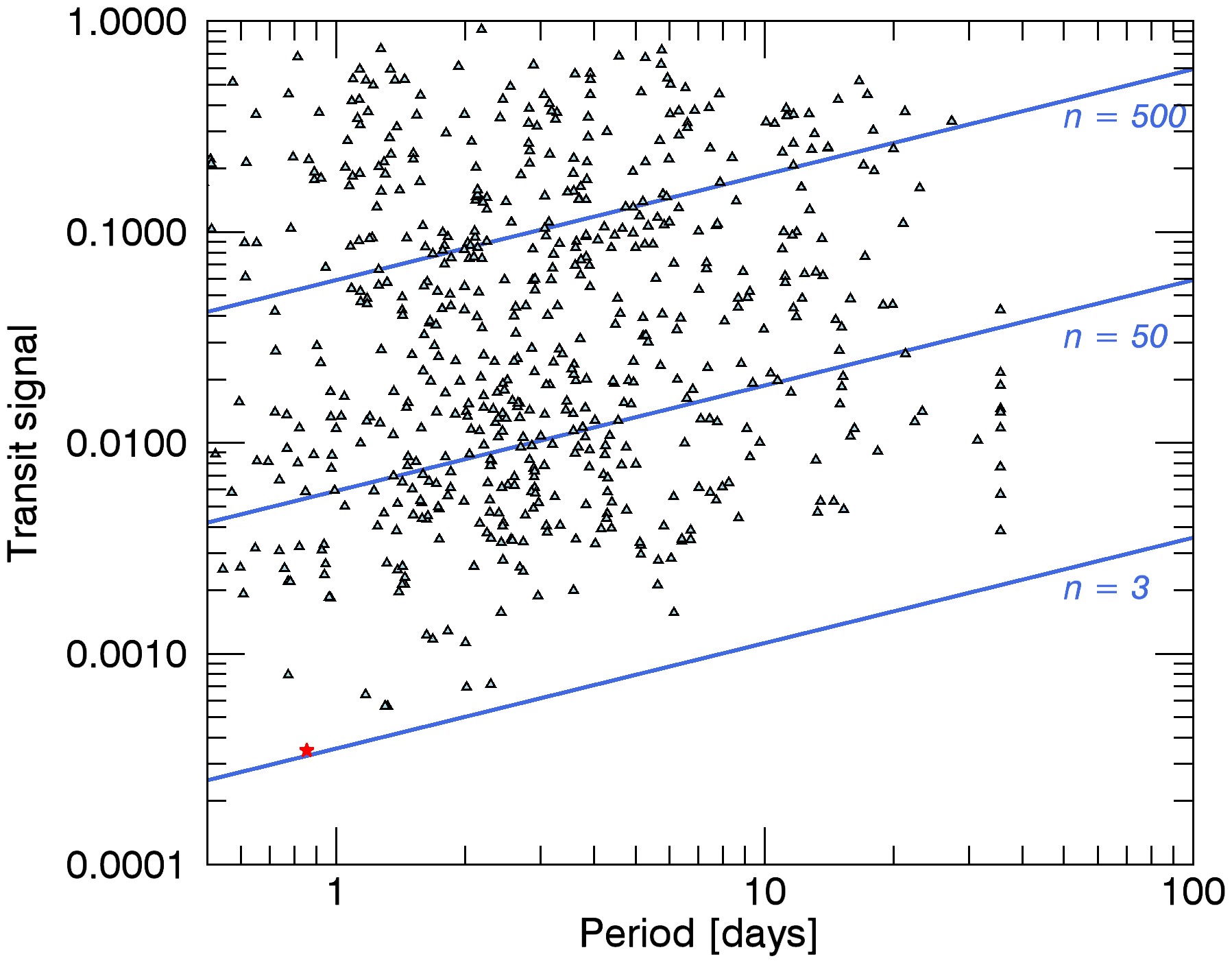}
\includegraphics[height=6.5cm,width=7.5cm]{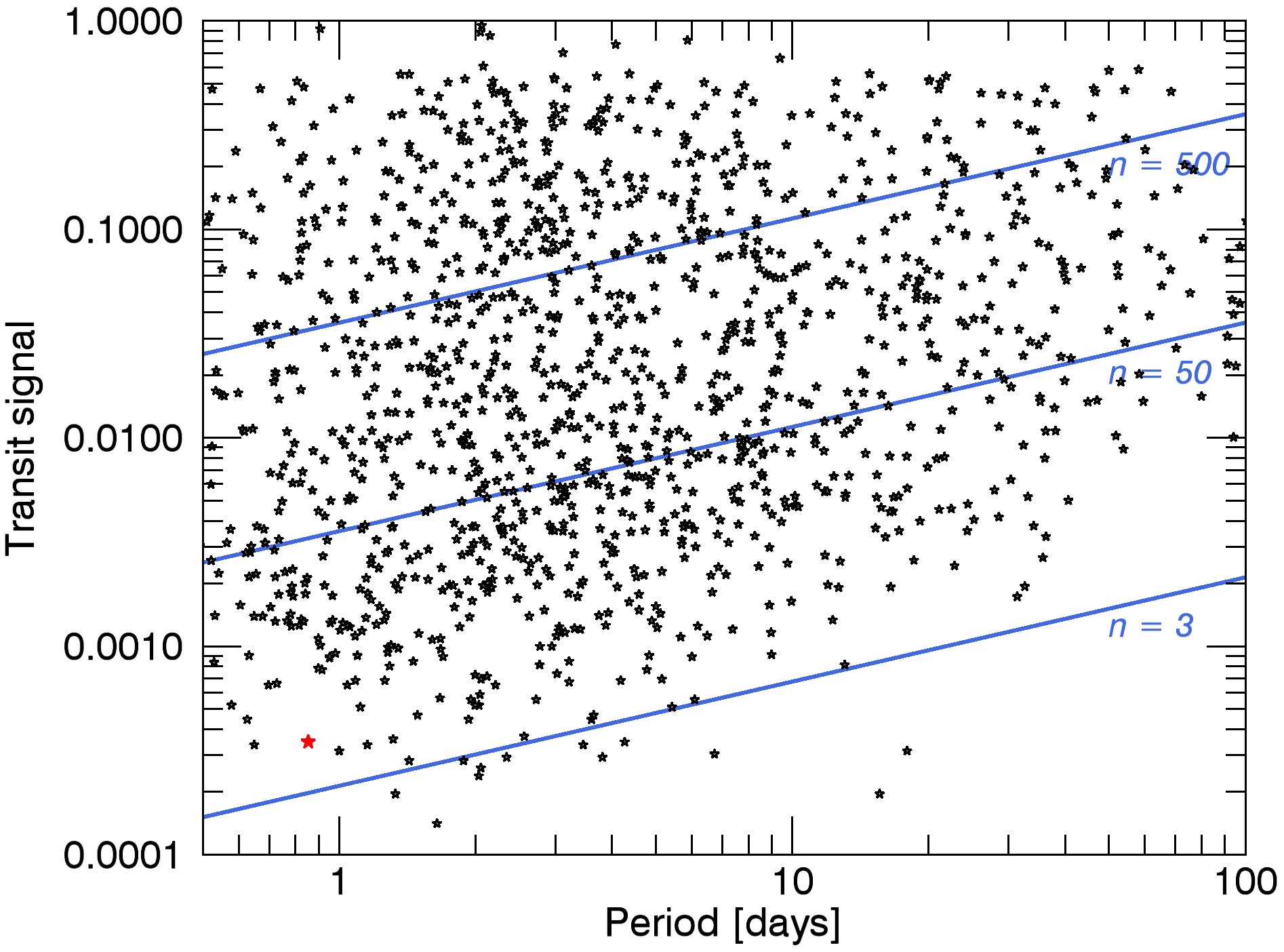}
\end{center}
\caption{Transit signal as a function of the period of all candidates and EB in fields whose duration was less than 40 days (top) and those whose duration was greater than 80 days (bottom). The plain blue lines show the expected transit S/N for three different noise levels (given by the n values). The position of CoRoT-7b is indicated by the red star.}
\label{Fig:duration}
\end{figure}

We also checked whether the distributions of depths and periods for the detections are consistent with what we might expect given the duration of the runs. We separated the short runs (duration less than 40 days: SRa01, SRa02, SRa03, SRa05, SRc01, SRc02, and LRa07) from the longest ones with a duration greater than 80 days, and compared in these two groups the candidates and detached eclipsing binaries to the expected transit signal that is the noise level over the transit duration, following the approach described by \cite{Pont2006}. It was calculated for a time sampling of 512~sec as the product of the depth of the transit and the square root of the number of points in the transit and using the median photometric precision at $R = 14$ given in Table~\ref{tab:fields}. Figure~\ref{Fig:duration} shows how the detection threshold varies as a function of the run duration. While the transit signal of a planet like CoRoT-7b appears at the limit of the detection threshold for short duration runs, it is well within CoRoT detection capacity when the duration of the run goes over 80 days. 
Longer run durations favour the detection of shallow or long orbital period transits, as expected.

\subsection{Follow-up observations of the candidates}
\label{ssec:fup_obs}

Ground-based photometric and spectroscopic follow-up observations formed a key part of the \corot\ exoplanet programme. Photometry taken during and just outside the transits (on-off photometry) with larger telescopes at higher spatial resolution, was used to confirm whether the transits occurred on the main target or a fainter nearby star. High contrast imaging helped identify background binaries or physical triple systems further. Radial velocity (RV) measurements allowed us to identify objects with multiple sets of spectral lines, and to measure the masses of any actual planets, together with the eccentricity of their orbits. These or additional spectroscopic data were also used to estimate the fundamental parameters (effective temperature, gravity, mass, radius, and age) of the target stars.
The role these ground-based observations in assessing the nature of the candidates has been already discussed in detail in previous run reports \citep[e.g.][]{Moutou2009,Cabrera2009} or planet discovery papers \citep[e.g.][]{Leger2009}, so we do not describe the full process in detail here. 

A total of \totalfu\ candidates were observed by at least one ground-based facility as part of the \corot\ follow-up programme, representing 70\% of all the candidates which had been deemed worthy of follow-up at one point or another. We note that the follow-up observations started as soon as possible after the end of each run, while the candidate vetting and light curve modelling process described in Section~\ref{sec:detvet} evolved and matured continuously during the mission. Thus, 88 of the candidates initially deemed worthy of follow-up were later identified as unambiguous EBs on the basis of their light curves. At that point, they were removed from the follow-up programme, but some had already been observed. This sample provides us with a valuable opportunity to check the validity of our vetting procedures.

\begin{figure}[h]
\begin{center} 
\includegraphics[height=9cm,width=8.5cm]{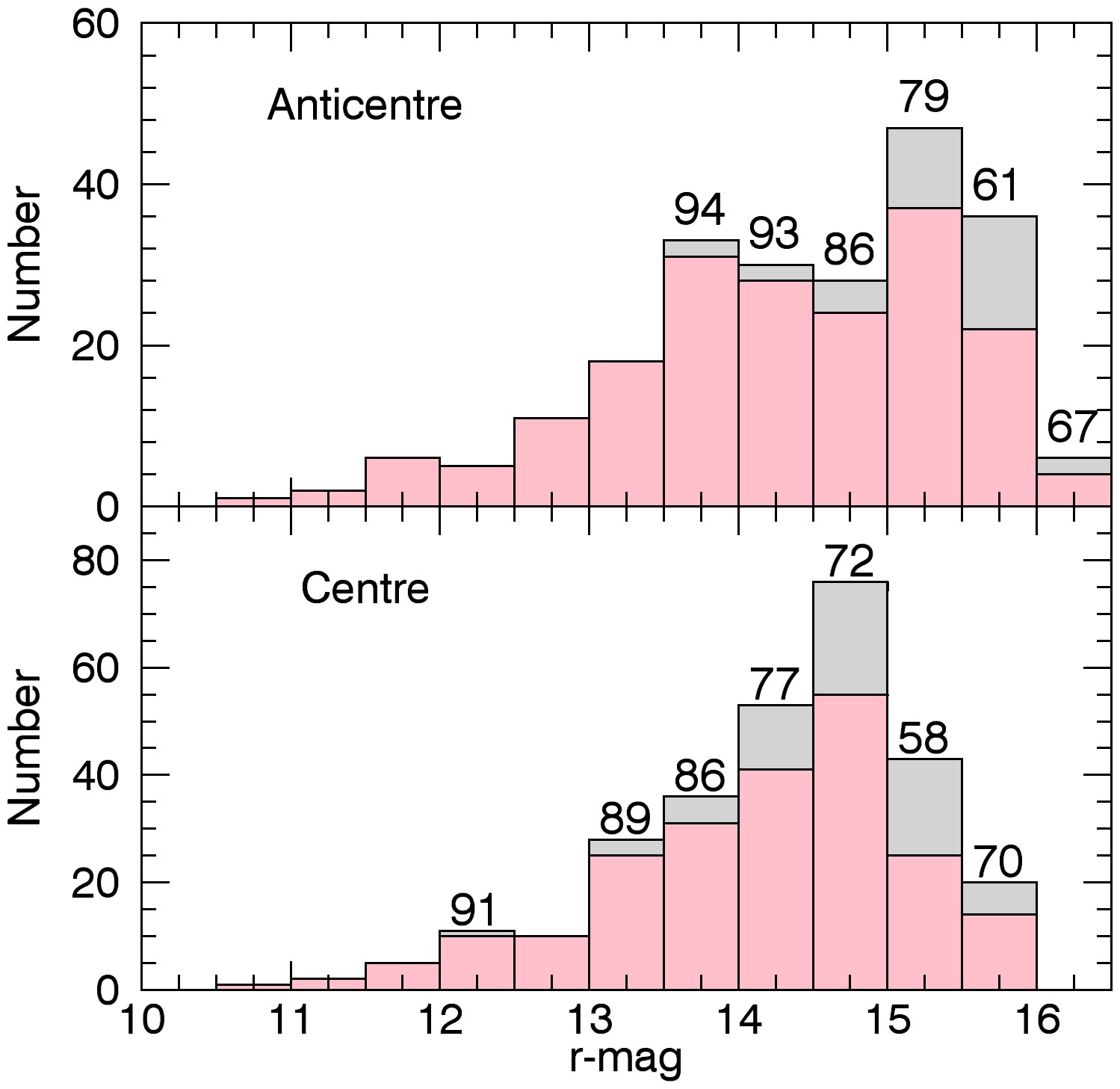}
\end{center}
\caption{$r$-magnitude distribution of the full set of candidates (grey) and those that received follow-up observations (pink). The numbers at the top of each bin give the percentage of candidates observed from the ground in that bin when different from 100\%.}
\label{fig:mag_dist_fup}
\end{figure}

Figure~\ref{fig:mag_dist_fup} shows the distribution of the $r$-band magnitude of the candidates, together with the fraction which received at least some follow-up observations in each magnitude bin. This figure excludes the candidates which were initially included in the follow-up programme but later discarded on the basis of a refined light curve analysis. The only exceptions to this rule are four cases from IRa01, the first pointing of the mission, because the follow-up for that run was completed before the re-analysis of any of the light curves began. As Fig.~\ref{fig:mag_dist_fup} shows, candidates spanning the full magnitude range were followed up, but brighter stars were given higher priority: nearly all the candidates with $r < 14$ received follow-up observations, while the fraction drops to 78\% for $14 \leq r \leq 15$, and 63\% for the faintest targets with $r>15$. Overall, almost 85\% of the candidates included in the figure were observed.

\subsection{Outcome of the follow-up observations}
\label{ssec:fup_res}

Based on the results of the follow-up observations, transit candidates were assigned to one of the following classes: 
\begin{enumerate}
\item[-] {\sl Spectroscopic binary (SB)}: Either the radial velocity cross-correlation function (CCF) of these candidates shows multiple, well separated peaks (indicating a double- or triple-lined spectrum), or the RV variations clearly indicate a stellar mass companion.
\item[-] {\sl Wide CCF (WCCF)}: The CCF of these candidates shows a very broad peak, preventing the measurement of precise RVs. This can occur either because the host star is hot, and its spectrum contains few abosorption lines, or is rapidly rotating, as is typical of A and early F-type stars. While planetary companions are not excluded for these more massive stars, their characterisation remains out of reach with standard methods, and these candidates are set aside. 
\item[-] {\sl Contaminating eclipsing binary (CEB)}: This category covers all cases where the \corot\ aperture contains an eclipsing binary whose light contaminates the target star, giving rise to a transit-like signal, independently if the EB is a background object, or is physically related to the brighter star (triple system). These configurations are identified through on-off photometry as described in detail in \cite{Deeg2009}, high contrast imaging, or the so-called RV mask effect, where the measured RV changes significantly depending on the cross-correlation mask used (indicating that stars of more than one spectral type contribute to the spectrum).
\item[-] {\sl Photometric eclipsing binary (PEB)}: These are candidates that were initially included in the follow-up programme, failing a clear identification as EB during the vetting tests described in Section~\ref{ssec:vet}, but were later identified as EBs based on a more thorough analysis of their light curves, after the start of the follow-up observations. In most cases, these objects were down-graded either because a secondary was detected at a phase other than 0.5, or on the basis of a more quantitative assessment of the eclipse depths in the three \corot\ band-passes. We expect these to be mostly EBs that are identical to the target stars, although instances of contaminating EBs may be present here as well.
\item[-] {\sl Unresolved}: This category comprises all the candidates whose nature remains unresolved, because the follow-up observations were either inconclusive, incomplete, or in some cases (for the lowest priority objects, mostly those at high magnitude) never started. The follow-up observations may remain inconclusive for a number of reasons: i) ground-based photometry demonstrates that the transit is on the main target, but the latter is too faint to allow RV measurements at the required precision; ii) repeated RV observations reveal no significant variation consistent with the ephemeris of the transits; iii) shallow transits for which on-off photometry are not precise enough to pinpoint the precise source of the photometric signal, and no RV measurements could be performed because of the faintness of the target.
\item[-] {\sl Planet}: Only candidates having passed a whole battery of tests, including unambiguous detection of the RV signal induced by the companion, or full statistical validation of the companion's planetary nature using the \corot\ light curve and all available ground-based data, are included in this category. All  but the most recent discoveries of these planets have been published in dedicated papers (see Bord\'e et al., {\it subm.}; Grziwa et al., {\it in prep.}; Gandolfi et al., {\it in prep.} for the most recent).  We keep them all in the final candidates catalogue for consistency and further assessment of the flag system. We note that we have included those that were reported as brown dwarfs in this category.    
\end{enumerate} 
These categories are reported for each candidate in Table~\ref{tab:candidates}, while the number of candidates in each category is summarised, run by run, in Table~\ref{tab:nb_cands}. 

\begin{figure}[h]
\includegraphics[height=9cm,width=8.5cm]{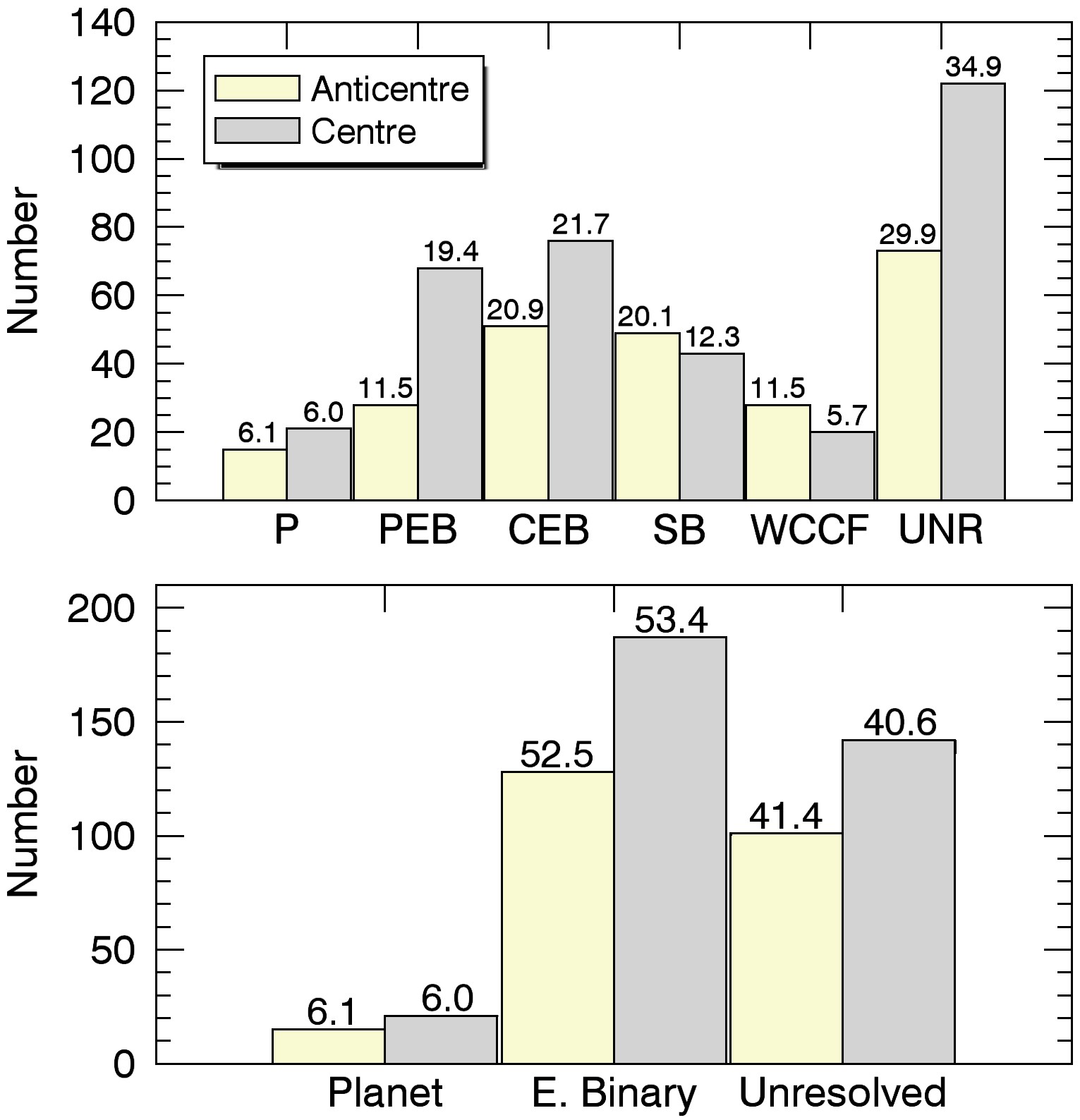}
\caption{Distribution of the candidates among the six categories defined on the basis of the follow-up observations (top), and the three larger groups obtained by merging some categories together (bottom) in centre and anticentre (see text for details). The number at the top of each bin gives the percentage of the corresponding category in the considered direction.}
\label{fig:cand_pie}
\end{figure}

Figure~\ref{fig:cand_pie} shows the distribution of the candidates among the categories defined above, for the Galactic centre and anti-centre fields separately. In both directions, about 40\% of the candidates remain unresolved (these are discussed further in Section~\ref{sec:unsolved}). Faint stars, which are challenging for precision RV measurements, account for a large part of this class, but it also contains some relatively bright targets, for which no clear RV variation could be detected. Confirmed planets account for only 6\% of all the candidates.

A more synthetic overview of the status of the candidates after follow-up can be obtained by merging categories which were distinct observationally, but are essentially the same in their underlying nature. The nature of the WCCF objects remains unknown: the transits could be caused by a small star, brown dwarf, or a Jupiter-sized planet, or by a contaminating EB. Indeed, one can not exclude the presence of a Jupiter-size planet orbiting a A-type star. Ultimately, WCCF objects can be merged into the `unresolved' class. The classes SB, PEB, and CEB can be merged in a single EB class. The SB class, identified as such through follow-up observations, consists indeed of undiluted EBs. In the same way, we only know for sure that the transit events identified as CEBs are from contaminating objects, whereas the other cases identified as PEBs may be identical to the target star or may be contaminators as well. Figure~\ref{fig:cand_pie}, bottom, shows the distribution of the candidates among this smaller set of classes. The unresolved cases now account for a little over 40\% of the candidates, while EBs are the main source of resolved false positives, at more than 50\% of the total. The confirmed planet fraction is unchanged at 6\%. Among the total number of resolved configurations in both directions, EBs be they diluted or not, account for 89.7\% and planets are 10.3\%.

\section{Evaluation of the candidate screening process}
\label{sec:hindsight}

While the flags described in Section~\ref{ssec:flags} were computed automatically for all candidates and detached EBs, they were not used in selecting or prioritising candidates for follow-up observations, or if so, only in an ad-hoc manner for individual cases. This is partly because the flags were not available during the early phase of the mission, and partly because we were wary of using an automated process in case we discarded good planet candidates. This could happen, for example, because they had non-standard properties (unusual host star spectral type, transit timing variations) or because the transit modelling failed to converge properly. In principle, however, the flags could have been used both to discriminate automatically between EBs and candidates (thereby avoiding, or considerably reducing, the visual vetting stage), and to prioritise the candidates for follow-up observations. Having performed the visual vetting and the follow-up, we can now use the benefit of hindsight to investigate, after the fact, to what extent the flag system could have been used in this manner. 

\subsection{Candidates versus EBs} 
\label{ssec:cands_vs_EBs}

\begin{figure}[ht]
\begin{center} 
\includegraphics[height=6cm,width=8.cm]{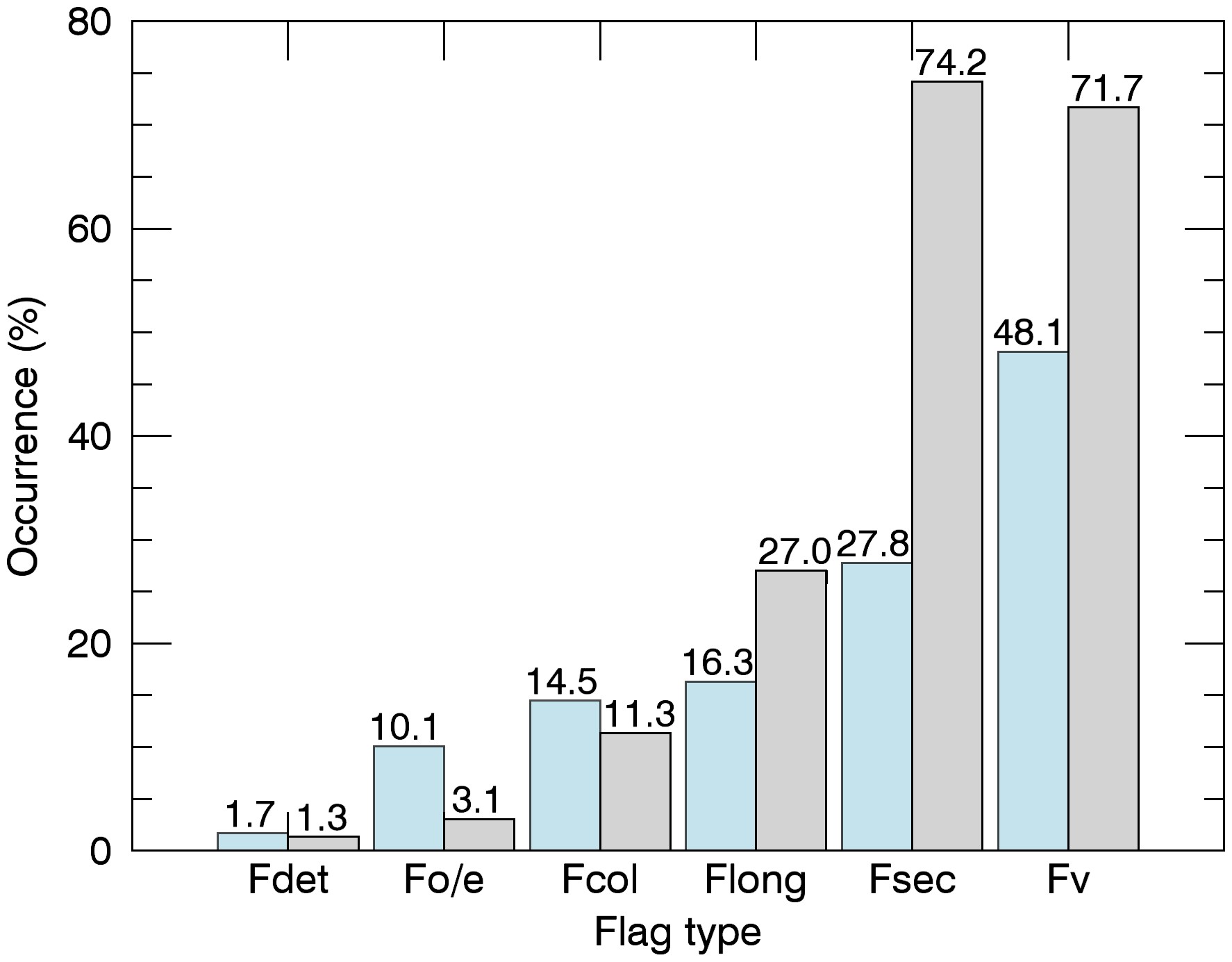}
\end{center}
\caption{Occurrence in percentage of each flag (see sect.~\ref{ssec:flags}) for the candidates (blue) and the EBs (grey).}
\label{fig:flag_dist}
\end{figure}

We first checked how useful the flag system could have been at the vetting stage, by comparing the occurrence of the different flags for the planet candidates and for the EBs, which are shown in Figure~\ref{fig:flag_dist}. 

The most striking feature of the figure is that the relative order of occurrence of the flags (most common to least common) is virtually identical for both candidates and EBs. $F_{\rm V}$ (V-shaped transit) and $F_{\rm sec}$ (secondary eclipse) are the most common, set in 48 and 28\% of the candidates, respectively, and in more than 70\% of the EBs. The next most common flag is $F_{\rm long}$ (long-duration transit), which is set in 16\% of the candidates and 27\% of the EBs. The remaining flags are all set in less than 15\% of the candidates or EBs. In particular, $F_{\rm det}$ (low detection significance) is set only for $<2$\% of the candidates and EBs, which shows that the detection teams were conservative when selecting transit-like events. The individual occurrence rates of individual flags for candidates and EBs do differ, so that -- for example -- a given object is more likely to be an EB than a planet candidate if either of $F_{\rm sec}$, $F_{\rm V}$ or $F_{\rm long}$ are set, but these differences are certainly not strong enough to allow a clear distinction to be made between the two groups. 

The relatively high number of candidates that have flags on secondary eclipse or long-duration transit is due to human decision during the vetting process. Indeed at the vetting stage, shallow secondaries that would trigger the $F_{\rm sec}$ flag could not necessarily be seen by eyes. Furthermore, real planets, such as CoRoT-1b or CoRoT-2b \citep{Alonso2009,Snellen2009}, can have detectable secondaries and thus we did not want to be excessively strict in excluding them. Similarly, while V-shaped transits are much rarer for companions that are small compared to the primary star, they are not entirely excluded (see Table~\ref{tab:planets1flag}), and therefore we refrained from using the shape of the transits at the vetting stage, though we did use the depth.

\begin{figure}[h]
\begin{center} 
\includegraphics[height=6cm,width=8.5cm]{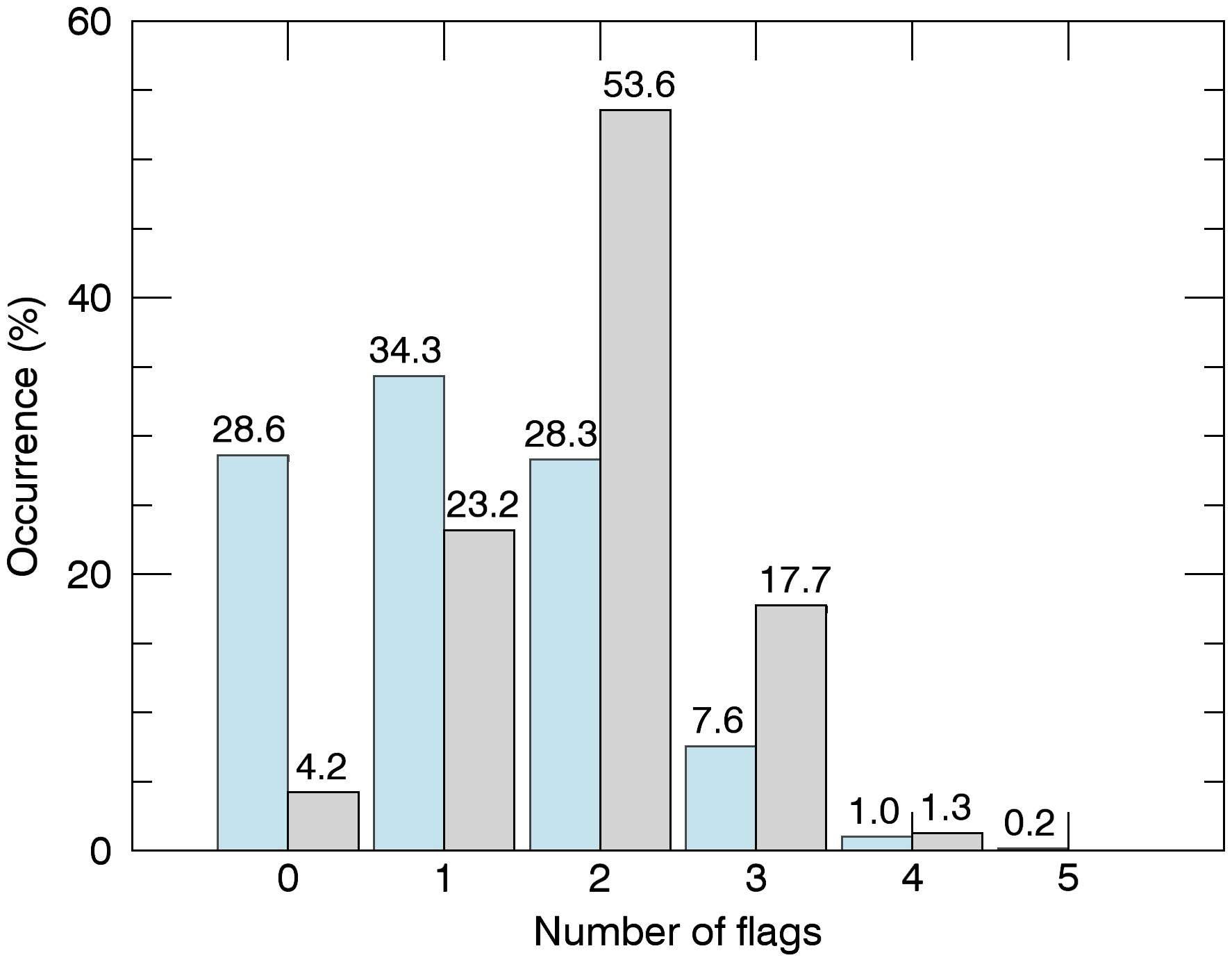}
\end{center}
\caption{Distribution of the number of flags occurrence of the candidates (blue) and the EB (grey). We note that none of the detected events are attributed six flags.}
\label{fig:nb_flags}
\end{figure}

We can go a step further by examining the number of flags set for individual candidates or EBs, and the combinations of flags which are set simultaneously. This is illustrated in Figure~\ref{fig:nb_flags}. 
28 \% of the candidates have no flag set at all, but the situation occurs for a few of the EBs as well. The vast majority of the two groups have one or more, while none have more than five out of the six flags set. Here we do see a clear difference between the candidates and the EBs: while 63\% of the candidates have $<2$ flags set, 72\% of the EBs have $\ge 2$. This suggests that the overall number of flags set may be a better, though still imperfect, indicator of whether a given object belongs in the EB or planet candidate group. We note also that all 37 of the \corot\ planets included in this analysis but one have fewer than two flags set. 


\begin{figure}[h]
\begin{center} 
\includegraphics[height=6cm,width=8.5cm]{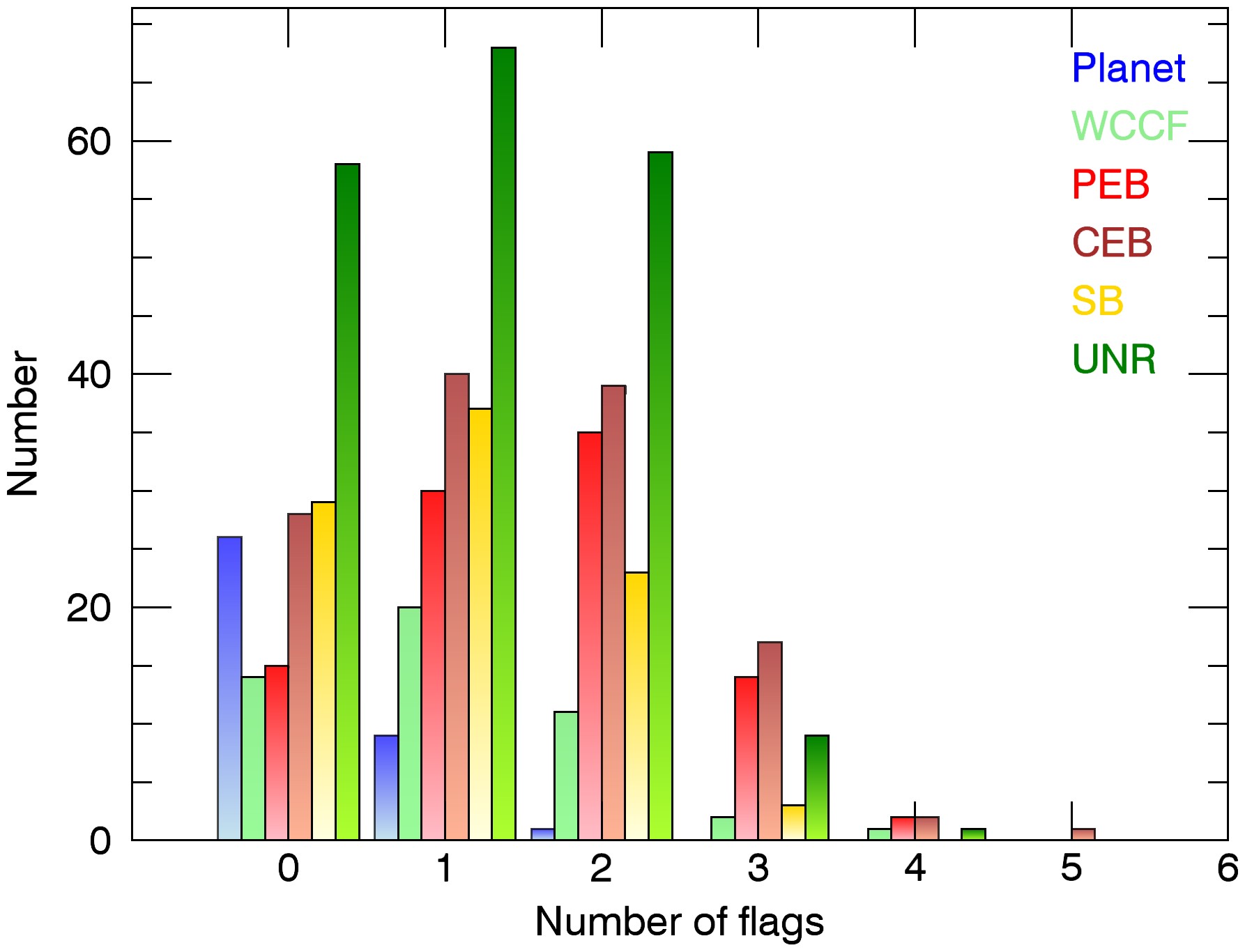}
\end{center}
\caption{ Histograms of flag occurrence of the candidates over the various categories defined following follow-up results (see Sect.~\ref{ssec:fup_res})}
\label{fig:flagsCand}
\end{figure}


\begin{table}
\begin{center}
\caption{\corot\ planets with one flag triggered}
\begin{tabular}{lccc}
\hline
\hline
Planet   &  Flag            &   Host-star  &  $R_{\star}$   \\
            &               &  spectral type &  $[R_{\odot}]$   \\
\hline
\Ctone   & secondary & G0V  & 1.11 $\pm$ 0.05\\
\Ctten &  V-shape & K1V & 0.79 $\pm$ 0.05 \\
\Ctsv &  duration & G5V & 1.19$_{-0.13}^{+0.14}$ \\
\Cttwt &  colour-depth & G0V &  1.136$^{+0.038}_{-0.09}$ \\
\Cttwf &  secondary  & K1V & 0.86 $\pm$ 0.09 \\
\Cttwfv &  V-shape & F9V & 1.19$^{+0.14}_{-0.03}$ \\
\Cttws &  duration & G5IV & 1.79$^{+0.18}_{-0.09}$\\
\Cttwh &  duration & G8/9IV & 1.78$\pm$0.11 \\
\hline
\end{tabular}
\label{tab:planets1flag}
\tablebib{(1) \cite{Barge2008};
(2) \cite{Bonomo2010}; (3) \cite{Csizmadia2011}; (4) \cite{Moutou2014};
(5) \cite{Alonso2014}; (6) \cite{Almenara2013}; (7) \cite{Cabrera2015}
}
\end{center}
\end{table}

\subsection{Follow-up results versus flagging}
\label{ssec:flags_vs_fup}

We took advantage of the large follow-up effort to check our flag system against the results obtained through complementary observations. To that end, we checked the distribution of the number of flags for each of the six classes of candidates defined in Sect.~\ref{ssec:fup_res}; the results are shown in Fig.~\ref{fig:flagsCand}. The EBs among candidates identified through follow-up observations follow the same trend in number of flags as those identified through the analysis of their light curve. For the other classes, the outcome is less obvious. The confirmed planets that have one flag set are highlighted in Table~\ref{tab:planets1flag}. In each case this flag can be explained by a real characteristic of the system. If a reliable spectral type had been available for the host-star before the automated transit modelling and flagging step, none of the planets would have had $F_{\rm long}$ set. Some would still have had $F_{\rm V}$ or $F_{\rm sec}$, but these are the few exceptions with genuinely high impact parameter or marginally detectable secondary eclipses.

On the other hand, taking at face value the observation that none of the confirmed planets had more than one flag set (except for CoRoT-36b/LRc07\_E2\_0307, Grziwa et al. {\it in prep.}), one might conclude that follow-up observations could have concentrated on the candidates that received at the most one flag. This would have reduced the number of follow-up targets by 36\%. However, the more salient conclusion of this exercise is that the flag system that we defined is too blunt a tool for reliably distinguishing between genuine planetary transits and astrophysical false positives reliably. A more nuanced approach is required, including for example a probabilistic assessment of the planetary nature of each candidate, as proposed by  \citet{morton2012} and implemented in the {\sc vespa} package \citep{vespa}.

\section{Unresolved candidates}
\label{sec:unsolved}

Following the conclusion of Section~\ref{ssec:flags_vs_fup}, planets could be searched for among unresolved candidates with one flag triggered at most, that is \totalunresfl\ candidates out of \totalunres\ whose status remains yet unresolved. About half of them have not been subject to ground-based complementary observations, and for those that were observed (56), the complementary observations remained inconclusive about their exact nature. Among those that were not observed, none is brighter than r-mag=13, five are in the range 13.4 to 14, and the remaining ones (51) are much fainter. As discussed in Sect.~\ref{ssec:fup_obs} follow-up observations were also driven by the magnitude of the targets. We checked the five brightest candidates without follow-up observations and we found that either the target was suspected being a giant or the USNO-A2 r-magnitude that was used in the input catalogue, mostly for the short runs, was much higher than the one from PPMXL, which is currently used in the last release of the input catalogue. In both case, the target ended up with a low priority for follow-up observations.

\begin{figure}[h]
\begin{center} 
\includegraphics[height=6.5cm,width=9cm]{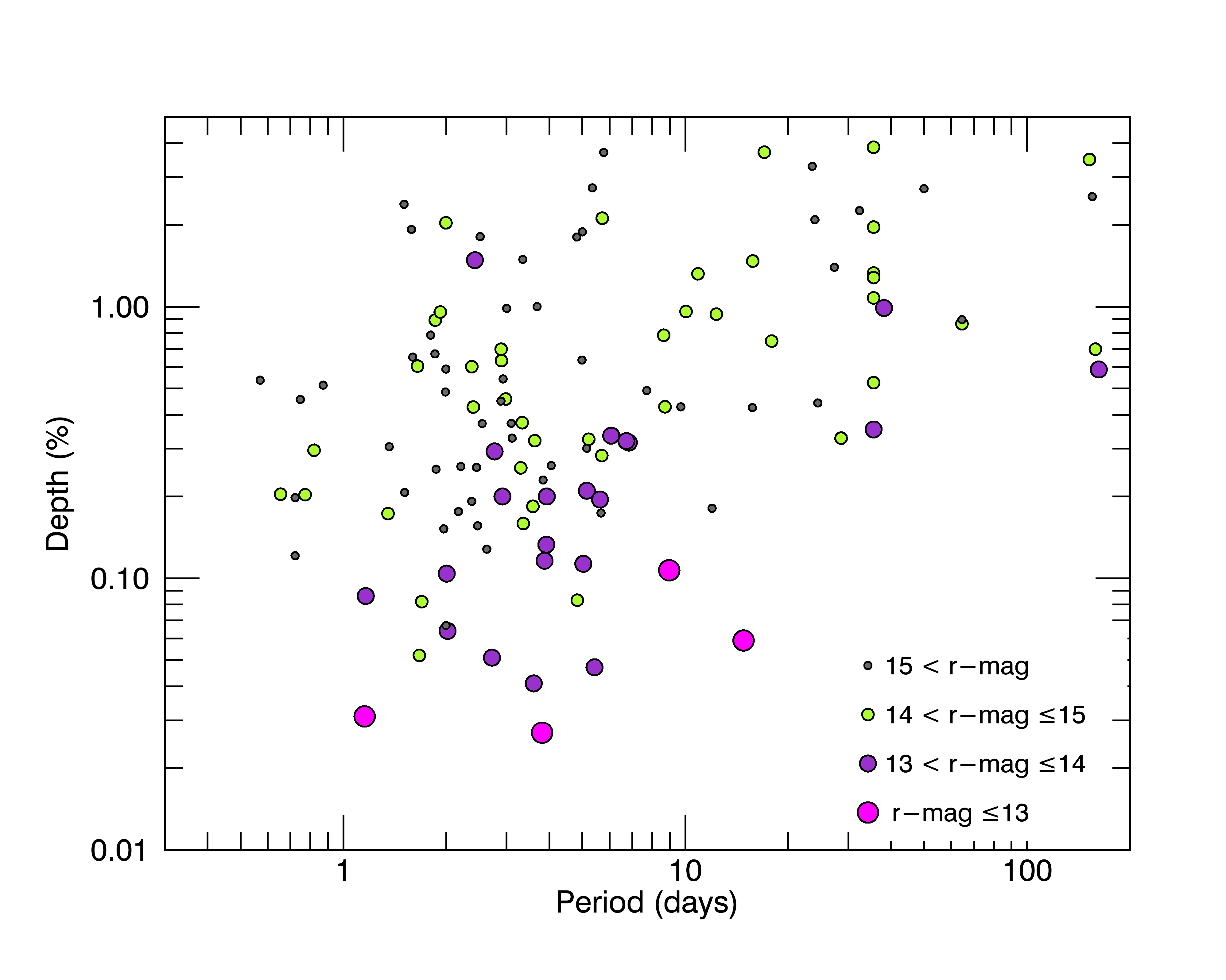}
\end{center}
\caption{Depth versus period of the unresolved candidates with at most one flag, according to their magnitude.}
\label{fig:depth_period_unres}
\end{figure}

Figure~\ref{fig:depth_period_unres} shows the transit depth of these \totalunresfl\ candidates as a function of their period and their magnitude. Five have a depth between 3 and 4\%. They are likely stellar systems but, in absence of a reliable spectral type of the host star, they are kept in the unresolved category. We note however that 68 of these unresolved events have a depth less than 0.05\% among which 12 a depth less than 0.01\%.
About 90\% have a r-mag $>$ 14. For Jupiter-size planets radial velocity measurements remain difficult  at the faint end of the \corot\ magnitude range (typically for r-mag $>$ 14.5).  For those whose host-star is brighter, the domain of Neptune-size planets and smaller is still challenging the current spectrograph performances. The later is well exemplified by CoRoT-22b \citep{Moutou2014} whose nature could not be fully secured by radial velocity measurements and required a complex process of planet validation, that also included ground-based images \citep{Diaz2014}.

According to current spectral classification \citep{Damiani2016}, 68 \% of these unresolved candidates are classified as dwarfs (luminosity class IV or V). The ExoDat spectral classification is not completely reliable but is at least indicative. As a result, selecting the unresolved events classified as dwarf and with one flag at the most, we get  81 good candidates. 
Assuming that the unresolved cases follow the same distribution as the resolved cases and would distribute over EB and planet classes, following the same proportions as derived in Sect.~\ref{ssec:fup_obs} from the resolved cases, we may expect that 8.4 $\pm$ 3 planets are lurking in this category, waiting to be confirmed as such. This highlights the challenge of establishing the planetary nature of \corot\ candidates. Even for Jupiter-sized planets, RV characterisation difficult at the faint end of the \corot\ magnitude range, while even bright stars hosting Neptune-sized planets require many RV observations. Examples of confirmed \corot\ planets, which were at the limits of the capabilities of RV facilities, include CoRoT-16b \citep{Ollivier2012}, CoRoT-19b \citep{Guenther2012}, and CoRoT-22b \citep{2014MNRAS.444.2783M}.

\section{\corot\ planet occurrences}
\label{sec:occurrence}
 
\begin{figure}[h]
\begin{center} 
\includegraphics[height=6.5cm,width=9cm]{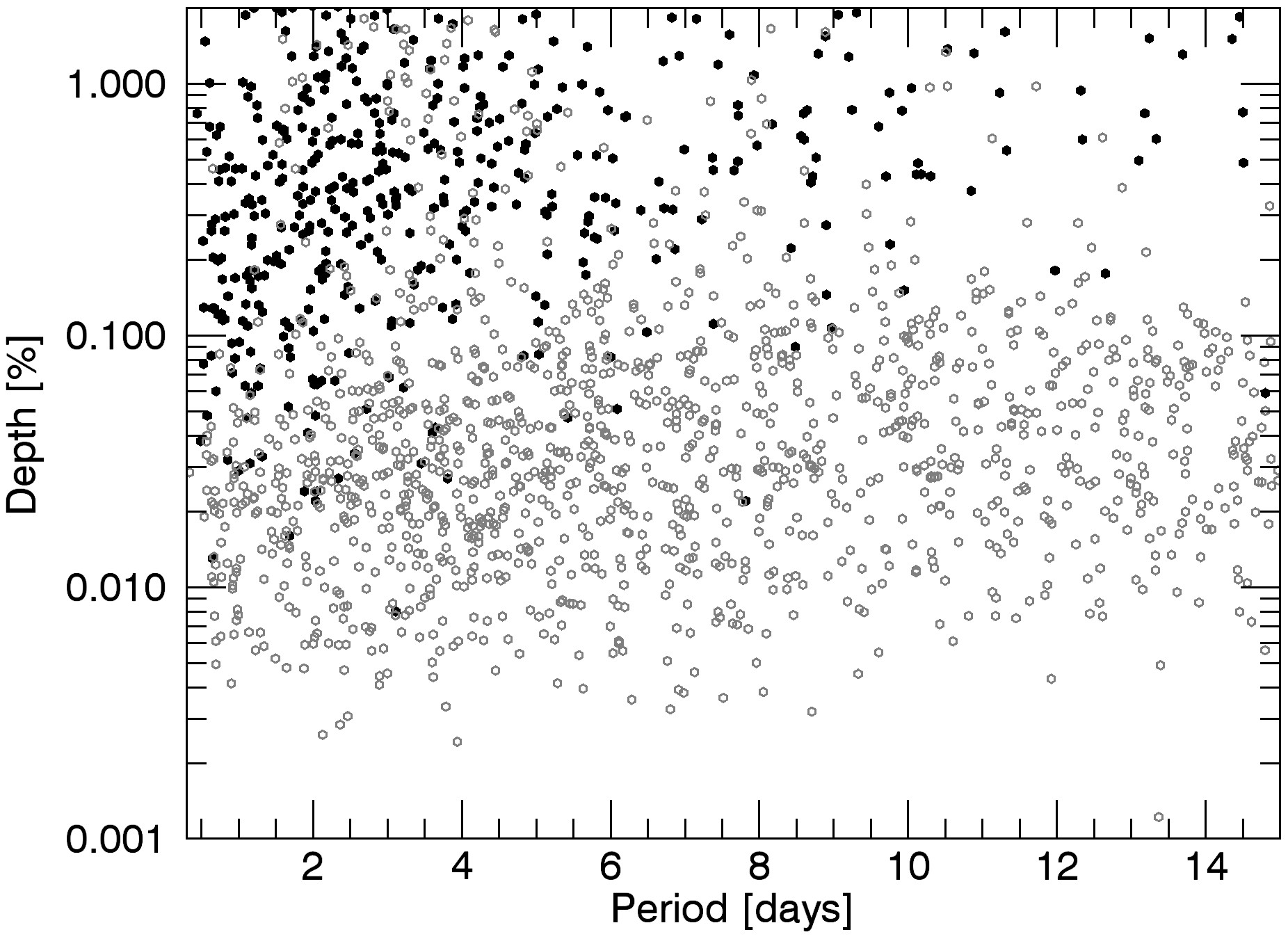}
\end{center}
\caption{Depth versus period of the unresolved candidates and confirmed planets detected by CoRoT (black dots) and by Kepler (grey dots).}
\label{fig:depth_Period}
\end{figure}

The number of planets detected is a function of the sensitivity and contamination of the complex \corot\ detection, vetting and follow-up process, as well as the underlying occurrence rate for different types of planets around different types of stars. A detailed estimate of the occurrence rate would require end-to-end simulations to measure the sensitivity, as well as a careful accounting of the various sources of false alarms among the candidates. A number of studies have done this, for example, for \emph{Kepler}, where the task is made somewhat more manageable by the extremely high precision of the light curves and the uniform detection and vetting process of the candidates \citep[see][e.g.]{Howard2012,Fressin2013,Christiansen2015}. Doing this for \corot\ is more tricky, not least because of the multiple and evolving pipelines used for candidate detection and selection over the mission lifetime. A proper completeness estimate is available for only one of the pipelines used \citep{Bonomo2012}, and then only for the special case of small (Neptune-size) planets. Additionally, the various pointings had different durations, and targeted regions of the sky with different stellar properties. Finally, the aging of the detectors over the mission lifetime affected the noise properties of the data \citep{Aigrain2009,Asensio-Torres2016} which might have had a significant impact on the detectability of the smaller planets in particular, but this effect has not been quantified. Proper completeness estimates of the \corot\ planet yield, or of the candidates catalogue, are thus beyond the scope of this paper. However, we can refine our calculations somewhat, if only to check to what extent the planet yield is consistent with occurrence rates derived from previous surveys. In support of the results presented in the following subsections,  Fig.~\ref{fig:depth_Period} shows candidates and confirmed planets as detected by the two missions in the period - depth diagram. Clearly there is a weak overlap only between the two missions detections in this parameters space. 

We consider the number of confirmed planets detected by \corot\, that is all published objects with a \corot\ planet number.  We do distinguish between  small ($R<5\,\RE$) and large (giant) planets. At least 2 of the latter, CoRoT-15b \citep{Bouchy2011} and CoRoT-33b \citep{Csizmadia2015}, with masses $>30\,\MJ$, are classified as brown dwarfs. Furthermore, \Ctthree\ \citep{Deleuil2008}, with a mass of $21.7\,\MJ$, could be considered either as a low-mass brown dwarf or massive planet.  For the purposes of the following discussion, we treat \Ctthree\ as a brown dwarf, and consider transiting brown dwarfs and hot Jupiters separately. However, we emphasise that this approach follows the standard classification and could be questioned as the distinction between planet and brown dwarf is unclear and increasingly controversial \citep[see][]{Schneider2011,Hatzes2015}. Finally, as the orbital periods of \corot\ planets fall in a relatively narrow range (from a few hours to just under 13 days, except for CoRoT-9b), we do not attempt to look for trends in period space. 

\subsection{Small size planet occurrences}
In the small planet group, \corot\ detected four confirmed or validated planets: CoRoT-7b, CoRoT-22b and the two planets of the only \corot\ transiting multi-planet system, CoRoT-24b and c. To convert this to a crude occurrence rate, we first consider only the 16 pointings surveyed for more than 70 days. Such long durations were needed to detect these small planets, apart from CoRoT-7b, which was detectable after only 20 days of observations, but is unusual in that it orbits one of the brightest K-dwarfs observed during the entire mission. Additionally, we restrict ourselves to the sample of G, K and M-type targets in these runs with a luminosity class of V and a magnitude  $r^\prime < 15.5$. This stellar sample corresponds to the one for which we have an estimate of the detection completeness, of $36.6 \pm 0.4\%$ for transiting planets in this size range from \cite{Bonomo2012}. 

We found a total of 19\,562 targets that satisfy these criteria. This number is corrected from targets that have been observed more than one time in different runs.  \citep{Damiani2016} assessed the statistical accuracy of the spectral classification based on broad-band multi-colour photometry in the \corot\ input catalogue, and found that in the worst case, 85\% of targets classified as dwarfs are indeed well-classified, and that the spectral type is generally good up to half a spectral class for the majority of \corot\ targets. We thus adopted an error of 15\% in the number of stars around which \corot\ could have found small transiting planets. Finally, following \cite{Howard2012}, we corrected for the probability that a given planet transits using the $a/R_\star$ parameter measured from the light curve. Combining all these factors, we obtained an approximate occurrence rate for planets with $R_p < 5\,\RE$ and periods of $P < 10$\,days around G, K, and M main sequence stars of $0.46 \pm 0.28 \%$. \\
Because the CoRoT planets validation is strongly driven by the follow-up process, to account for its incompleteness we added candidates in the small size domain. Among the candidates whose nature remains unknown and depth is compatible with being a Neptune-size or smaller planet (i.e. transit depth less than 0.20\%), we selected those detected in the same reference sample as the three planets, and with one or zero flags. We ended up with 12 candidates. All these candidates have an orbital period less than 15 days. Assuming all these candidates are real planets, this gives an occurrence rate of $2.2 \pm  0.6 \%$, a value which is still more than three times less than that the percentage of stars with at least one planet of $7.43 \pm 0.5 \%$ derived by \cite{Fressin2013} for the class of small Neptunes and super Earth in the Kepler field. Our result confirms the discrepancy pointed out by \cite{Bonomo2012} between the \corot\ and \kepler\ detection rates for this class of small-size planets.  

\subsection{Giant planet occurrences}

For the giant planets, we followed the same methodology, but we enlarged the stellar sample of reference to all spectral types from F5 to K5 and with no restriction on the run duration. In order to mitigate incompleteness in the follow-up process in the planet occurrence, as few candidates fainter than $r^\prime \simeq$ 15.5 were followed-up from the ground, we adopted $r^\prime < 15.1$ as a conservative limit in magnitude. This gives  26\,391 targets without duplication. Considering only planets that have orbital periods of ten days or less in this range of magnitudes (21 planets), we obtained an approximate occurrence rate of $0.88 \pm 0.2\%$, without correcting for the detection completeness. The later has never been properly estimated for \corot, but in this range of transit depths, it is reasonable to assume that very few transits are missed. Assuming a conservative value of 90\% for the detection completeness gives an occurrence rate of 0.98 $\pm 0.26\%$. This result is in agreement with the occurrence rates estimated for Hot Jupiters from radial velocity surveys \citep{Wright2012,Mayor2011} but not with the estimates for \emph{Kepler} data \citep{Howard2012,Fressin2013,Santerne2016}. The later range from 0.4 to 0.5\% for F, G, and K dwarfs, that is less than our finding for \corot\ data. We note that separating giants detected in the anticentre fields from those in the centre ones, we found no significant difference in the occurrence rates of the close-in giants which are 0.96 $\pm$ 0.37 and 0.99 $\pm$0.31 respectively.   

Additionally, three giant planets discovered by \corot\, CoRoT-10b \citep{Bonomo2010}, CoRoT-37b (Gandolfi et al. {\sl in prep.}), and CoRoT-9b \citep{Deeg2010} have periods in the range 10 to 100 days, with 13.24, 20.05, and 95.27 days respectively. Considering these planets in a single bin gives an occurrence rate of $0.61 \pm 0.36 \%$ assuming here again a detection completeness of 90\% and keeping the same reference stellar sample as for the giants.  While of very low precision, this number is in agreement with the recent estimate of $0.9 \pm 0.24$  \% of \cite{Santerne2016} for the Period-valley giants. We emphasise however that this estimate should be seen as a lower limit. Indeed, the follow-up did not concentrate on single transits and its completeness in this period domain must be lower. As a first order correction attempt, we revised this occurrence rate adding single event candidates and some at longer orbital period whose nature remains unknown and which didn't get more than one flag tune on. To that purpose, we selected those that appear compatible with a giant planet in this range of orbital period with the following criteria: a transit depth between 0.4 and 3.0\% and a duration between 3.0 and 10.0 hours. We selected 6 candidates which match these criteria. This increases the occurrence rate to 1.86 $\pm$ 0.68\%. 

\subsection{Brown dwarf occurrences}
Finally, we obtained an approximate occurrence rate of $0.07 \pm 0.05\%$ for brown dwarfs, from CoRoT-3b and CoRoT-33b only, as CoRoT-15b's magnitude is 15.47, and using the same stellar population and detection completeness as for the hot Jupiters. This value is about four times smaller than the estimate by \cite{Santerne2016} of $0.29 \pm 0.17\%$ for the \kepler\ field, but the latter covers periods up to 400 days of orbital period, while the three \corot\ brown dwarfs all have $P<6$\,days. In addition, this number assumes a follow-up completeness of 100\% which might not reflect the reality and may explain part of the differences. However, our results indicate that short-period brown dwarfs are significantly less numerous than hot Jupiters in the same period range, a trend already noticed by  \cite{Csizmadia2015} and \cite{Santerne2016}.

\section{Summary}
\label{sec:concl}

We provide the full catalogue of all the transit-like events identified in the \corot\ light curves during the mission lifetime by the \corot\ collaboration. It includes planet candidates that were identified after the release of each run, their status once follow-up observations were completed, and also binaries. The later were separated in two groups: detached eclipsing binaries and contact binaries. All these events, whose detection was performed by different teams in a first stage, have been re-analysed with a unified procedure and using the same softwares so that to derive, in a homogeneous and consistent way, the complete set of transit parameters, using fixed limb-darkening coefficients because of the uncertainties on the spectral classification. In addition to this modelling, we also carried out a vetting process of both candidates and binaries. It is based on a simple binary flag system over basic tests: detection significance, presence of a secondary, odd or even depth differences, colour dependence, V-shape transit, and duration of the transit. Among the \totalfeat\ transit-like signals that were analysed, it allowed us to identify periodic signals that are instrumental false positives and false detections. In total, the presented catalogue comes up with \totalCandfilt\ events initially flagged as planet candidates which include published planets and brown dwarfs, \totalEBfilt\ EBs, \totalCB\ contact binaries, and \totalFA\ false alarms. For all candidates, including published planets or brown dwarfs and those whose status was later resolved through subsequent detailed analyses of their light curves or follow-up observations, the catalogue provides the complete set of transit fitted parameters with their associated errors, the validation assessment, and a summary of the outcome of follow-up observations when they were carried out.  For the detached eclipsing binaries,  the catalogue provides the basic transit fitted parameters and the validation assessment as for the planet candidates, and it indicates those that are clear eccentric binaries. For the contact binaries, only the basic parameters are given: ephemeris: transit epoch and period, and an estimate of the amplitude of the modulated signal. Finally, the catalogue provides the list of false alarms identified through a careful analysis of \corot\ light curves. For the unresolved candidates, we expect that the provided flagging will help in deciding which may be the highest  priority in future observations or in tuning automated classification softwares.

The \corot\ exoplanet programme was supported by a large accompanying ground-based observation programme, whose completion required several years for some of the candidates. Owing to the limited amount of telescope time available and to the relatively faint nature of the target stars of  \corot's exoplanet programme, we purposely did not carry out the follow-up observations of all candidates but concentrated efforts on the brightest ones, typically those brighter than the 15th magnitude. We resolved the nature of about two third of the candidates but could not firmly conclude for the remaining. 

Follow-up completeness is close to 100\% for the brightest candidates and about 78\% in the range $r$-mag 14 to 15.  
As a result of both follow-up and light curves detailed analysis, all instances of EB account for 89.5\% of the candidates whose nature has been resolved, which gives a false positive rate close to 90\%. \totalunres\ candidates remain unresolved among which \totalunresfl\ have some chances being of planetary nature. For the latter, the limitation comes from the target's magnitude that does not allow a proper characterisation of the transiting body. They may be in the giant domain but around a faint star, or in the small-size one around a bright star. A simple scaling of our analysis of the candidates whose nature has been secured, allows us to estimate that about eight (within an error of three) planets among them  should still be still confirmable.  Although a slightly higher number of dwarfs in the anticentre stellar fields suggests at least similar occurrence rates in both directions as for the EB, we found that \corot\ planets are detected at a 50\% higher rate in the centre direction than in the anticentre one. 
The complexity and changes in time of the detection and candidates screening process, together with the lack of precise and complete parameters of the underlying stellar populations prevent us from drawing firm conclusions. This abundance behaviour could be related to different properties of stellar populations in the two opposite galactic directions that \corot\ probed. It could also simply be related to a detection favoured by a much longer duration on average of the runs in the centre direction.

Finally, using this first complete catalogue, we attempted to provide estimates of the occurrence rates. 
By distinguishing the various populations in the \corot\ detections, we find that the occurrence of small-size planets with $R_p < 5\RE$ orbiting GKM dwarfs within ten days is 0.46 $\pm$ 0.28 \%. Including the best candidates in the small-size domain in the occurrence rate estimate, assuming these candidates are planets that could not be confirmed by radial velocity observations, increases the occurrence rate to 2.2 $\pm$ 0.6 \%. This still a very low occurrence rate compared to \kepler\ estimates (7.43 $\pm$ 0.5 \%) confirms the disagreement previously obtained by \cite{Bonomo2012} from the analysis of six \corot\ fields only.  Either small-size planets escaped the detection process likely due to unproperly corrected instrumental noises or this discrepancy points to a trend in the planet formation process as a function of the Galactic properties. Indeed, \corot\ fields and the \kepler\ one probe very different Galactic regions. Other sources of discrepancy might be the poor characterisation of the stellar population in the various \corot\ fields but also the lack of an unique magnitude reference system between the two mission targets catalogues.

For giant planets the situation is reversed: the occurrence rate of hot Jupiters (P $< 10$ days) in the \corot\ fields is 0.98 $\pm$ 0.26\% that is about twice the estimate derived by \cite{Santerne2016} for the \kepler\ field based on planet candidates follow-up observations. For longer orbital periods, between 10 and 100 days, the occurrence rate of the secured planets in this range is 0.61 $\pm$ 0.36 \%. As for the small size planets, to account for the likely incompleteness of follow-up observations in this period domain, observations that have led the \corot\ planets identification process, we revised this value including our best candidates, single or periodic. This increases the occurrence rate to 1.86 $\pm$ 0.68\%, a value which is more in agreement with previous studies \citep{Mayor2011,Fressin2013,Santerne2016} and support the idea that some planets are still awaiting for confirmation in the candidates list. Assuming the separation between giant planets and brown dwarfs is real, we finally derived a brown dwarf occurrence rate of 0.07 $\pm$ 0.05 \%
 
We do, however, emphasise that these planet occurrence calculations are first order estimates for different reasons. First, this paper presents results that have been achieved before the final processing of all the \corot\ light curves in their version N2-4.4. Not only does this release provide an homogeneous processing of the light curves but it also corrects for the discontinuities produced by hot pixels and pointing displacements, and for the systematics \citep{Ollivier2016,Guterman2016}, effects that may have prevented the detection of small planets \citep{Bonomo2012}. Second, the detections are the product of different software programmes that have evolved over the years, preventing a precise estimate of the detection completeness for the various planet populations. Third, the follow-up process has led the targets selection and somehow biased the planets identification process. Its incompleteness may lead us to underestimate some of the occurrences. Finally, one of the main issues when assessing occurrence rates is not only the candidates detection and the planet validation but also the good knowledge we have of the stellar population that is observed. For the faint \corot\ targets that were observed in various pointings whose selection was decided according to a wide diversity of criteria that changed during the mission lifetime, the incomplete knowledge of \corot\ targets remains a limitation, despite recent improvements of the targets spectral classification. In particular, the stellar multiplicity is completely ignored in the targets characterisation process. The forthcoming release of GAIA catalogues and in particular of the binaries and the parallaxes will certainly greatly improve the situation. Using them to characterise the stellar population in both \corot\ and \kepler\ stellar fields would provide stellar reference samples with accurate parameters and homogeneously determined. This, completed by an homogeneous reanalysis of the recently released \corot\ light curves, including the transit detections is now required to check for any dependency of planet occurrences with the stellar Galactic properties as suggested from the current comparison with \kepler\ ones. In the future, we expect that the upcoming TESS and PLATO missions will answer this question on a much larger scale.

\begin{acknowledgements}
The French team thanks the CNES for its continuous support on the \corot\ Exoplanet programme, that included grants 0879, 98761, 426808, and 251091. The authors wish to thank the staff at OHP for their support during their numerous nights on SOPHIE, the staff at ESO La Silla Observatory for their support and for their contribution to the success of the HARPS project and operation. The team at the IAC acknowledges support by grants ESP2007-65480-C02-02 and AYA2010-20982-C02-02, AYA2012-39346-C02-02 and ESP2015-65712-C5-4-R, all of the Spanish Secretary of State for R\&D\&i (MINECO). The CoRoT/Exoplanet catalogue (Exodat) was made possible by observations collected for years at the Isaac Newton Telescope (INT), operated on the island of La Palma by the Isaac Newton group in the Spanish Observatorio del Roque de Los Muchachos of the Instituto de Astrophysica de Canarias. The German CoRoT team (TLS and University of Cologne) acknowledges DLR grants 50OW0204, 50OW0603, and 50QP0701. The Swiss team acknowledges the ESA PRODEX programme and the Swiss National Science Foundation for their continuous support on CoRoT ground follow-up.. S. Aigrain acknowledges STFC grant ST/G002266 Office in the form of a Return Grant.
\end{acknowledgements}

\bibliographystyle{aa.bst}
\bibliography{MegaPaper}
\newpage 
\begin{table}
\caption{Discarded transit-like signals. FD: false detection, G: ghost signal; V: stellar variability.  {\it (Sample only, the full table is given on line)}}
\begin{tabular}{rcc}
\hline
\hline
 CoRoT-ID & Run & Type  \\
 \hline
\hline
 102855348 &                  IRa01 & FD \\
 102814334 &                  IRa01 & FD \\
 102823343 &                  IRa01 & FD \\
 102870852 &                  IRa01 & FD \\
 102817472 &                  IRa01 & FD \\
 102940723 &                  IRa01 & FD \\
 102973070 &                  IRa01 & FD \\
 102913574 &                  IRa01 & FD \\
 102939944 &                  IRa01 &  G \\
 102855409 &                  IRa01 &  G \\
 102835817 &                  IRa01 &  G \\
 102805893 &                  IRa01 &  G \\
 102777119 &                  IRa01 &  G \\
 102588918 &                  LRa01 & FD \\
 102776522 &                  LRa01 & FD \\
 102717012 &                  LRa01 & FD \\
 102733319 &                  LRa01 & FD \\
 102754051 &                  LRa01 & FD \\
 102790592 &            LRa01 LRa06 & FD \\
 102746008 &                  LRa01 & FD \\
 102682858 &            LRa01 LRa06 & FD \\
 102692686 &                  LRa01 & FD \\
 102633553 &                  LRa01 & FD \\
 102663890 &                  LRa01 & FD \\
 102709133 &                  LRa01 & FD \\
 102634420 &            LRa01 LRa06 &  G \\
 102613782 &                  LRa01 & FD \\
 102754163 &                  LRa01 &  G \\
 102617334 &                  LRa01 &  G \\
 102609031 &            LRa01 LRa06 & FD \\
 102584786 &            LRa01 LRa06 & FD \\
 102595682 &            LRa01 LRa06 & FD \\
 102582070 &            LRa01 LRa06 & FD \\
 102795835 &      LRa01 IRa01 LRa06 & FD \\
 102717803 &            LRa01 LRa06 & FD \\
 102779171 &      LRa01 IRa01 LRa06 &  G \\
 102697826 &            LRa01 LRa06 &  G \\
 102674894 &            LRa01 LRa06 & FD \\
 102577194 &                  LRa01 & FD \\
 223923921 &                  SRa01 & FD \\
 500007038 &                  SRa01 & FD \\
 223980568 &                  SRa01 & FD \\
 224005945 &            SRa01 SRa05 & FD \\
  223951052 &                  SRa01 & FD \\
 223979111 &                  SRa01 & FD \\
 223963003 &                  SRa01 & FD \\
 223958368 &                  SRa01 & FD \\
 223932273 &            SRa01 SRa05 & FD \\
 221672091 &                  SRa02 & FD \\
 221703782 &                  SRa02 & FD \\
 221660085 &                  SRa02 & FD \\
 221649080 &                  SRa02 & FD \\
 221658515 &                  SRa02 & FD \\
 221688572 &                  SRa02 & FD \\
 221664141 &                  SRa02 &  G \\
 221625707 &                  SRa02 & FD \\
 221640500 &                  SRa02 & FD \\
 221616045 &                  SRa02 & FD \\
 221649266 &                  SRa02 & FD \\
 221663085 &                  SRa02 & FD \\
 221652902 &            SRa02 LRa07 & FD \\
 221652242 &                  SRa02 & FD \\
 221620551 &                  SRa02 &  G \\
 221627739 &                  SRa02 & FD \\
 110674478 &                  LRa02 & FD \\
 110831843 &                  LRa02 & FD \\
 103013798 &                  LRa02 & FD \\
\hline
\label{tab:discarded}
\end{tabular}
\hfill\
\end{table}

\newpage 
\onecolumn
\begin{landscape}
\begin{table*}
\caption{Planetary candidates parameters and status after follow-up observations or the second step of the vetting process: P: planet; unr: unresolved; SB: spectroscopic binary; PEB: photometric eclipsing binary; CEB: contaminating eclipsing binary; WCCF: wide CCF that indicates the host star is hot and/or is a fast rotator.  {\it (Sample only, the full table is given on line)}} 
\scalebox{0.8}{
\begin{tabular}{llcrrrccccccccc}
\hline
\hline
 CoRoT-ID & Run &  Epoch & Period  & depth  & dur  & R$_p$/R$_\star$  & a/R$_\star$ &b  & rho$_\star$ & r-mag & Spectral & Flags  &  Nature & FU \\
                    &        & (BJD)   &  (days) &   (\%)  &  (h) &                              &                     &     & ($\rho_\odot$) &  & type   &  &  &   \\
\hline
\hline
 100468104 &                 LRc01 &   2454236.053457$\pm$  0.000868 &   1.043600$\pm$0.00001 &     0.460$\pm$    0.009 &     2.004$\pm$    0.040 &    0.0751$\pm$   0.002 &   2.45$\pm$  0.08 &   0.91$\pm$  0.01 &   0.18$\pm$  0.02 & 15.27 &    G2 & 010010 &   PEB &  no \\
 100567226 &                 LRc01 &   2454236.650441$\pm$  0.002378 &   7.711616$\pm$0.00023 &     0.822$\pm$    0.042 &     2.221$\pm$    0.110 &    0.1003$\pm$   0.006 &  14.82$\pm$  1.23 &   0.90$\pm$  0.03 &   0.73$\pm$  0.18 & 14.35 &    K5 & 000000 &    SB & yes \\
 100589010 &                 LRc01 &   2454236.335234$\pm$  0.004144 &   0.780472$\pm$0.00004 &     0.167$\pm$    0.009 &     3.859$\pm$    0.173 &    0.5404$\pm$   0.000 &   1.60$\pm$  0.00 &   1.50$\pm$  0.00 &   0.09$\pm$  0.00 & 15.81 &    K3 & 010011 &   CEB & yes \\
 100609705 &                 LRc01 &   2454236.251308$\pm$  0.015360 &   3.303820$\pm$0.00061 &     0.112$\pm$    0.012 &     6.978$\pm$    0.664 &    0.3625$\pm$  78.144 &   1.71$\pm$ 65.18 &   1.33$\pm$ 79.88 &   0.01$\pm$  0.70 & 13.69 &    G2 & 000001 &   CEB & yes \\
 100654833 &                 LRc01 &   2454236.278830$\pm$  0.003079 &   0.531066$\pm$0.00002 &     0.129$\pm$    0.008 &     2.473$\pm$    0.137 &    0.2790$\pm$ 128.372 &   1.36$\pm$119.91 &   1.24$\pm$133.03 &   0.12$\pm$ 31.54 & 13.56 &    K0 & 001001 &    SB & yes \\
 100725706 &                 LRc01 &   2454233.624803$\pm$  0.000982 &  13.240160$\pm$0.00016 &     1.514$\pm$    0.023 &     2.542$\pm$    0.042 &    0.1483$\pm$   0.023 &  22.36$\pm$  1.10 &   0.92$\pm$  0.05 &   0.85$\pm$  0.13 & 14.76 &    G2 & 000001 &     P & yes \\
 100768215 &                 LRc01 &   2454234.903972$\pm$  0.006775 &   2.940266$\pm$0.00024 &     0.332$\pm$    0.013 &     8.609$\pm$    0.303 &    0.2978$\pm$  24.927 &   1.64$\pm$ 19.46 &   1.23$\pm$ 26.40 &   0.01$\pm$  0.24 & 15.42 &    F5 & 011001 &   unr & yes \\
 100773735 &                 LRc01 &   2454234.735644$\pm$  0.000607 &   4.973831$\pm$0.00004 &     0.992$\pm$    0.005 &     5.252$\pm$    0.028 &    0.0950$\pm$   0.001 &   6.45$\pm$  0.17 &   0.61$\pm$  0.03 &   0.15$\pm$  0.01 & 15.85 &    F5 & 010010 &    SB & yes \\
 100834293 &                 LRc01 &   2454230.666796$\pm$  0.001538 &  11.301173$\pm$0.00019 &     1.607$\pm$    0.051 &     2.387$\pm$    0.067 &    0.1447$\pm$   0.027 &  24.19$\pm$  2.01 &   0.89$\pm$  0.06 &   1.48$\pm$  0.37 & 15.49 &    K4 & 000001 &   PEB & yes \\
 100905839 &                 LRc01 &   2454221.357894$\pm$  0.002770 &  15.741157$\pm$0.00046 &     1.472$\pm$    0.038 &     3.778$\pm$    0.116 &    0.1299$\pm$   0.004 &  19.24$\pm$  0.91 &   0.86$\pm$  0.02 &   0.38$\pm$  0.05 & 15.42 &    F5 & 000000 &   unr & yes \\
 100980128 &                 LRc01 &   2454232.869230$\pm$  0.004602 &   6.001884$\pm$0.00030 &     0.082$\pm$    0.006 &     3.169$\pm$    0.210 &    0.2545$\pm$ 139.191 &   4.05$\pm$  3.14 &   1.22$\pm$143.55 &   0.02$\pm$  0.06 & 14.66 &    K5 & 000101 &   CEB & yes \\
 100982006 &                 LRc01 &   2454220.305756$\pm$  0.002675 &  16.674931$\pm$0.00042 &     0.408$\pm$    0.011 &     4.002$\pm$    0.108 &    0.5690$\pm$   0.000 &  12.76$\pm$  0.00 &   1.50$\pm$  0.00 &   0.10$\pm$  0.00 & 16.03 &    A0 & 000111 &    SB & yes \\
 101026464 &                 LRc01 &   2454235.485009$\pm$  0.001721 &   1.321907$\pm$0.00003 &     0.598$\pm$    0.026 &     1.904$\pm$    0.076 &    0.3692$\pm$  94.674 &   2.85$\pm$ 59.59 &   1.27$\pm$100.55 &   0.18$\pm$ 11.14 & 15.44 &    G0 & 010001 &   PEB & yes \\
 101031117 &                 LRc01 &   2454205.892551$\pm$  0.002615 &  53.794205$\pm$0.00123 &     0.810$\pm$    0.013 &     5.164$\pm$    0.091 &    0.0961$\pm$   0.002 &  50.26$\pm$  2.01 &   0.86$\pm$  0.01 &   0.59$\pm$  0.07 & 15.90 &    G2 & 000100 &    SB & yes \\
 101055792 &                 LRc01 &   2454235.732251$\pm$  0.002908 &   1.938394$\pm$0.00007 &     0.127$\pm$    0.009 &     1.768$\pm$    0.122 &    0.0399$\pm$   0.003 &   4.08$\pm$  0.91 &   0.92$\pm$  0.04 &   0.24$\pm$  0.16 & 15.63 &    G0 & 000000 &   CEB & yes \\
 101086161 &                 LRc01 &   2454232.820704$\pm$  0.000735 &   6.212380$\pm$0.00006 &     0.742$\pm$    0.009 &     2.684$\pm$    0.036 &    0.0892$\pm$   0.001 &  11.93$\pm$  0.44 &   0.82$\pm$  0.02 &   0.59$\pm$  0.07 & 14.49 &    A5 & 000000 &     P & yes \\
 101095286 &                 LRc01 &   2454231.140619$\pm$  0.001680 &  10.105790$\pm$0.00018 &     0.436$\pm$    0.005 &     6.791$\pm$    0.071 &    0.3616$\pm$  19.729 &   4.98$\pm$ 10.95 &   1.28$\pm$ 20.76 &   0.02$\pm$  0.11 & 15.53 &    F8 & 010001 &   CEB & yes \\
 101106246 &                 LRc01 &   2454230.080584$\pm$  0.004697 &  11.324195$\pm$0.00058 &     0.543$\pm$    0.021 &     5.436$\pm$    0.213 &    0.3287$\pm$  27.443 &   7.27$\pm$  9.64 &   1.24$\pm$ 29.38 &   0.04$\pm$  0.16 & 14.76 &    K2 & 000001 &   CEB & yes \\
 101115531 &                 LRc01 &   2454236.513689$\pm$  0.000708 &   0.520945$\pm$0.00000 &     0.237$\pm$    0.005 &     1.800$\pm$    0.032 &    0.3120$\pm$  35.261 &   1.50$\pm$ 30.89 &   1.25$\pm$ 36.89 &   0.17$\pm$ 10.30 & 13.50 &    A0 & 000101 &   unr & yes \\
 101123916 &                 LRc01 &   2454236.493089$\pm$  0.003387 &   1.138595$\pm$0.00005 &     0.349$\pm$    0.013 &     4.755$\pm$    0.152 &    0.3365$\pm$  78.650 &   1.50$\pm$ 71.09 &   1.27$\pm$ 82.63 &   0.03$\pm$  4.93 & 15.32 &    K2 & 001001 &   unr & yes \\
 101175376 &                 LRc01 &   2454233.769489$\pm$  0.005467 &   5.278815$\pm$0.00031 &     0.793$\pm$    0.014 &    15.203$\pm$    0.237 &    0.3752$\pm$  37.259 &   1.89$\pm$ 29.50 &   1.26$\pm$ 39.97 &   0.00$\pm$  0.15 & 15.34 &    A5 & 011001 &   CEB & yes \\
  101186644 &                 LRc01 &   2454199.661244$\pm$  0.010585 &  40.663250$\pm$0.00380 &     0.460$\pm$    0.031 &     6.246$\pm$    0.338 &    0.0612$\pm$   0.002 &  49.68$\pm$  1.74 &   0.00$\pm$  2.78 &   0.99$\pm$  0.10 & 15.91 &    G2 & 000000 &   CEB & yes \\
 101206560 &                 LRc01 &   2454235.792435$\pm$  0.000175 &   1.742992$\pm$0.00000 &     3.236$\pm$    0.006 &     2.259$\pm$    0.039 &    0.1626$\pm$   0.001 &   6.76$\pm$  0.24 &   0.02$\pm$  1.39 &   1.36$\pm$  0.14 & 14.95 &    K5 & 000000 &     P & yes \\
 101206989 &                 LRc01 &   2454236.202786$\pm$  0.002918 &   0.577831$\pm$0.00002 &     0.048$\pm$    0.005 &     1.220$\pm$    0.122 &    0.0271$\pm$   0.002 &   1.61$\pm$  0.24 &   0.96$\pm$  0.02 &   0.17$\pm$  0.07 & 13.35 &    G5 & 011000 &   unr & yes \\
 101218359 &                 LRc01 &   2454236.453673$\pm$  0.001821 &   1.154835$\pm$0.00002 &     0.165$\pm$    0.008 &     1.885$\pm$    0.079 &    0.4077$\pm$  30.492 &   2.06$\pm$ 24.00 &   1.36$\pm$ 31.21 &   0.09$\pm$  3.06 & 15.22 &    F5 & 010001 &   PEB &  no \\
 101259269 &                 LRc01 &   2454236.540337$\pm$  0.009925 &   4.429458$\pm$0.00049 &     0.387$\pm$    0.013 &    14.245$\pm$    0.419 &    0.3572$\pm$  54.821 &   1.64$\pm$ 47.11 &   1.28$\pm$ 57.61 &   0.00$\pm$  0.26 & 13.50 &    A5 & 010001 &   PEB &  no \\
 101332685 &                 LRc01 &   2454225.683591$\pm$  0.005109 &  13.101103$\pm$0.00076 &     0.495$\pm$    0.024 &     4.402$\pm$    0.218 &    0.3563$\pm$ 111.147 &   9.67$\pm$ 74.42 &   1.27$\pm$117.59 &   0.07$\pm$  1.63 & 14.14 &    F5 & 010001 &   CEB & yes \\
 101351899 &                 LRc01 &   2454217.077826$\pm$  0.002647 &  21.970001$\pm$0.00059 &     1.594$\pm$    0.039 &     5.062$\pm$    0.107 &    0.1355$\pm$   0.005 &  23.82$\pm$  0.93 &   0.86$\pm$  0.02 &   0.37$\pm$  0.04 & 14.61 &    G2 & 010100 &    SB & yes \\
 101368192 &                 LRc01 &   2454236.316716$\pm$  0.000513 &   4.256670$\pm$0.00002 &     0.505$\pm$    0.003 &     3.678$\pm$    0.022 &    0.0661$\pm$   0.001 &   8.23$\pm$  0.43 &   0.48$\pm$  0.09 &   0.41$\pm$  0.06 & 13.63 &    A5 & 000000 &     P & yes \\
 101424939 &                 LRc01 &   2454235.599705$\pm$  0.002355 &   2.176965$\pm$0.00006 &     0.193$\pm$    0.009 &     2.403$\pm$    0.108 &    0.3176$\pm$  89.158 &   2.66$\pm$ 46.49 &   1.27$\pm$ 92.58 &   0.05$\pm$  2.79 & 15.36 &    K1 & 010101 &   PEB &  no \\
 101434308 &                 LRc01 &   2454200.676696$\pm$  0.004950 &  79.967743$\pm$0.00297 &     1.448$\pm$    0.015 &    10.748$\pm$    0.132 &    0.1140$\pm$   0.002 &  50.84$\pm$  2.95 &   0.55$\pm$  0.08 &   0.27$\pm$  0.05 & 15.79 &    K0 & 010000 &   PEB & yes \\
 101436549 &                 LRc01 &   2454236.548922$\pm$  0.001160 &   0.734881$\pm$0.00001 &     0.410$\pm$    0.012 &     2.152$\pm$    0.055 &    0.3521$\pm$  62.724 &   1.73$\pm$ 51.35 &   1.27$\pm$ 66.02 &   0.13$\pm$ 11.36 & 14.05 &    K3 & 010001 &   CEB & yes \\
 101439653 &                 LRc01 &   2454228.646801$\pm$  0.000469 &  17.497412$\pm$0.00009 &     4.825$\pm$    0.041 &     3.024$\pm$    0.026 &    0.2333$\pm$   0.007 &  36.77$\pm$  0.54 &   0.80$\pm$  0.02 &   2.17$\pm$  0.10 & 15.64 &    G8 & 000001 &   PEB &  no \\
 101482707 &                 LRc01 &   2454220.386021$\pm$  0.000917 &  39.889950$\pm$0.00033 &     1.712$\pm$    0.013 &     6.127$\pm$    0.046 &    0.3972$\pm$   7.009 &  31.06$\pm$ 19.96 &   1.22$\pm$  7.79 &   0.25$\pm$  0.49 & 13.93 &    K3 & 000001 &    SB & yes \\
 101614469 &                 LRc01 &   2454235.188527$\pm$  0.010845 &   6.865639$\pm$0.00083 &     0.220$\pm$    0.010 &    11.534$\pm$    0.447 &    0.5478$\pm$   0.000 &   2.26$\pm$  0.00 &   1.50$\pm$  0.00 &   0.00$\pm$  0.00 & 15.76 &    G0 & 010011 &   CEB & yes \\
 101653417 &                 LRc01 &   2454233.259854$\pm$  0.002575 &   5.782229$\pm$0.00017 &     0.245$\pm$    0.012 &     2.775$\pm$    0.114 &    0.3355$\pm$  76.063 &   5.79$\pm$ 31.87 &   1.28$\pm$ 79.22 &   0.08$\pm$  1.28 & 14.81 &    F8 & 010001 &   CEB & yes \\
 101675703 &                 LRc01 &   2454236.207777$\pm$  0.002648 &   1.793128$\pm$0.00005 &     0.359$\pm$    0.013 &     3.012$\pm$    0.116 &    0.0644$\pm$   0.003 &   2.80$\pm$  0.29 &   0.87$\pm$  0.03 &   0.09$\pm$  0.03 & 13.91 &    F8 & 000000 &   CEB & yes \\
 101708400 &                 LRc01 &   2454232.770660$\pm$  0.001721 &   4.221396$\pm$0.00008 &     1.298$\pm$    0.012 &     8.308$\pm$    0.076 &    0.1136$\pm$   0.001 &   3.24$\pm$  0.09 &   0.75$\pm$  0.02 &   0.03$\pm$  0.00 & 12.78 &    A5 & 010010 &    SB & yes \\
 102282722 &                 LRa03 &   2455105.449152$\pm$  0.000843 &   3.551001$\pm$0.00003 &     2.060$\pm$    0.035 &     2.557$\pm$    0.040 &    0.1698$\pm$   0.027 &   7.49$\pm$  0.37 &   0.90$\pm$  0.06 &   0.45$\pm$  0.07 & 14.27 &    G5 & 010001 &   unr &  no \\
 102294953 &                 LRa03 &   2455108.210538$\pm$  0.004060 &   0.915302$\pm$0.00004 &     0.071$\pm$    0.007 &     2.021$\pm$    0.192 &    0.2477$\pm$ 191.372 &   1.52$\pm$147.72 &   1.22$\pm$197.05 &   0.06$\pm$ 16.26 & 15.52 &    G8 & 000101 &   CEB & yes \\
 102296945 &                 LRa03 &   2455104.553952$\pm$  0.001298 &   3.874284$\pm$0.00006 &     0.116$\pm$    0.003 &     2.232$\pm$    0.062 &    0.0316$\pm$   0.003 &  11.57$\pm$  4.66 &   0.55$\pm$  0.52 &   1.38$\pm$  1.67 & 14.91 &    F8 & 000000 &   unr & yes \\
 102304094 &                 LRa03 &   2455102.709660$\pm$  0.003074 &   6.191418$\pm$0.00021 &     0.738$\pm$    0.041 &     2.675$\pm$    0.135 &    0.3888$\pm$  83.543 &   8.81$\pm$ 74.09 &   1.28$\pm$ 89.01 &   0.24$\pm$  6.02 & 15.07 &    K3 & 000001 &    SB & yes \\
 102312073 &                 LRa03 &   2455101.288171$\pm$  0.003727 &  36.260544$\pm$0.00132 &     2.756$\pm$    0.112 &     3.481$\pm$    0.145 &    0.1762$\pm$   0.017 &  58.26$\pm$  4.40 &   0.83$\pm$  0.05 &   2.01$\pm$  0.46 & 15.25 &    G5 & 000001 &  WCCF & yes \\
 102327039 &                 LRa03 &   2455103.797122$\pm$  0.000378 &   7.926828$\pm$0.00003 &     1.081$\pm$    0.003 &     7.610$\pm$    0.019 &    0.0951$\pm$   0.000 &   8.19$\pm$  0.13 &   0.33$\pm$  0.04 &   0.12$\pm$  0.01 & 15.68 &    G8 & 010010 &    SB & yes \\
 102347809 &                 LRa03 &   2455075.557353$\pm$  0.004282 &  35.231670$\pm$0.00128 &     1.366$\pm$    0.033 &     6.088$\pm$    0.171 &    0.1182$\pm$   0.003 &  30.77$\pm$  1.93 &   0.82$\pm$  0.03 &   0.31$\pm$  0.06 & 15.50 &    G2 & 000000 &   CEB & yes \\
\hline
\end{tabular}
\label{tab:candidates}
}
\end{table*}

\newpage
\begin{table*}
\caption{Single transit events among the planet candidates.}
\begin{tabular}{llccccccccc}
\hline
\hline
CoRoT-ID & Run &  Epoch & period  & depth  & dur  & r-mag & Spectral & Flags  &  Nature  & FU \\
                &        &  (BJD)   &(days) &  (\%)     &   (h)   &           &     type    &     &     & \\
\hline
 100887662 &             LRc01 &   2454322.270238$\pm$0.000619 &   $\gtrsim$142 &  3.483$\pm$ 0.026 &  11.677$\pm$ 0.082 & 15.93 &    A0 & 000001 & unr &  no \\
 101068850 &             LRc01 &   2454294.506093$\pm$0.000429 &   $\gtrsim$142 &  4.471$\pm$ 0.039 &   9.330$\pm$ 0.083 & 15.82 &    A5 & 000001 &  SB & yes \\
 101157411 &             LRc01 &   2454293.936860$\pm$0.002052 &   $\gtrsim$142 &  0.587$\pm$ 0.080 &   6.220$\pm$ 0.000 & 15.06 &    G2 & 000000 & unr & yes \\
 102387834 &             LRa03 &   2455159.583271$\pm$0.002346 &   $\gtrsim$148 &  0.697$\pm$ 0.034 &   3.039$\pm$ 0.218 & 13.42 &    K3 & 000000 & unr & yes \\
 102802996 &             IRa01 &   2454163.426543$\pm$0.001772 &   $\gtrsim$ 54 &  1.691$\pm$ 0.069 &   3.711$\pm$ 0.141 & 15.56 &    F5 & 000001 & WCC & yes \\
 102822869 &             IRa01 &   2454188.928345$\pm$0.002384 &   $\gtrsim$ 54 &  0.896$\pm$ 0.055 &   2.474$\pm$ 0.000 & 16.08 &    K1 & 000000 & unr &  no \\
 102874481 &             IRa01 &   2454156.802266$\pm$0.000462 &   $\gtrsim$ 54 &  3.212$\pm$ 0.024 &   6.184$\pm$ 0.053 & 14.18 &    K2 & 000000 & WCC & yes \\
 102895957 &             IRa01 &   2454162.953333$\pm$0.001705 &   $\gtrsim$ 54 &  0.706$\pm$ 0.018 &   5.180$\pm$ 0.149 & 12.67 &    K0 & 000101 & WCC & yes \\
 102973379 &             IRa01 &   2454171.736236$\pm$0.002136 &   $\gtrsim$ 54 &  0.865$\pm$ 0.015 &   8.659$\pm$ 0.141 & 14.82 &    B6 & 000000 & unr &  no \\
 104768853 &       LRc05 LRc06 &   2455312.976240$\pm$0.379440 &   $\gtrsim$ 20 &  4.222$\pm$ 0.037 &  10.696$\pm$ 0.085 & 15.37 &    A5 & 000011 & PEB &  no \\
 106015624 &             LRc02 &   2454577.574564$\pm$0.001207 &   $\gtrsim$145 &  2.543$\pm$ 0.027 &   8.927$\pm$ 0.109 & 15.81 &    G2 & 000000 & unr &  no \\
 110659798 &             LRa02 &   2454880.959203$\pm$0.001458 &   $\gtrsim$114 &  0.335$\pm$ 0.007 &   4.787$\pm$ 0.127 & 15.06 &    G2 & 000100 &  SB & yes \\
 211616889 &             SRc01 &   2454216.010564$\pm$0.012741 &   $\gtrsim$ 25 &  0.525$\pm$ 0.019 &  30.708$\pm$ 1.163 & 15.86 &    F8 & 000000 & unr &  no \\
 211621528 &             SRc01 &   2454220.954299$\pm$0.004186 &   $\gtrsim$ 25 &  0.353$\pm$ 0.025 &   6.141$\pm$ 0.365 & 15.50 &    F6 & 000000 & unr &  no \\
 211631779 &             SRc01 &   2454213.818543$\pm$0.002793 &   $\gtrsim$ 25 &  1.963$\pm$ 0.043 &  10.236$\pm$ 0.257 & 15.90 &    G2 & 000000 & unr &  no \\
 211634383 &             SRc01 &   2454215.659145$\pm$0.002352 &   $\gtrsim$ 25 &  3.863$\pm$ 0.012 &  64.785$\pm$ 0.201 & 13.04 &    G8 & 000000 & unr &  no \\
 211641087 &             SRc01 &   2454211.393121$\pm$0.001907 &   $\gtrsim$ 25 &  1.330$\pm$ 0.056 &   3.412$\pm$ 0.192 & 15.06 &    K0 & 000000 & unr &  no \\
 211647475 &             SRc01 &   2454224.349735$\pm$0.003131 &   $\gtrsim$ 25 &  0.707$\pm$ 0.010 &  22.163$\pm$ 0.236 & 15.65 &    A5 & 000000 & WCC & yes \\
 211649312 &             SRc01 &   2454212.404953$\pm$0.004319 &   $\gtrsim$ 25 &  1.280$\pm$ 0.056 &   8.530$\pm$ 0.399 & 14.48 &    G0 & 000000 & unr &  no \\
 211650063 &             SRc01 &   2454227.869986$\pm$0.002899 &   $\gtrsim$ 25 &  1.710$\pm$ 0.032 &  15.354$\pm$ 0.288 & 15.62 &    G2 & 000001 &  SB & yes \\
 211666578 &             SRc01 &   2454208.394234$\pm$0.003307 &   $\gtrsim$ 25 &  1.077$\pm$ 0.027 &  10.236$\pm$ 0.277 & 15.59 &    G2 & 000000 & unr &  no \\
 604178606 &             LRa04 &   2455503.788410$\pm$0.001063 &   $\gtrsim$ 77 &  1.648$\pm$ 0.018 &   6.723$\pm$ 0.085 & 12.40 &    A5 & 000000 &  SB & yes \\
 631571789 &             LRc07 &   2455691.326890$\pm$0.001910 &   $\gtrsim$ 81 &  2.787$\pm$ 0.044 &  10.508$\pm$ 0.186 & 15.87 &    G2 & 000000 &  SB & yes \\
 633483984 &       LRc10 LRc07 &   2454110.229903$\pm$0.004037 &   $\gtrsim$ 20 &  2.057$\pm$ 0.008 &  83.186$\pm$ 0.425 & 15.06 &    G0 & 000100 & PEB & yes \\
\hline
\end{tabular}
\label{tab:mono}
\end{table*}
\newpage

\begin{table*}
\caption{Likely detached eclipsing binaries of which only a single eclipse was observed. The last column indicates whether a secondary was observed (EB+sec) or not (EB).  {\it (Sample only, the full table is given on line)}}
\begin{tabular}{llcccccccc}
\hline
\hline
 CoRoT-ID & Run &  Epoch & period  & depth  & dur  & r-mag & Spectral & Flags  &  Nature  \\
                 &         & (BJD)  &  (days)  &   (\%)   &  (h)  &          &     type    &           &                \\
\hline
\hline
 102919036 &             IRa01 &  2454174.4537940$\pm$0.004703 &   $\gtrsim$ 54 &  1.641$\pm$ 0.044 &  16.187$\pm$ 0.432 & 15.60 &   O9V & 000001 &     EB \\
 102819749 &             IRa01 &  2454098.1059830$\pm$0.000232 &   $\gtrsim$ 54 & 13.354$\pm$ 0.052 &   7.419$\pm$ 0.030 & 14.38 &  F5IV & 000000 &     EB \\
 102801672 &             IRa01 &  2454155.1336450$\pm$0.008315 &   $\gtrsim$ 54 &  4.076$\pm$ 0.317 &   7.825$\pm$ 0.565 & 15.28 &  A5IV & 000001 & EB+sec \\
 102829388 &             IRa01 &  2454163.8502930$\pm$0.000776 &   $\gtrsim$ 54 &  8.569$\pm$ 0.055 &  18.552$\pm$ 0.108 & 15.78 &   A5V & 000001 &     EB \\
 102647266 &             LRa01 &  2454465.4379690$\pm$0.013505 &   $\gtrsim$131 &  7.424$\pm$ 0.154 &   5.431$\pm$ 0.146 & 16.56 &   A5V & 000001 &     EB \\
 102574444 &             LRa01 &  2454422.3187830$\pm$0.000420 &   $\gtrsim$131 &  6.159$\pm$ 0.066 &   5.432$\pm$ 0.077 & 13.97 &  F8IV & 000001 &     EB \\
 102582649 &       LRa01 LRa06 &  2454510.9758390$\pm$0.000244 &   $\gtrsim$131 & 17.030$\pm$ 0.151 &   5.432$\pm$ 0.051 & 15.85 &   F6V & 000001 &     EB \\
 102586624 &       LRa01 LRa06 &  2454473.1756410$\pm$0.000548 &   $\gtrsim$131 & 15.797$\pm$ 0.117 &  10.864$\pm$ 0.092 & 14.99 &  A5IV & 000001 &     EB \\
 300003789 &             LRa02 &  2454842.0261040$\pm$0.000189 &   $\gtrsim$114 & 11.874$\pm$ 0.117 &   4.671$\pm$ 0.053 & 14.44 &  F8IV & 000001 &     EB \\
 110851732 &             LRa02 &  2454865.9032800$\pm$0.000592 &   $\gtrsim$114 &  7.685$\pm$ 0.033 &  14.013$\pm$ 0.057 & 15.92 &   A5V & 000000 &     EB \\
 110679817 &             LRa02 &  2454811.8894470$\pm$0.000216 &   $\gtrsim$114 &  7.400$\pm$ 0.040 &   4.787$\pm$ 0.030 & 15.62 &  G2IV & 000000 &     EB \\
 110827324 &             LRa02 &  2454875.3730270$\pm$0.000863 &   $\gtrsim$114 &  6.005$\pm$ 0.057 &  11.966$\pm$ 0.103 & 14.67 &   G8V & 000001 &     EB \\
 110680774 &             LRa02 &  2454824.9235410$\pm$0.000337 &   $\gtrsim$114 &  6.894$\pm$ 0.020 &  14.360$\pm$ 0.042 & 15.29 &   A5V & 000000 & EB+sec \\
 110851928 &             LRa02 &  2454825.3592330$\pm$0.007684 &   $\gtrsim$114 & 10.926$\pm$ 0.161 &   4.786$\pm$ 0.079 & 14.38 &  A5IV & 000001 &     EB \\
 102299893 &             LRa03 &  2455234.1879060$\pm$0.000396 &   $\gtrsim$148 & 37.611$\pm$ 0.116 &  15.195$\pm$ 0.046 & 15.40 &   A7V & 000001 & EB+sec \\
 605308520 &             LRa04 &  2455498.7186640$\pm$0.001499 &   $\gtrsim$ 77 &  3.872$\pm$ 0.041 &  15.132$\pm$ 0.150 & 14.48 &   G0V & 000001 &     EB \\
 605087784 &             LRa04 &  2455516.0710810$\pm$0.001359 &   $\gtrsim$ 77 & 12.866$\pm$ 0.085 &  15.131$\pm$ 0.110 & 15.22 &   G2V & 000001 & EB+sec \\
 605144096 &             LRa04 &  2455499.1986420$\pm$0.000278 &   $\gtrsim$ 77 & 20.975$\pm$ 0.117 &  10.087$\pm$ 0.051 & 15.44 &   A0V & 000001 &     EB \\
 602013995 &             LRa05 &  2455585.4083140$\pm$0.000297 &   $\gtrsim$ 90 &  9.620$\pm$ 0.049 &   5.786$\pm$ 0.041 & 15.08 &   F5V & 000000 &     EB \\
 738643355 &             LRa07 &  2456209.4010750$\pm$0.001202 &   $\gtrsim$ 29 & 32.354$\pm$ 0.073 &  11.312$\pm$ 0.039 & 14.30 & G5III & 000000 & EB+sec \\
 737917418 &             LRa07 &  2456216.9849030$\pm$0.001026 &   $\gtrsim$ 29 &  1.301$\pm$ 0.052 &   3.016$\pm$ 0.104 & 15.25 &   O9V & 000101 &     EB \\
 737553986 &             LRa07 &  2456187.1860230$\pm$0.278301 &   $\gtrsim$ 29 &  3.131$\pm$ 0.115 &   1.589$\pm$ 0.000 & 15.50 & G0III & 000000 &     EB \\
 738645307 &             LRa07 &  2456204.8067820$\pm$0.014339 &   $\gtrsim$ 29 & 33.847$\pm$ 0.160 &  11.681$\pm$ 0.050 & 13.90 &   A7V & 000011 &     EB \\
 221635112 &       LRa07 SRa02 &  2456217.8125120$\pm$0.002304 &   $\gtrsim$ 20 &  1.594$\pm$ 0.060 &   5.279$\pm$ 0.131 & 13.68 &  G2IV & 000000 &     EB \\
 737198603 &             LRa07 &  2456217.0058110$\pm$0.004343 &   $\gtrsim$ 29 &  2.429$\pm$ 0.151 &   6.787$\pm$ 0.392 & 14.47 &  G0IV & 000001 &     EB \\
 737552479 &             LRa07 &  2456219.4950660$\pm$0.000872 &   $\gtrsim$ 29 & 14.613$\pm$ 0.154 &   8.295$\pm$ 0.083 & 15.25 &  F5IV & 000001 & EB+sec \\
 737198435 &             LRa07 &  2456230.0395590$\pm$0.001043 &   $\gtrsim$ 29 & 15.634$\pm$ 0.090 &   9.426$\pm$ 0.048 & 15.32 &  G0IV & 000001 &     EB \\
 737556177 &             LRa07 &  2456226.1645000$\pm$0.001026 &   $\gtrsim$ 29 & 19.965$\pm$ 0.268 &  13.197$\pm$ 0.092 & 15.17 &  F8IV & 000001 &     EB \\
 737916449 &             LRa07 &  2456208.5834500$\pm$0.003738 &   $\gtrsim$ 29 &  5.258$\pm$ 0.245 &   4.525$\pm$ 0.137 & 15.54 &  B6IV & 000000 &     EB \\
 221649269 &       LRa07 SRa02 &  2456215.2371030$\pm$0.004336 &   $\gtrsim$ 20 &  1.474$\pm$ 0.103 &   6.033$\pm$ 0.408 & 15.43 &  A5IV & 000000 &     EB \\
 737200242 &             LRa07 &  2456212.6561430$\pm$0.001559 &   $\gtrsim$ 29 &  7.726$\pm$ 0.197 &   6.033$\pm$ 0.128 & 15.10 &   A0V & 000001 &     EB \\
 737916519 &             LRa07 &  2456229.0730350$\pm$0.003177 &   $\gtrsim$ 29 &  3.347$\pm$ 0.106 &   6.033$\pm$ 0.237 & 15.29 &   A7V & 000000 &     EB \\
 100758671 &             LRc01 &  2454238.0985440$\pm$0.000542 &   $\gtrsim$142 & 12.859$\pm$ 0.038 &   8.756$\pm$ 0.034 & 15.48 &  A5IV & 000000 & EB+sec \\
 101044188 &             LRc01 &  2454279.1111380$\pm$0.005101 &   $\gtrsim$142 &  1.939$\pm$ 0.050 &   8.759$\pm$ 0.300 & 15.13 & K0III & 000001 &     EB \\
 100542479 &             LRc01 &  2454120.2318630$\pm$1.565207 &   $\gtrsim$142 &  4.255$\pm$ 0.079 &   8.275$\pm$ 0.150 & 13.97 &   F8V & 000001 &     EB \\
 105921588 &             LRc02 &  2454676.9168170$\pm$0.000243 &   $\gtrsim$145 & 35.041$\pm$ 0.093 &  20.831$\pm$ 0.052 & 15.86 &   A5V & 000001 &     EB \\
 310126749 &             LRc03 &  2454980.8846320$\pm$0.001609 &   $\gtrsim$ 89 & 29.685$\pm$ 0.203 &   7.616$\pm$ 1.579 & 14.89 &   A0V & 000100 &     EB \\
 310241881 &             LRc03 &  2454929.3697270$\pm$0.002790 &   $\gtrsim$ 89 & 11.648$\pm$ 0.226 &  13.330$\pm$ 0.266 & 15.39 &   B1V & 000001 & EB+sec \\
 104232556 &             LRc04 &  2455043.7478260$\pm$0.001310 &   $\gtrsim$ 84 &  3.529$\pm$ 0.053 &  10.845$\pm$ 0.137 & 15.15 & G5III & 000001 &     EB \\
 104223201 &             LRc04 &  2455063.8665080$\pm$0.003248 &   $\gtrsim$ 84 &  2.488$\pm$ 0.086 &   9.037$\pm$ 0.330 & 14.66 &   G2V & 000001 &     EB \\
 105592930 &             LRc05 &  2455354.9217980$\pm$0.001795 &   $\gtrsim$ 87 &  7.333$\pm$ 0.058 &   9.337$\pm$ 0.063 & 15.16 &  F8IV & 000001 &     EB \\
 104476149 &             LRc06 &  2455430.3165200$\pm$0.000338 &   $\gtrsim$ 77 & 29.740$\pm$ 0.091 &  16.781$\pm$ 0.043 & 15.99 &  F8IV & 000101 & EB+sec \\
 104726248 &       LRc06 LRc05 &  2455426.1837940$\pm$0.001351 &   $\gtrsim$ 20 &  3.303$\pm$ 0.110 &   3.356$\pm$ 0.107 & 15.64 &  F8IV & 000001 &     EB \\
 223956963 &             SRa01 &  2454543.8735990$\pm$0.001571 &   $\gtrsim$ 23 &  8.474$\pm$ 0.115 &  11.224$\pm$ 0.142 & 14.73 &   M2I & 000000 &     EB \\
 223929762 &       SRa01 SRa05 &  2454547.1971660$\pm$0.000574 &   $\gtrsim$ 20 &  5.798$\pm$ 0.111 &   3.848$\pm$ 0.074 & 13.84 &  A5IV & 000001 &     EB \\
 224015696 &             SRa01 &  2454546.5039710$\pm$0.000215 &   $\gtrsim$ 23 &  3.403$\pm$ 0.045 &   2.564$\pm$ 0.035 & 15.50 &  A5IV & 000101 &     EB \\
\hline
\label{tab:monoEB}
\end{tabular}
\end{table*}
\newpage 
\begin{table*}
\caption{Detached eclipsing binaries parameters and their categories (last column): 0.0 a secondary is detected at phase 0.5, 0.1 a secondary is seen but not at phase 0.5, 0.2 no secondary is detected or seen (see sect.~3.4)  {\it (Sample only, the full table is given on line)}}
\begin{tabular}{llcrlrrcccc}
\hline
\hline
  CoRoT-ID & Run &  Epoch &  \multicolumn{2}{c}{period}    & depth  & dur  & r-mag & Spectral & Flags  &  Nature  \\
                   &        & (BJD)   & \multicolumn{2}{c}{(days)} &  (\%)   &  (h)      &           &      type   &            &   \\
\hline
 102900859 &                  IRa01 &  2454131.6769320$\pm$0.000102 &   4.853439 & $\pm$0.000014 &   9.703$\pm$  0.020 &  8.620$\pm$0.015 & 15.27 &    F6V & 010011 & 0.0 \\
 102902696 &                  IRa01 &  2454134.6788270$\pm$0.000482 &   1.980958 & $\pm$0.000030 &  31.816$\pm$  0.151 &  4.469$\pm$0.057 & 14.35 &   A5IV & 010001 & 0.0 \\
 102844383 &                  IRa01 &  2454133.9285540$\pm$0.000608 &   1.527205 & $\pm$0.000027 &   0.855$\pm$  0.008 &  2.639$\pm$0.027 & 15.81 &   F8IV & 000000 & 0.0 \\
 102896719 &                  IRa01 &  2454134.7506080$\pm$0.000629 &   1.231210 & $\pm$0.000023 &   1.098$\pm$  0.014 &  2.305$\pm$0.018 & 13.69 &   A5IV & 010000 & 0.0 \\
 102836138 &                  IRa01 &  2454133.6538860$\pm$0.000088 &   3.558188 & $\pm$0.000009 &  15.208$\pm$  0.034 &  4.270$\pm$0.008 & 15.42 &   F8IV & 010001 & 0.0 \\
 102774523 &                  IRa01 &  2454129.7628170$\pm$0.000116 &   5.917754 & $\pm$0.000020 &  26.976$\pm$  0.083 &  3.977$\pm$0.012 & 15.49 &    K0V & 010001 & 0.0 \\
 102733170 &                  IRa01 &  2454133.8829030$\pm$0.000142 &   1.979428 & $\pm$0.000008 &   5.390$\pm$  0.011 &  5.701$\pm$0.011 & 15.42 &    A0V & 010001 & 0.0 \\
 102879429 &                  IRa01 &  2454133.8162850$\pm$0.000252 &   4.030543 & $\pm$0.000030 &   6.871$\pm$  0.038 &  2.902$\pm$0.015 & 14.66 &   F8IV & 010001 & 0.0 \\
 102884662 &                  IRa01 &  2454131.2742580$\pm$0.000333 &   3.848470 & $\pm$0.000036 &  34.714$\pm$  0.068 & 13.854$\pm$0.024 & 16.03 &  G0III & 010011 & 0.0 \\
 102824749 &                  IRa01 &  2454131.1953310$\pm$0.000105 &   8.097553 & $\pm$0.000023 &  18.376$\pm$  0.020 &  7.774$\pm$0.008 & 15.44 &    A5V & 010010 & 0.0 \\
 102870524 &                  IRa01 &  2454134.6168130$\pm$0.000286 &   1.867599 & $\pm$0.000016 &   1.112$\pm$  0.014 &  1.165$\pm$0.014 & 15.90 &  G0III & 010001 & 0.0 \\
 102741994 &                  IRa01 &  2454133.4916110$\pm$0.000172 &   4.622043 & $\pm$0.000023 &   8.594$\pm$  0.028 &  4.437$\pm$0.013 & 15.82 &    G2V & 010001 & 0.0 \\
 102785724 &                  IRa01 &  2454134.8991220$\pm$0.000140 &   4.716288 & $\pm$0.000019 &  21.118$\pm$  0.047 &  6.565$\pm$0.013 & 14.49 &    F6V & 010011 & 0.0 \\
 102892869 &                  IRa01 &  2454131.3794870$\pm$0.000735 &   4.075875 & $\pm$0.000081 &   1.542$\pm$  0.014 &  3.913$\pm$0.031 & 14.76 &  G0III & 010101 & 0.0 \\
 102882044 &                  IRa01 &  2454127.8463820$\pm$0.000193 &   9.073544 & $\pm$0.000043 &  22.010$\pm$  0.093 &  3.920$\pm$0.017 & 13.50 &    K2V & 000001 & 0.0 \\
 102793963 &                  IRa01 &  2454134.1519300$\pm$0.000481 &   1.242180 & $\pm$0.000018 &   1.749$\pm$  0.029 &  1.371$\pm$0.022 & 15.06 &    K2V & 010001 & 0.0 \\
 102819692 &                  IRa01 &  2454134.8486840$\pm$0.000053 &   1.382680 & $\pm$0.000002 &  28.895$\pm$  0.038 &  3.916$\pm$0.005 & 15.34 &   F1IV & 010001 & 0.0 \\
 102724646 &                  IRa01 &  2454134.9375770$\pm$0.000099 &   0.352998 & $\pm$0.000001 &   9.378$\pm$  0.029 &  1.542$\pm$0.004 & 15.91 &    K0V & 010001 & 0.0 \\
 102844991 &                  IRa01 &  2454134.4677200$\pm$0.000083 &   1.074246 & $\pm$0.000003 &  15.220$\pm$  0.027 &  2.888$\pm$0.005 & 14.95 &    G2V & 010001 & 0.0 \\
 102842466 &                  IRa01 &  2454133.8322120$\pm$0.000150 &   4.917104 & $\pm$0.000022 &  63.946$\pm$  0.085 & 14.397$\pm$0.017 & 13.50 &    K0V & 010011 & 0.0 \\
 102826984 &                  IRa01 &  2454133.8828030$\pm$0.000056 &   1.476765 & $\pm$0.000002 &  25.243$\pm$  0.037 &  3.544$\pm$0.005 & 14.14 &   A5IV & 010001 & 0.0 \\
 102803023 &                  IRa01 &  2454133.1874480$\pm$0.000150 &   2.320584 & $\pm$0.000010 &  14.903$\pm$  0.032 &  5.458$\pm$0.010 & 15.36 &   A5IV & 010011 & 0.0 \\
 102813089 &                  IRa01 &  2454134.1055700$\pm$0.000063 &   1.306273 & $\pm$0.000002 &  20.790$\pm$  0.023 &  3.198$\pm$0.003 & 14.05 &   A5IV & 010000 & 0.0 \\
 102826074 &                  IRa01 &  2454091.6607660$\pm$0.001122 &  48.577117 & $\pm$0.000704 &  15.541$\pm$  0.086 &  9.327$\pm$0.048 & 15.64 &    A7V & 000001 & 0.0 \\
 102889458 &                  IRa01 &  2454134.2981250$\pm$0.000160 &   2.019839 & $\pm$0.000010 &  21.091$\pm$  0.034 &  6.981$\pm$0.010 & 13.93 &   A5IV & 010011 & 0.0 \\
 102886012 &                  IRa01 &  2454134.5784140$\pm$0.001820 &   1.584459 & $\pm$0.000085 &   0.857$\pm$  0.026 &  2.890$\pm$0.081 & 15.76 &   A5IV & 010001 & 0.0 \\
 102841939 &                  IRa01 &  2454133.4509710$\pm$0.000090 &   2.377641 & $\pm$0.000006 &  29.610$\pm$  0.056 &  3.538$\pm$0.006 & 13.91 &    A0V & 010001 & 0.0 \\
 102801922 &                  IRa01 &  2454133.4725990$\pm$0.001210 &   5.458740 & $\pm$0.000193 &   0.931$\pm$  0.014 &  3.668$\pm$0.053 & 14.27 &    B8V & 000001 & 0.1 \\
 102904593 &                  IRa01 &  2454125.1574550$\pm$0.000277 &  16.896029 & $\pm$0.000128 &  17.949$\pm$  0.083 &  5.677$\pm$0.025 & 14.91 &    G2V & 000001 & 0.1 \\
 102735257 &                  IRa01 &  2454133.2567510$\pm$0.001183 &  23.698302 & $\pm$0.000763 &  11.504$\pm$  0.115 &  4.550$\pm$0.042 & 15.25 &    G8V & 000001 & 0.1 \\
 102931335 &                  IRa01 &  2454134.8680840$\pm$0.000113 &   3.979278 & $\pm$0.000013 &  22.983$\pm$  0.034 &  4.775$\pm$0.006 & 15.68 &   A5IV & 000000 & 0.1 \\
 102760888 &                  IRa01 &  2454134.1962500$\pm$0.002251 &   1.904146 & $\pm$0.000124 &   0.155$\pm$  0.006 &  2.833$\pm$0.101 & 15.50 &    A0V & 000011 & 0.1 \\
 102901962 &                  IRa01 &  2454134.6515990$\pm$0.000472 &   2.444625 & $\pm$0.000035 &   9.990$\pm$  0.037 &  7.275$\pm$0.023 & 14.95 &    G2V & 000011 & 0.2 \\
 102870613 &                  IRa01 &  2454131.7150160$\pm$0.002800 &   7.137310 & $\pm$0.000544 &  55.538$\pm$  0.092 &  8.907$\pm$0.015 & 14.39 &    F6V & 010011 & 0.0 \\
 102943073 &                  IRa01 &  2454137.9825880$\pm$0.000107 &   1.644106 & $\pm$0.000006 &   4.722$\pm$  0.016 &  2.683$\pm$0.010 & 15.77 &    B8V & 010101 & 0.0 \\
 102912741 &                  IRa01 &  2454137.2699780$\pm$0.000225 &   1.245128 & $\pm$0.000008 &  10.081$\pm$  0.039 &  2.809$\pm$0.010 & 15.28 &    A5V & 010001 & 0.0 \\
 102879375 &                  IRa01 &  2454137.8072520$\pm$0.000051 &   0.977191 & $\pm$0.000002 &   7.804$\pm$  0.013 &  2.767$\pm$0.004 & 13.42 &   F5IV & 010001 & 0.0 \\
 102811578 &                  IRa01 &  2454137.3639640$\pm$0.000112 &   1.668697 & $\pm$0.000006 &   2.627$\pm$  0.008 &  2.884$\pm$0.008 & 15.57 &   A5IV & 010011 & 0.0 \\
 102872646 &                  IRa01 &  2454137.0472510$\pm$0.000274 &   1.882877 & $\pm$0.000016 &   3.489$\pm$  0.014 &  4.609$\pm$0.016 & 15.63 &   F8IV & 010011 & 0.0 \\
 102806377 &                  IRa01 &  2454134.4150360$\pm$0.000218 &   3.816674 & $\pm$0.000023 &  11.265$\pm$  0.012 &  4.397$\pm$0.086 & 14.25 &   A5IV & 000000 & 0.1 \\
 102790392 &                  IRa01 &  2454134.9865150$\pm$0.000164 &   4.910225 & $\pm$0.000024 &  14.116$\pm$  0.040 &  6.128$\pm$0.016 & 15.25 &    G8V & 010011 & 0.0 \\
 102828417 &                  IRa01 &  2454137.7317330$\pm$0.000316 &   9.594495 & $\pm$0.000096 &  11.402$\pm$  0.049 &  6.448$\pm$0.025 & 15.41 &    G8V & 010011 & 0.0 \\
 102776386 &                  IRa01 &  2454137.2178760$\pm$0.001072 &   2.206876 & $\pm$0.000072 &   5.724$\pm$  0.036 &  4.343$\pm$0.031 & 15.89 &    A5V & 010000 & 0.0 \\
 102853429 &                  IRa01 &  2454136.8619940$\pm$0.000949 &   1.638118 & $\pm$0.000048 &   1.558$\pm$  0.020 &  3.538$\pm$0.042 & 15.82 &    A5V & 010101 & 0.0 \\
 102932176 &                  IRa01 &  2454137.3148150$\pm$0.000202 &   0.872226 & $\pm$0.000006 &  22.059$\pm$  0.040 &  3.935$\pm$0.013 & 15.62 &    A7V & 010001 & 0.0 \\
 102818428 &                  IRa01 &  2454131.2076630$\pm$0.000232 &   7.455240 & $\pm$0.000045 &   5.461$\pm$  0.019 &  6.799$\pm$0.021 & 14.31 &   A5IV & 010011 & 0.0 \\
 102849586 &                  IRa01 &  2454124.1927030$\pm$0.047179 &  29.134695 & $\pm$0.030159 &   0.372$\pm$  0.036 &  5.034$\pm$0.125 & 15.12 &    G8V & 000000 & 0.1 \\
 102982347 &                  IRa01 &  2454136.8916670$\pm$0.000080 &   2.977598 & $\pm$0.000007 &  11.313$\pm$  0.027 &  4.431$\pm$0.009 & 15.67 &   A5IV & 010001 & 0.0 \\
\hline
\end{tabular}
\label{tab:EBs}
\end{table*}
\newpage 
\begin{table}
\caption{Contact binaries parameters. {\it (Sample only, the full table is given on line)}} 
\begin{tabular}{llcrlrcr}
\hline
\hline
  CoRoT-ID & Run &  Epoch & \multicolumn{2}{c}{period}  & depth & r-mag & Spectral   \\
                   &        & (BJD)    & \multicolumn{2}{c}{(days)}  & (\%)   &  & type   \\
 \hline
 102794135 &                  IRa01 &  2454134.8450650$\pm$0.000006 &   0.263893 &       &   7.474 &  14.14 &   F6V \\
 102846142 &                  IRa01 &  2454134.6527860$\pm$0.000267 &   0.410900 & $\pm$0.000003 &  21.024 &  15.00 &  A5IV \\
 102725806 &                  IRa01 &  2454134.7549890$\pm$0.000016 &   0.341804 & $\pm$0.000001 &  11.416 &  15.22 &   F5V \\
 102897917 &                  IRa01 &  2454134.8277380$\pm$0.000006 &   0.445615 & $\pm$0.000001 &  27.168 &  15.72 &   G2V \\
 102924081 &                  IRa01 &  2454137.8810640$\pm$0.000285 &   0.373480 & $\pm$0.000003 &  33.998 &  12.36 &   A7V \\
 102826085 &                  IRa01 &  2454137.0883200$\pm$0.000021 &   1.026050 & $\pm$0.000008 &   6.208 &  12.96 &   A0V \\
 102808511 &                  IRa01 &  2454137.9855420$\pm$0.000005 &   0.239670 &       &  39.942 &  13.18 & K0III \\
 102955089 &                  IRa01 &  2454137.8401450$\pm$0.000153 &   0.571610 & $\pm$0.000003 &  43.028 &  15.03 &   G2V \\
 102910432 &                  IRa01 &  2454137.9430470$\pm$0.000011 &   0.312172 &       &   6.608 &  14.89 &   F6V \\
 102821683 &                  IRa01 &  2454136.8721520$\pm$0.001160 &   1.810740 & $\pm$0.000065 &   2.723 &  14.77 &   A0V \\
 102961901 &                  IRa01 &  2454137.8887300$\pm$0.000536 &   0.420895 & $\pm$0.000007 &  42.194 &  15.43 &  G2IV \\
 102794063 &                  IRa01 &  2454137.9682360$\pm$0.000241 &   0.381919 & $\pm$0.000003 &  29.092 &  15.57 &   G8V \\
 102806220 &                  IRa01 &  2454137.8889950$\pm$0.000241 &   0.349696 & $\pm$0.000003 &  47.727 &  15.70 &  G2IV \\
 102806409 &                  IRa01 &  2454137.9001890$\pm$0.000398 &   0.647349 & $\pm$0.000008 &  33.212 &  15.50 &   A0V \\
 102861060 &                  IRa01 &  2454134.0349720$\pm$0.000035 &   1.531843 & $\pm$0.000061 &   1.481 &  15.40 &   A0V \\
 102738837 &                  LRa01 &  2454397.3612970$\pm$0.000011 &   0.846498 & $\pm$0.000003 &   2.336 &  13.83 &  A5IV \\
 102745707 &                  LRa01 &  2454397.3160450$\pm$0.000128 &   0.288539 &       &  63.638 &  12.80 & O8III \\
 102594134 &                  LRa01 &  2454396.6893470$\pm$0.000049 &   0.789553 &       &   0.596 &  13.96 &   A7V \\
 102634660 &            LRa01 LRa06 &  2454396.6475250$\pm$0.000019 &   0.850373 & $\pm$0.000001 &  26.130 &  13.69 &  A5IV \\
 102786829 &            LRa01 IRa01 &  2454396.7969970$\pm$0.000088 &   0.751067 & $\pm$0.000012 &   0.669 &  15.03 &  A5IV \\
 102788679 &      LRa01 IRa01 LRa06 &  2454397.3766250$\pm$0.000005 &   0.243943 &       &   8.476 &  13.40 &   F5V \\
 102634388 &            LRa01 LRa06 &  2454396.8577070$\pm$0.000013 &   0.635896 & $\pm$0.000001 &   2.975 &  13.84 &   A5V \\
 102686255 &            LRa01 LRa06 &  2454397.2253130$\pm$0.000387 &   0.319579 & $\pm$0.000002 &   4.120 &  12.94 & G5III \\
 102630432 &            LRa01 LRa06 &  2454397.3402030$\pm$0.000033 &   0.525376 & $\pm$0.000002 &   0.426 &  12.90 &  A5IV \\
 102619636 &            LRa01 LRa06 &  2454395.3626110$\pm$0.001408 &   2.622253 & $\pm$0.000048 &   3.052 &  14.40 &   A5V \\
 102738068 &            LRa01 LRa06 &  2454397.1967580$\pm$0.000010 &   0.362463 & $\pm$0.000001 &   0.911 &  12.95 &  A5IV \\
 102674076 &            LRa01 LRa06 &  2454396.8782410$\pm$0.001148 &   2.437481 & $\pm$0.000036 &   5.028 &  14.47 &   B9V \\
 102773399 &      LRa01 IRa01 LRa06 &  2454397.1120390$\pm$0.000193 &   0.605546 & $\pm$0.000002 &   6.117 &  14.49 &   A5V \\
 102752468 &      LRa01 IRa01 LRa06 &  2454396.9455540$\pm$0.000028 &   1.197437 & $\pm$0.000011 &   0.950 &  15.01 &   A5V \\
 102638060 &            LRa01 LRa06 &  2454397.1716330$\pm$0.000031 &   0.283966 &       &  10.563 &  15.17 &   G2V \\
 102760539 &      LRa01 IRa01 LRa06 &  2454397.2917380$\pm$0.000007 &   0.227565 &       &  17.603 &  15.52 &  G2IV \\
 102661163 &            LRa01 LRa06 &  2454396.9803400$\pm$0.000069 &   1.510652 & $\pm$0.000021 &   0.321 &  13.42 &   A0V \\
 102798366 &            LRa01 IRa01 &  2454135.4774940$\pm$0.000024 &   0.395632 &       &   2.764 &  15.10 &  F8IV \\
 110666679 &                  LRa02 &  2454787.0909840$\pm$0.000009 &   0.358187 &       &   4.982 &  12.82 &   A5V \\
 300002276 &                  LRa02 &  2454786.2503730$\pm$0.001636 &   1.482793 & $\pm$0.000038 &   0.388 &  12.59 &  A5IV \\
 102915357 &                  LRa02 &  2454787.1513500$\pm$0.000158 &   0.541430 & $\pm$0.000010 &   0.109 &  13.53 &  A5IV \\
 110833565 &                  LRa02 &  2454786.7797200$\pm$0.002498 &   1.984232 & $\pm$0.000077 &   1.144 &  13.62 &  F8IV \\
 110753852 &                  LRa02 &  2454787.1485570$\pm$0.004852 &   0.393419 & $\pm$0.000029 &   8.060 &  13.91 &  G0IV \\
 110672896 &                  LRa02 &  2454786.6975110$\pm$0.000492 &   1.198219 & $\pm$0.000101 &   0.070 &  13.91 &  F5IV \\
 110688556 &                  LRa02 &  2454786.3391320$\pm$0.000948 &   1.344077 & $\pm$0.000020 &   1.973 &  13.43 &   F5V \\
  110663705 &                  LRa02 &  2454786.7562770$\pm$0.000014 &   0.808755 & $\pm$0.000002 &  28.583 &  13.63 &  G2IV \\
 110766755 &                  LRa02 &  2454787.1030840$\pm$0.000060 &   0.761902 & $\pm$0.000007 &   0.633 &  13.74 &   A5V \\
 110741580 &                  LRa02 &  2454787.1133000$\pm$0.000409 &   0.644416 &       &   0.834 &  14.12 &   A0V \\
 110750397 &                  LRa02 &  2454787.1746930$\pm$0.000008 &   0.364125 &       &   6.724 &  14.90 & K1III \\
 110834449 &                  LRa02 &  2454786.5668850$\pm$0.000034 &   1.214411 & $\pm$0.000018 &   1.752 &  14.57 & K2III \\
 110756288 &                  LRa02 &  2454786.3727370$\pm$0.000205 &   1.242877 & $\pm$0.000043 &   0.553 &  15.32 & K0III \\
 110844293 &                  LRa02 &  2454786.9822300$\pm$0.000035 &   1.071085 & $\pm$0.000011 &   1.383 &  14.46 &  F2II \\
 300001585 &                  LRa02 &  2454787.1861850$\pm$0.000150 &   0.387078 & $\pm$0.000004 &   0.304 &  13.81 &  G5IV \\
 110656884 &                  LRa02 &  2454786.0741520$\pm$0.000197 &   1.241899 & $\pm$0.000044 &   0.927 &  14.25 &  A5IV \\
\hline
\label{tab:CBs}
\end{tabular}
\end{table}
\end{landscape}
\end{document}